\documentclass[aps,pra,reprint,10pt,twocolumn,superscriptaddress,floatfix]{revtex4-1}

\usepackage{times}
\usepackage{amsmath,amssymb,bm,graphicx}
\usepackage[breaklinks,colorlinks=true,urlcolor=blue,citecolor=blue,linkcolor=blue]{hyperref}
\usepackage{mathtools}
\usepackage{xcolor}
\usepackage{enumitem}
\usepackage{braket}
\usepackage{xspace}

\begin{document}

\title{Dephasing and leakage dynamics of noisy Majorana-based qubits: \\ Topological versus Andreev}

\date{\today}

\author{Ryan V. Mishmash}
\affiliation{Department of Physics, University of California, Berkeley, California 94720, USA}
\affiliation{Department of Physics, Princeton University, Princeton, New Jersey 08540, USA}
\affiliation{Department of Physics and Institute for Quantum Information and Matter, California Institute of Technology, Pasadena, California 91125, USA}
\affiliation{Walter Burke Institute for Theoretical Physics, California Institute of Technology, Pasadena, California 91125, USA}

\author{Bela Bauer}
\affiliation{Station Q, Microsoft Research, Santa Barbara, California 93106, USA}

\author{Felix von Oppen}
\affiliation{Dahlem Center for Complex Quantum Systems and Fachbereich Physik,
Freie Universit{\"a}t Berlin, 14195, Berlin, Germany}

\author{Jason Alicea}
\affiliation{Department of Physics and Institute for Quantum Information and Matter, California Institute of Technology, Pasadena, California 91125, USA}
\affiliation{Walter Burke Institute for Theoretical Physics, California Institute of Technology, Pasadena, California 91125, USA}

\begin{abstract}
Topological quantum computation encodes quantum information nonlocally by nucleating non-Abelian anyons separated by distances $L$, typically spanning the qubit device size. This nonlocality renders topological qubits exponentially immune to dephasing from \emph{all} sources of classical noise with operator support local on the scale of $L$. We perform detailed analytical and numerical analyses of a time-domain Ramsey-type protocol for noisy Majorana-based qubits that is designed to validate this coveted topological protection in near-term devices such as the so-called `tetron' design. By assessing dependence of dephasing times on tunable parameters, e.g., magnetic field, our proposed protocol can clearly distinguish a bona fide Majorana qubit from one constructed from semilocal Andreev bound states, which can otherwise closely mimic the true topological scenario in local probes. In addition, we analyze leakage of the qubit out of its low-energy manifold due to classical-noise-induced generation of quasiparticle excitations; leakage limits the qubit lifetime when the bulk gap collapses, and hence our protocol further reveals the onset of a topological phase transition.  This experiment requires measurement of two nearby Majorana modes for both initialization and readout---achievable, for example, by tunnel coupling to a nearby quantum dot---but no further Majorana manipulations, and thus constitutes an enticing pre-braiding experiment. Along the way, we address conceptual subtleties encountered when discussing dephasing and leakage in the context of Majorana qubits.
\end{abstract}

\maketitle

\section{Introduction} \label{sec:intro}

Developing a scalable quantum computing architecture that can withstand decoherence to the extent necessary for real-world applications poses an enormous scientific and technological undertaking. One path forward pursues Majorana-based topological qubits. These promise to provide robust protection against decoherence at the hardware level by storing quantum information nonlocally, specifically in a fermion-parity degree of freedom in a topological phase \cite{Kitaev01_PhysU_44_131,sarma_majorana_2015}. Since 2012, remarkable experimental progress \cite{mourik_signatures_2012,deng_anomalous_2012,das_zero-bias_2012,churchill_superconductor-nanowire_2013,lee_spin-resolved_2013,finck_anomalous_2013,albrecht_exponential_2016,deng_majorana_2016,zhang_quantized_2018,vaitiekenas_effective_2018,deng_nonlocality_2018,moor_electric_2018,suominen_zero-energy_2017,nichele_scaling_2017} (see Ref.~\cite{lutchyn_majorana_2018} for a recent review) has established the potential existence of Majorana zero modes in devices following the popular recipe \cite{lutchyn_majorana_2010,oreg_helical_2010} of engineering topological superconductivity \cite{alicea_new_2012} by interfacing $s$-wave superconductors with spin-orbit-coupled semiconducting nanowires subjected to modest magnetic  fields.

Most of these laboratory efforts have, however, focused on end-of-wire tunneling conductance spectroscopy, a local measurement which may have difficulty differentiating bona fide well-separated Majorana zero modes in a topological phase \cite{Kitaev01_PhysU_44_131} from near-zero-energy Andreev bound states (ABSs) arising in a trivial phase. One might naively expect that the latter should not `stick' near zero energy upon variation of system parameters such as magnetic field, chemical potential, etc.  Important theoretical and numerical work \cite{kells_near-zero-energy_2012,prada_transport_2012,liu_andreev_2017,moore_two-terminal_2018,setiawan_electron_2017,moore_quantized_2018,liu_distinguishing_2018,vuik_reproducing_2018,penaranda_quantifying_2018,reeg_zero-energy_2018,stanescu_illustrated_2018} has nevertheless shown that seemingly quite reasonable spatial variations in the potentials can trap ABSs---even deep in the trivial phase---that exhibit a high propensity to reside near zero energy (at least within the energy resolution of transport experiments). Andreev states, i.e., ordinary (complex) fermionic modes, can always be described in terms of two Majorana operators.  Oftentimes near-zero-energy ABSs are \emph{semilocal} in that they consist of two Majorana operators separated in space by a finite distance, albeit one much less than the physical wire length \cite{moore_two-terminal_2018,moore_quantized_2018,vuik_reproducing_2018,penaranda_quantifying_2018,stanescu_illustrated_2018}; the corresponding coupling strength between the constituent Majorana operators can thus be very small. This property endows semilocal ABSs with the (rather unfortunate) ability to masquerade as true Majorana zero modes even with respect to more detailed local-probe characteristics such as $2e^2/h$ zero-bias-peak conductance quantization \cite{setiawan_electron_2017,moore_quantized_2018} and $4\pi$ Josephson periodicity \cite{chiu_fractional_2018}. ABSs induced by local potential inhomogeneities arguably describe much of the experimental data as compellingly as true Majorana zero modes.  While states encoded through semilocal ABSs might still furnish qubits with some degree of protection \cite{vuik_reproducing_2018}, such a situation is clearly suboptimal.

Various schemes have been proposed to distinguish Majorana zero modes from Andreev levels: (1) Detecting the bulk topological phase transition accompanying the onset of Majorana modes---e.g., through bulk-gap closure and reopening \cite{stanescu_close_2012,mishmash_approaching_2016}---surely provides an unambiguous fingerprint; see Ref.~\cite{grivnin_concomitant_2018} for an experiment in this direction.  Recently studied multi-terminal N-S-N devices provide an appealing platform for revealing the bulk phase transition via non-local conductance measurements \cite{AndreevRectifier,danon_nonlocal_2019,menard_conductance-matrix_2019,zhang_quantum_2019}.  (2) Majorana modes generate correlations in conductance measured at the two ends of a wire, whereas ABSs localized to each end generically produce no such correlations \cite{moore_two-terminal_2018}. (3) Sensitivity of the energy to local perturbations, such as a quantum dot \cite{clarke_experimentally_2017,prada_measuring_2017,deng_nonlocality_2018} or sharp edge potential \cite{liu_distinguishing_2018} is expected to differ sharply for ABSs and Majorana modes due to the nonlocal, and hence energetically rigid, nature of the latter (see also Ref.~\cite{yavilberg_differentiating_2019}). (4) The current-phase relation in a Cooper-pair transistor has been shown to also differ qualitatively for these two scenarios \cite{schrade_andreev_2018}. (5) On the more challenging end, demonstration of inherently robust non-Abelian braiding operations is possible only with true Majorana zero modes.

We add to this list by analyzing in detail a time-domain protocol that not only unambiguously differentiates ABSs from Majorana modes, but further reveals important device characteristics (notably qubit lifetimes and precise Majorana-hybridization energies), exposes the topological phase transition, and clearly elucidates the enormous benefit of Majorana-based qubits insofar as robustly \emph{protecting} quantum information. The main idea is to use \emph{noise sensitivity} of a prototype qubit to validate (or invalidate) its topological nature. Similar ideas were originally put forth in Sec.~IV of Ref.~\cite{aasen_milestones_2016} in a related context.  
As we will see, in our framework ABSs that arise over a restricted parameter regime before the onset of well-separated Majorana modes in fact become a feature rather than a bug; their presence allows one to benchmark the quality of a bona fide topological qubit within a single device.  

The remainder of this paper is organized as follows.  Section~\ref{sec:executive} provides an executive summary of our main results.  Section~\ref{sec:models} introduces the microscopic models that we employ and also outlines our numerical simulation strategy.  The Ramsey-like protocol that we use to probe the qubit dynamics is detailed in Sec.~\ref{sec:protocol}.  Section~\ref{sec:analytics} analytically estimates dephasing and leakage times for the qubit, while Sec.~\ref{NumericalResults} extracts these quantities from explicit time-dependent numerical simulations.  We conclude with a discussion in Sec.~\ref{sec:discussion}.  Appendices provide supplemental calculational details.  

\begin{figure}[t]
\begin{center} 
\includegraphics[width=0.7\columnwidth]{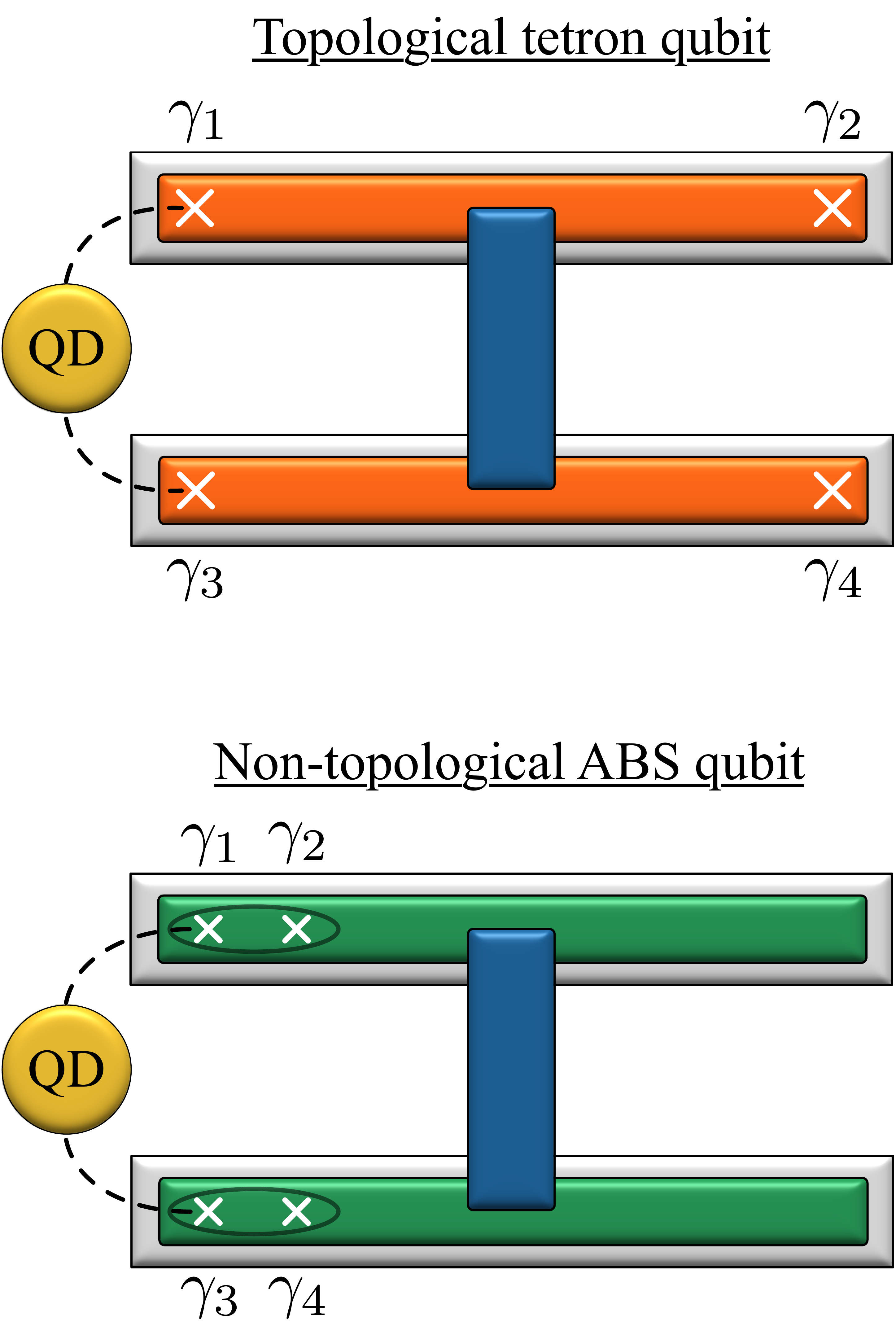}
\end{center}
\caption{Schematics of the topological tetron qubit (top) and a non-topological qubit arising from Andreev bound states (ABSs) residing near zero energy (bottom). Both cases consist of two parallel semiconducting nanowires (gray) proximity coupled to $s$-wave superconductors (orange and green, in the topological and non-topological scenarios, respectively); a trivial superconducting bridge connects the two parallel superconducting segments. As described in the text, the spatial separation of the low-energy Majorana modes $\gamma_{1,2,3,4}$ is qualitatively distinct in the two scenarios. The two leftmost such modes, i.e., $\gamma_1$ and $\gamma_3$, are coupled to a nearby quantum dot (QD), allowing initialization and readout of the respective Majorana parity $i\gamma_1\gamma_3 = \pm1$.
\label{fig:qubits}}
\end{figure}

\section{Executive Summary} \label{sec:executive}

For concreteness, we focus on the so-called `tetron' Majorana qubit design shown in Fig.~\ref{fig:qubits} \cite{plugge_majorana_2017,karzig_scalable_2017,knapp_dephasing_2018} (equivalent for our purposes to the `loop qubit' \cite{knapp_dephasing_2018}). This prototype topological qubit consists of two parallel semiconducting nanowires (gray) each proximitized by an $s$-wave superconductor and tuned to the topological phase with an external magnetic field (orange denotes topological superconductors).  The concomitant four Majorana zero modes---defined through \emph{instantaneous} eigenstates of the microscopic, time-dependent Hamiltonian---are denoted $\gamma_{1,2,3,4}$.  A trivial superconductor (blue) bridges the two wires. In a real physical implementation, the entire mesoscopic device is floating so as to protect against quasiparticle poisoning events from the outside via charging energy \cite{karzig_scalable_2017}.  For the non-topological ABS qubit \footnote{For more parallel terminology one can view the ABS qubit as a superficial, hardly topological tetron (SHT-tetron).} that we wish to contrast against, $\gamma_{1,2,3,4}$ instead represent a maximally localized set of Majorana operators that weakly hybridize at the left end of the device as illustrated in the bottom of Fig.~\ref{fig:qubits}.  Unless specified otherwise our discussion below pertains to both types of qubits.  

Assuming that we are working on time scales much less than the characteristic poisoning time, and additionally that the system is confined to the low-energy, nearly degenerate ground-state manifold with high probability---e.g., the temperature and characteristic noise frequencies are well below the bulk excitation gap (see below)---we can encode logical qubit states by $|0\rangle \equiv |i\gamma_1\gamma_2 = +1\rangle$ and $|1\rangle \equiv |i\gamma_1\gamma_2 = -1\rangle$. [Here global fermion parity is fixed to $(i\gamma_1\gamma_2)(i\gamma_3\gamma_4) = +1$, so that specifying $i\gamma_1\gamma_2$ automatically specifies $i\gamma_3\gamma_4$.] In this basis, we can identify Pauli operators $Z \equiv i\gamma_1\gamma_2$, $X \equiv i\gamma_1\gamma_3$, and $Y \equiv -i\gamma_1\gamma_4$. Suppose that $\gamma_1$ and $\gamma_2$ couple with hybridization energy $\varepsilon_{12}(t)$ that is time-dependent due to noise, and that $\gamma_3$ and $\gamma_4$ similarly couple with energy $\varepsilon_{34}(t)$. The time-averaged couplings decay asymptotically with the intra-wire Majorana separation as an oscillatory exponential: $\varepsilon_{12} \sim \cos(k_\mathrm{top} L_\mathrm{top})e^{-L_\mathrm{top}/\xi_\mathrm{top}}$ and $\varepsilon_{34} \sim \cos(k_\mathrm{bot} L_\mathrm{bot})e^{-L_\mathrm{bot}/\xi_\mathrm{bot}}$, where $k$ is related to the Fermi wave vector of the wire, $\xi$ is the effective superconducting coherence length~\cite{das_sarma_splitting_2012}, and `top'/`bot' refers to the top and bottom wires of the tetron.  For the topological qubit, $L_{\rm top/bot}$ are given by the wire length $L$---thus taking full advantage of the exponential suppression.  By contrast, for the ABS qubit $L_{\rm top/bot}$ is on the scale of the coherence length or smaller, and hence the splitting need not conform to such an exponential. In principle, couplings between Majoranas on different wires, such as $\varepsilon_{14}$, $\varepsilon_{23}$, etc., can also be present. However, since the separation between these Majorana modes  generally exceeds that between Majorana modes on the same wire, these couplings will be significantly smaller than the intra-wire hybridization energies---at least if the superconducting bridge connecting the two wires is sufficiently long and well-gapped. In this case, their primary effect is to reduce the amplitude of qubit oscillations and slightly shift the qubit precession frequency while leaving the qualitative features unaffected. We can therefore safely ignore such subdominant terms and write the Hamiltonian governing the dynamics of the qubit as
\begin{eqnarray}
      H(t) &=& \frac{i}{2}[\varepsilon_{12}(t)\gamma_1\gamma_2 + \varepsilon_{34}(t) \gamma_3\gamma_4]
      \nonumber \\
      &=& \frac{1}{2}[\varepsilon_{12}(t) + \varepsilon_{34}(t)]Z \equiv \frac{1}{2}E(t) Z.
      \label{qubitH}
\end{eqnarray}
Finally, a quantum dot (ignored thus far) sits proximate to the left ends of the nanowires, thereby allowing a joint parity readout \cite{gharavi_readout_2016,karzig_scalable_2017} of nearby Majoranas $\gamma_1$ and $\gamma_3$, i.e., the Pauli $X$ operator.

Our proposed protocol is conceptually very simple: (1) Initialize the system into an $X$ eigenstate by measuring $i\gamma_1\gamma_3$; (2) let the system evolve unitarily for a wait time $t$; and finally (3) re-measure $X$. Assuming the unitary evolution is governed by Eq.~\eqref{qubitH}, which ignores inter-wire Majorana hybridization as well as higher-energy excitations, the qubit precesses about the $z$ axis on the equator of the Bloch sphere. Let us write the instantaneous qubit splitting defined through Eq.~\eqref{qubitH} as $E(t) = \hbar \omega_0 + \delta E(t)$. Here  $\omega_0$ is the \emph{time-averaged} qubit precession frequency while $\delta E(t)$ encodes the effects of classical noise and is responsible for dephasing. Note that we assume here that $\delta E(t)$ fluctuates around zero mean.  Taking the initial state to be an eigenstate of $X = i\gamma_1\gamma_3$ with eigenvalue $+1$, 
noise averaging the $X$ readout measurement gives
\begin{align}
\overline{\langle \psi(t) | X | \psi(t) \rangle} = \cos(\omega_0 t) f(t/T_2), \label{eq:X_decay}
\end{align}
where the envelope function $f(t/T_2)$ decays on a time scale given by the \emph{dephasing time} $T_2$. This experiment therefore simultaneously probes $\omega_0$ (and thus reveals the mean Majorana hybridization energy) as well as $T_2$---both critical device characteristics that differ starkly for topological and ABS qubits. We note that the described protocol is similar in spirit to the so-called Ramsey sequence \cite{vion_manipulating_2002,ithier_decoherence_2005}. The main differences are that we (1) do not include an external driving field transverse to $Z$ and (2) initialize and read out the qubit with projective $X$ measurements in contrast to utilizing $\pi/2$ pulses and $Z$ initialization/readout \footnote{A related sequence involving $Z$ initialization, two $\pi/2$ rotations about the $x$ axis buttressing a wait time $t$, followed by a final $Z$ measurement was proposed in Ref.~\cite{aasen_milestones_2016}, as appropriate for the qubit design presented therein.}.

A particularly advantageous feature of Majorana-based topological qubits is that, due to their inherently nonlocal encoding, the splitting $E$ remains exponentially small---and thus also `exponentially flat'---in response to changes in \emph{all} local Hamiltonian couplings (e.g., chemical potential, Zeeman field, etc.). This property endows topological qubits with exponential protection from all classical noise sources, yielding the following nontrivial scaling relation that connects time-averaged splitting and dephasing \cite{aasen_milestones_2016}:
\begin{align}
\frac1{\omega_0} \sim T_2^a \sim e^{L/\xi}. \label{eq:scaling}
\end{align}
Here $a$ is an order-one number dependent on details of the noise.  In other words, as the topological qubit becomes `perfect', the splitting vanishes and the dephasing time diverges in a correlated fashion.  
For comparison, the transmon qubit \cite{koch_charge-insensitive_2007} benefits similarly from an exponential protection from charge noise, but not other noise sources; additionally, there the qubit splitting need not be exponentially small.  The pre-braiding protocol sketched above and analyzed in detail below is capable of probing the scaling relation in Eq.~\eqref{eq:scaling}---which can for practical purposes be taken as a \emph{definition} of a topologically protected quantum memory. Note also that as the qubit quality improves (i.e., as $L/\xi$ increases) the splitting $\hbar \omega_0$ becomes more difficult to resolve in transport \cite{albrecht_exponential_2016,vaitiekenas_flux-induced_2018} yet easier to resolve in the time-domain measurements employed in our protocol.

The trivial ABS qubit exhibits qualitatively different behavior. Firstly, since the requisite `accidental' low-energy Andreev bound states would emerge due to some local, nonuniversal features in the potential landscapes near the ends of the wires, the ABS qubit would clearly not exhibit exponential protection against arbitrary local noise sources.  Moreover, neither $\omega_0$ nor $T_2$ will vary appreciably with wire length $L$, in sharp contrast to the scaling relation in Eq.~\eqref{eq:scaling} that uniquely identifies the topological qubit.  

While it would be ideal to perform experiments at different $L$, keeping all other parameters fixed in this process has its practical challenges. We nevertheless argue that one can compellingly distinguish the two scenarios in a \emph{single} prototype qubit at fixed $L$ by carrying out the protocol at different field strengths $B$. Numerous numerical simulations \cite{liu_andreev_2017,moore_two-terminal_2018,moore_quantized_2018,vuik_reproducing_2018,penaranda_quantifying_2018,stanescu_illustrated_2018} indicate a tendency for low-energy Andreev bound states to form over an extended field interval below the onset of a topological phase hosting true Majorana zero modes. In such a scenario, upon increasing $B$ from zero, the system first realizes an ABS qubit, then encounters a topological phase transition at a critical field $B_c$, and finally forms a topological qubit.

The behavior of $\omega_0$ and $T_2$ during this evolution reveals a wealth of information. In the ABS regime there is likely no clear universal behavior present, though the measured coherence times would provide a useful baseline.  When $B$ approaches $B_c$, the bulk gap becomes close to zero.  Here even low-frequency noise can efficiently excite the qubit out of the computational subspace, causing a precipitous reduction in the qubit lifetime.  (Below we denote the characteristic `leakage' time associated with such excitations by $T_\mathrm{leak}$.) Hence, our protocol provides a novel means of detecting the topological phase transition. As $B$ increases beyond $B_c$, the gap re-opening and concomitant appearance of robust Majorana zero modes rapidly boost the qubit's coherence time---ideally to values exceeding the ABS-qubit lifetime by orders of magnitude. The scaling relation in Eq.~\eqref{eq:scaling}, which should now be viewed in terms of field-induced variation of $\xi$ at fixed $L$, becomes operative in this topological regime.  The topological-qubit frequency $\omega_0$ additionally exhibits characteristic oscillations with magnetic field, reflecting oscillatory overlap of the Majorana wave functions \cite{das_sarma_splitting_2012}, in turn yielding out-of-phase oscillations in the dephasing time $T_2$ \cite{aasen_milestones_2016}. 
Further increasing $B$ eventually suppresses the gap for the topological phase and hence increases $\xi$, thereby diminishing $1/\omega_0$ and $T_2$ in accordance with Eq.~\eqref{eq:scaling}. Note also that for sufficiently long topological wires, the exponentially long dephasing time $T_2$ is eventually cut off by the leakage time $T_\mathrm{leak}$ (which is set by the $L$-independent excitation gap).

\begin{figure}[t]
\begin{center} 
\includegraphics[width=\columnwidth]{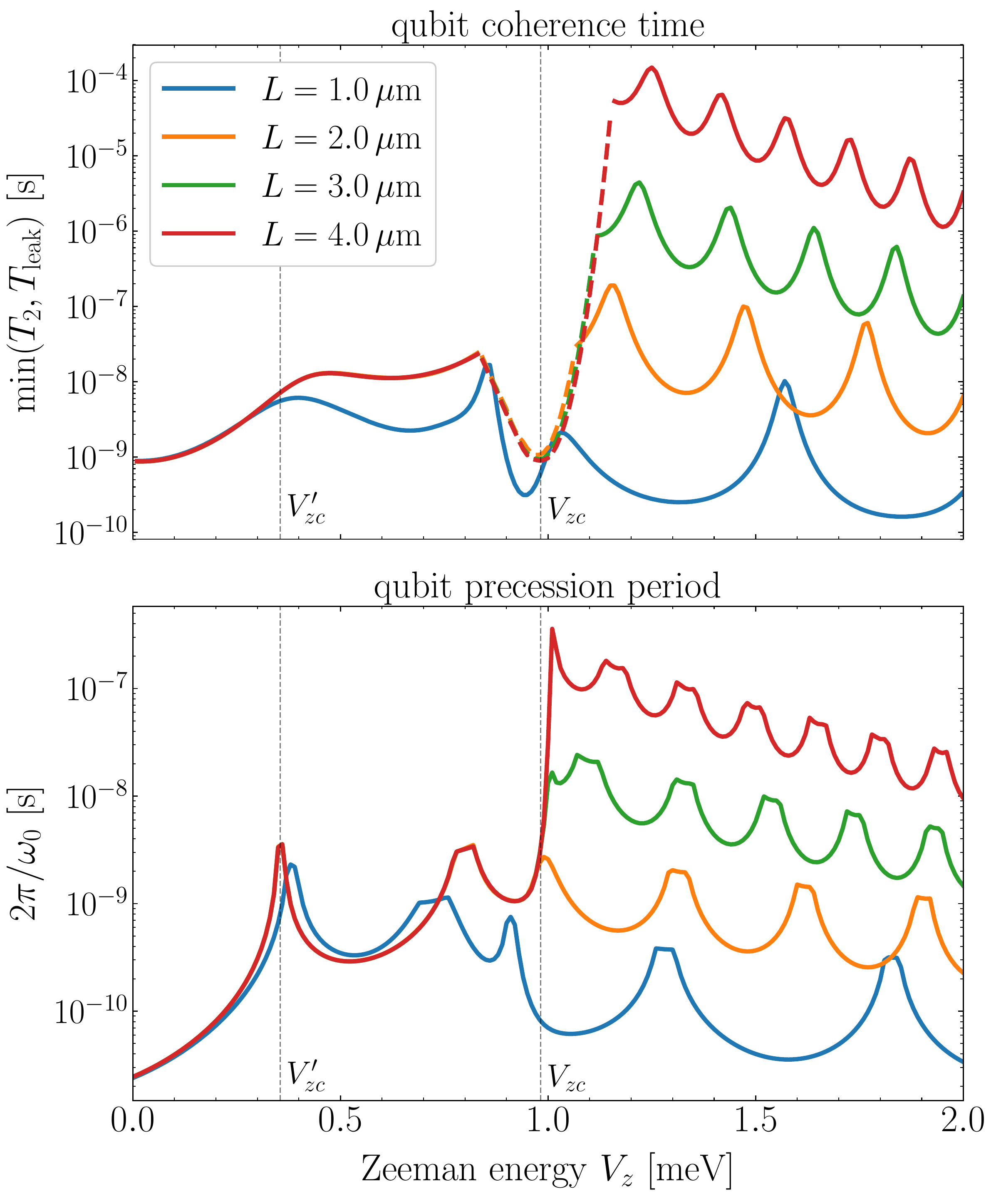}
\end{center}
\caption{Qubit coherence time defined as $\min(T_2, T_\mathrm{leak})$ (top) and qubit precession period $2\pi/\omega_0$ (bottom) as a function of Zeeman energy $V_z$ for tetron qubits constructed from the spinful nanowire model of Eq.~\eqref{eq:nanowire} at different wire lengths $L$. For $V_{zc}' < V_z < V_{zc}$, the qubit is of the non-topological ABS variety (Fig.~\ref{fig:qubits}, bottom panel), while for $V_z > V_{zc}$, the qubit is topological (Fig.~\ref{fig:qubits}, top panel). In the top panel, solid curves indicate that the qubit lifetime is \emph{dephasing limited} ($T_2 < T_\mathrm{leak}$), while dashed curves indicate that the qubit lifetime is \emph{leakage limited} ($T_\mathrm{leak} < T_2$). These calculations were performed using the quasi-analytic methods of Sec.~\ref{sec:analytics}; for more details, including the specific parameters chosen, please see Sec.~\ref{sec:spinful}.
\label{fig:Vzscan_summary}}
\end{figure}

The above points are demonstrated in Fig.~\ref{fig:Vzscan_summary}, where we show concrete calculations of the qubit coherence time defined as $\min(T_2, T_\mathrm{leak})$ and qubit precession period $2\pi/\omega_0$ for noisy tetron qubits modeled within the canonical single-band description of proximitized spin-orbit-coupled nanowires \cite{lutchyn_majorana_2010,oreg_helical_2010}. Here, the horizontal axis denotes the Zeeman energy, related to the magnetic field $B$ through $V_z = \frac12 g \mu_B B$, with $g$ the effective Land\'e $g$-factor (assumed equivalent for both wires of the tetron) and $\mu_B$ the Bohr magneton. 
In these plots, and in all simulations in this paper, we neglect dependence of the noise on magnetic field and other parameters.  We stress that although this assumption will undoubtedly be violated to some degree in experiment, the non-monotonicity and sheer magnitude of the effects illustrated in Fig.~\ref{fig:Vzscan_summary} are unlikely to be modified significantly by such non-universal noise properties.  
For more details of the model Hamiltonian, means of calculation, and chosen physical parameters for the data in Fig.~\ref{fig:Vzscan_summary}, see Secs.~\ref{sec:Hamis}, \ref{sec:analytics}, and \ref{sec:spinful} below.

Many previous studies have explored the influence of noise on Majorana-based qubits from various perspectives (see, e.g., Refs.~\cite{goldstein_decay_2011,rainis_majorana_2012,schmidt_decoherence_2012,pedrocchi_majorana_2015,hu_majorana_2015,pedrocchi_monte_2015,rahmani_optimal_2017,ritland_optimal_2018,knapp_dephasing_2018,knapp_modeling_2018,li_four-majorana_2018,munk_fidelity_2019}).  We emphasize that here we are proposing to use noise to our \emph{advantage} in verifying the topological nature of a prototype Majorana qubit, and as a byproduct are able to extract crucial device characteristics that would otherwise be difficult to obtain. Time-domain experiments of the type that we propose certainly entail significant challenges, but they are arguably a prerequisite for successful demonstration of exponentially accurate braiding gates.

\section{Microscopic models and theoretical formulation} \label{sec:models}

\subsection{Microscopic Hamiltonians} \label{sec:Hamis}

We exploit two different microscopic Hamiltonians for the one-dimensional (1D) topological superconductors comprising our prototype tetron qubits. For a truly minimal description we appeal to Kitaev's model \cite{Kitaev01_PhysU_44_131} for a 1D spinless $p$-wave superconductor, given by the Hamiltonian
\begin{align}
H_{\rm K} = \sum_{i=1}^L\left[-\mu c_i^\dagger c_i - \frac12\left(J c_i^\dagger c_{i+i} + \Delta c_i c_{i+1} + \mathrm{H.c.}\right)\right]. \label{eq:Kitaev}
\end{align}
Here $\mu$ is the chemical potential, $J \geq 0$ is the nearest-neighbor hopping strength, and $\Delta$ is the $p$-wave pairing strength (assumed real). The topological phase occurs for $|\mu| < J$, wherein a chain with open boundary conditions harbors a pair of Majorana zero modes (MZMs), one at each end.

For a more realistic treatment, we consider the one-band spinful model of a Rashba spin-orbit coupled nanowire proximitized with an $s$-wave superconductor and subjected to a Zeeman field \cite{lutchyn_majorana_2010,oreg_helical_2010}.  The Hamiltonian reads
\begin{align}
  H_{\rm NW} &= \int_0^L dx\bigg{[}\psi^\dagger \bigg{(}-\frac{\hbar^2\partial_x^2}{2m} - \mu + V(x) \nonumber \\
  &- i \alpha \sigma^y \partial_x + V_z \sigma^z\bigg{)}\psi + \Delta(x) (\psi_\uparrow \psi_\downarrow + \mathrm{H.c.})\bigg{]},
  \label{eq:nanowire}
\end{align}
where $m$ is the effective electron mass, $\mu$ is the chemical potential, $\alpha$ is the Rashba spin-orbit strength, $V_z = \frac12 g \mu_B B$ is the Zeeman energy,
$\Delta(x)$ is the induced $s$-wave pairing amplitude, and Pauli matrices $\sigma^{y,z}$ act in spin space. 
Note that we explicitly included a spatially varying external electric potential $V(x)$ and also allowed the pairing potential $\Delta(x)$ to depend on position---these features are essential for creating conditions that trap near-zero-energy ABS's that mimic MZM's. A uniform system with $V(x) = 0$ and $\Delta(x) = \Delta_0$ resides in the topological phase provided $h > \sqrt{\Delta_0^2 + \mu^2}$. For numerical evaluation we discretize Eq.~\eqref{eq:nanowire} in a standard way; when essential, we will specify the lattice spacing used in this discretization procedure below.

At fixed global fermion parity, we microscopically model tetron qubits via
\begin{align}
    H = H_\mathrm{top} + H_\mathrm{bot}, \label{eq:microQubit}
\end{align}
where $H_{\mathrm{top}}$ and $H_{\mathrm{bot}}$ take the form of either Eqs.~\eqref{eq:Kitaev} or \eqref{eq:nanowire} and respectively describe the top and bottom topological superconductors in the qubit.  Classical noise is readily included by endowing parameters in $H$ with stochastic time dependence.  In principle, one should include a `bridge' term that couples the two  superconductors (see Fig.~\ref{fig:qubits}). We assume for simplicity that the bridge yields only  small quantitative effects and thus ignore its presence in our calculations.  
Both of our microscopic models describe noninteracting electrons in a non-number-conserving formulation and thus fail to capture charging-energy effects that would be present in a laboratory realization of the tetron (see Sec.~\ref{sec:executive}). This simplification is assumed for obvious computational tractability reasons; however, the effects of $1/f$ charge noise, as was argued to be the dominant physical source of tetron dephasing in Ref.~\cite{knapp_dephasing_2018}, can be captured at a rough qualitative level by taking the global chemical potentials to fluctuate stochastically in Eqs.~\eqref{eq:Kitaev} and \eqref{eq:nanowire}.

\subsection{Majorana-operator reformulation} \label{sec:MajoranaReform}

In our microscopic numerical simulations of tetrons, we find it most convenient to work in the language of \emph{local} Majorana operators $a_i$---with $\{a_i,a_j\} = 2\delta_{ij}$, $a_i^\dagger = a_i$, and $(a_i)^2 = 1$---by decomposing the local Dirac fermions in the usual way, e.g.,  $c_i = (a_{2i-1} - ia_{2i})/2$ in the Kitaev model context \cite{Kitaev01_PhysU_44_131}. In this basis, the (quadratic) microscopic tetron Hamiltonian is represented by an $N\times N$ anti-symmetric matrix $A = -A^T$:
\begin{equation}
    H = \frac{i}{4}\sum_{i,j=1}^N A_{ij} a_i a_j = \frac{i}{4}\,\mathbf{a}^T\,A\, \mathbf{a}, \label{eq:H-A}
\end{equation}
with $N$ the total number of local Majorana operators; for spinless (spin-1/2) models with $N_\mathrm{sites}$ physical sites, $N = 2N_\mathrm{sites}$ ($N = 4N_\mathrm{sites}$). It is a relatively straightforward numerical exercise (see, e.g., Refs.~\cite{wimmer_algorithm_2012,bravyi_complexity_2017,bauer_dynamics_2018}) to bring $A$ into so-called `canonical form' via an orthogonal rotation $U$:
\begin{equation}
    B = U^T\,A\,U = \bigoplus_{k=1}^{N/2} \varepsilon_k i \sigma^y, \label{eq:B}
\end{equation}
such that
\begin{equation}
    H = \frac{i}{4}\,\mathbf{b}^T\,B\,\mathbf{b} = \frac{i}{2} \sum_{k=1}^{N/2} \varepsilon_k b_{2k-1} b_{2k}. \label{eq:Hcanon}
\end{equation}
The canonical Majorana modes $b_i = [\mathbf{b}]_i$ are related to the original local Majoranas $a_i = [\mathbf{a}]_i$ via the orthogonal transformation $\mathbf{b} = U^T\,\mathbf{a}$, and $\varepsilon_k \geq 0$ are the energies (or instantaneous energies when the Hamiltonian depends on time).

Due to Wick's theorem, any quadratic system can be completely described by the real, anti-symmetric covariance matrix
\begin{equation}
    M_{ij} = \frac{-i}{2}\langle[a_i, a_j]\rangle, \label{eq:Mdef}
\end{equation}
where the square brackets denote a commutator and $\langle \dots \rangle = {\rm Tr} \{\rho \dots\}$ a quantum expectation value taken with respect to density matrix $\rho$; if $\rho = |\psi\rangle\langle\psi|$ represents a pure state, one finds $M^2=-1$. As an example, the ground state of $H$ corresponds to a covariance matrix
\begin{equation}
    M_\mathrm{g.s.} = U\,M_y\,U^T \label{eq:Mgs},
\end{equation}
with
\begin{equation}
    M_y = \bigotimes_{k=1}^{N/2} i\sigma^y \label{eq:M0}
\end{equation}
a reference covariance matrix in the canonical basis of the modes $b_i$ which encodes $i b_{2k-1} b_{2k} = -1~\forall\,k=1,\dots,N/2$. It is straightforward to construct more general states by appropriately toggling elements of $M_y$ (see Refs.~\cite{bravyi_complexity_2017,bauer_dynamics_2018} and Sec.~\ref{sec:protocol}).

Finally, the Schr\"odinger equation for Gaussian states takes the form of the following ordinary matrix differential equation for the covariance matrix:
\begin{equation}
    \frac{dM(t)}{dt} = [A(t), M(t)], \label{eq:dMdt}
\end{equation}
where $A(t)$ specifies the (possibly time-dependent) Hamiltonian [cf.~Eq.~\eqref{eq:H-A}] and the square brackets denote the matrix commutator. Any physical observable can then be computed with knowledge of $M(t)$; relevant examples of such measurements 
will be described in the next section.

\section{Ramsey-type protocol} \label{sec:protocol}

\subsection{Qubit definition} \label{sec:qubitdef}

In what follows we assume for concreteness that the devices in Fig.~\ref{fig:qubits} possess even global fermion parity.  
We will encode the qubit in the subspace consisting of the two lowest-lying \emph{instantaneous} even-parity 
eigenstates of the microscopic Hamiltonian $H(t)$ that depends on time due to noise.  These presumed nearly degenerate states could arise because the system either realizes  bona fide topological phases yielding four well-separated Majorana modes (upper panel of Fig.~\ref{fig:qubits}), or exhibits a pair of `accidentally' low-energy Andreev bound states (lower panel of Fig.~\ref{fig:qubits}).
Since we will explore platforms that contain both types of near-degeneracies depending on parameters, it is useful to treat them in a common framework.  
To this end we introduce a quartet of \emph{maximally localized} (in real space) Majorana operators $\gamma_{1,2,3,4}(t)$ that span the instantaneous qubit subspace at time $t$.  In the topological case residual overlap between Majorana operators splits the degeneracy by an energy that is exponentially small in their separation; in the low-energy Andreev-bound-state scenario, pairs of Majorana operators sit in close proximity yet happen to  hybridize weakly.  

For a more precise definition of the qubit, suppose that 
\begin{equation}
      P(t) \equiv [i\gamma_1(t)\gamma_2(t)][i\gamma_3(t)\gamma_4(t)] = +1
\end{equation}
in the lowest-lying even-fermion-parity states, and define Pauli operators 
\begin{equation}
      Z_t = i \gamma_1(t)\gamma_2(t),~~~ X_t = i \gamma_1(t) \gamma_3(t).
\end{equation}
Our logical qubit states are then the minimum-energy many-body states with $Z_t = \pm 1$, i.e.,
\begin{equation}
    \ket{0_t} \equiv \ket{Z_t = +1},~~~\ket{1_t} \equiv \ket{Z_t = -1}.
\end{equation}
The subscript $t$ (suppressed in Sec.~\ref{sec:executive}) is included here and below as a reminder that these states are defined with respect to Majorana operators extracted from the instantaneous Hamiltonian at time $t$, rather than with respect to a fixed basis. 

Some discussion is warranted regarding the definition of the maximally localized Majorana operators $\gamma_{1,2,3,4}(t)$ in our microscopic numerical simulations. The canonical modes $b_i$ of the quadratic Hamiltonian $H$ defined above [cf.~Eqs.~\eqref{eq:B} and \eqref{eq:Hcanon}] are not unique since arbitrary SO(2) rotations $O_k^{2\times2}=\exp(i\theta_k \sigma^y)$ within each $2\times2$ block of Eq.~\eqref{eq:B} leave the matrix $B$ invariant:
\begin{equation}
    B = O\,B\,O^T,\quad O = \bigoplus_{k=1}^{N/2} O_k^{2\times2}. \label{eq:Btilde}
\end{equation}
Thus, Majorana modes $\tilde{b}_i = [\tilde{\mathbf{b}}]_i$ defined through
\begin{equation}
    \tilde{\mathbf{b}} = O^T\,\mathbf{b} = \tilde{U}^T\mathbf{a}, \label{eq:btilde}
\end{equation}
with $\tilde{U} = U\,O$, also form a canonical set with the same instantaneous energies, i.e.,
\begin{equation}
    H = \frac{i}{4}\,\tilde{\mathbf{b}}^T\,B\,\tilde{\mathbf{b}} = \frac{i}{2} \sum_{k=1}^{N/2} \varepsilon_k \tilde{b}_{2k-1} \tilde{b}_{2k}. \label{eq:Hcanontilde}
\end{equation}
The Majorana modes $\gamma_{1,2,3,4}(t)$ depicted in Fig.~\ref{fig:qubits} refer to $\tilde{b}_{1,2,3,4}$ with $O_{k=1,2}^{2\times2}$ chosen such that the associated wave functions maximally localize at a given time $t$. Specifically, we first find a set of canonical near-zero-energy modes $b_{1,2,3,4}$ by bringing $A$ into canonical form using the software package presented in Ref.~\cite{wimmer_algorithm_2012}; in the local basis of the $a_i$, these modes are represented by column vectors of the matrix $U$, say $U_{i,j=1,2,3,4}$. We then optimize the $O_{k=1,2}^{2\times2}$ (each characterized by an angle $0\leq\theta_k<2\pi$) such that for the resulting $\tilde{U} = U\,O$ the 4\textsuperscript{th} moments $\sum_{j=1}^2\sum_{i=1}^N |\tilde{U}_{ij}|^4$ and $\sum_{j=3}^4\sum_{i=1}^N |\tilde{U}_{ij}|^4$ (for $k=1,2$, respectively) are maximized; now, in the local $a_i$ basis, these maximally localized modes $\tilde{b}_{1,2,3,4}$ are given by column vectors of $\tilde{U}$, i.e.,  $\tilde{U}_{i,j=1,2,3,4}$. We order these modes according to their real-space locations as shown in Fig.~\ref{fig:qubits}, thereby giving the desired $\gamma_{1,2,3,4}(t)$ \footnote{This method for obtaining maximally localized near-zero-energy Majorana modes working entirely in the local Majorana representation parallels that described in Ref.~\cite{moore_two-terminal_2018} using the more traditional Bogoliubov-de Gennes (BdG) framework.}.

As an aside, the numerical procedure spelled out above is tailored to the situation where the two superconductors comprising the qubit decouple---which again we assume for simplicity is the case throughout.   The Majorana modes $\gamma_{1,2}(t)$ ($\gamma_{3,4}(t)$) then have support entirely on the top (bottom) wire of the tetron. In the more realistic scenario in which a nonzero `bridge' Hamiltonian couples the wires, a physically plausible definition of $\gamma_{1,2,3,4}(t)$ involves instead maximizing the single 4\textsuperscript{th} moment $\sum_{j=1}^4\sum_{i=1}^N |\tilde{U}_{ij}|^4$ via a single orthogonal rotation $O^{4\times4}$ acting on the $b_{1,2,3,4}$.  In the latter case $\gamma_{1,2,3,4}(t)$ are not generally eigenmodes of the Hamiltonian.

\subsection{Protocol details}

Section~\ref{sec:executive} summarized our time-domain Ramsey-type protocol for the physical implementation of the tetron. In the `ideal' case the protocol involves (1) initializing the system into, say, the $X_{t = 0} = +1$ eigenstate at time $t = 0$ via energy-level spectroscopy on the nearby quantum dot which tunnel couples to $\gamma_1$ and $\gamma_3$, (2) letting the system evolve freely for time $t$ under the influence of classical noise, and (3) reading out $X_t = \pm1$ with the same quantum dot used for initialization. (For practical purposes, we will sometimes depart from this `ideal' protocol by initializing and measuring with respect to a fixed basis.)  In the remainder of this section we elaborate on these steps to set the stage for our analytic treatment of dephasing and noise in Sec.~\ref{sec:analytics} and our full time-dependent microscopic simulations of tetron qubits in Sec.~\ref{NumericalResults}.

\subsubsection{Initialization}

We start with initialization via the proximate quantum dot. At the level of our analysis, we assume that the quantum dot is `perfect' in the sense that a measurement projects exactly and instantaneously onto the qubit state 
\begin{equation}
    \ket{\psi(0)} = \frac{1}{\sqrt{2}}(\ket{0_{t = 0}} +  \ket{1_{t = 0}}) \label{eq:psit0}
\end{equation}
with $X_{t = 0} = +1$ at time $t = 0$. Under this highly idealized assumption, we need not explicitly include the quantum dot in our analysis. In our microscopic simulations of noisy tetron qubits in Sec.~\ref{NumericalResults}, we use the \emph{time-averaged} Hamiltonian---specified by an anti-symmetric matrix $A_0$---to define the relevant $\gamma_{1,2,3,4}$ used for initialization, which we denote $\gamma^{(0)}_{1,2,3,4}$. [Obtaining the modes instead using the initial Hamiltonian $A(t=0) \neq A_0$ for each noise realization, as implied by Eq.~\eqref{eq:psit0}, leads to only negligible quantitative differences in the results presented there; see Sec.~\ref{sec:fixedvinst} for more discussion on this point.] Specifically, we bring $A_0$ into canonical form as described above in Sec.~\ref{sec:models} [see Eq.~\eqref{eq:Hcanon}]; we subsequently find the corresponding set of maximally localized near-zero-energy Majorana modes, which are encoded in column vectors of an orthogonal matrix $\tilde{U}_0$. Next, we construct a reference covariance matrix $M_y(t=0)$ [cf.~Eqs.~\eqref{eq:Mgs} and \eqref{eq:M0}] in the basis of these modes $\tilde{b}_i^{(0)}$ corresponding to, say, $X_0 = i\gamma^{(0)}_1\gamma^{(0)}_3 = +1$ and $i\gamma^{(0)}_2\gamma^{(0)}_4 = -1$. The global fermion parity thus reads $(i\gamma^{(0)}_1\gamma^{(0)}_2)(i\gamma^{(0)}_3\gamma^{(0)}_4) = +1$, which for all presented simulations coincides with the parity of the absolute ground state of $A_0$. (Recall that modes $\tilde{b}_{i=1,2,3,4}$ within the low-energy manifold correspond to the maximally localized modes $\gamma_{1,2,3,4}$, while the remainder $\tilde{b}_{i\neq1,2,3,4}$ are equivalent to the respective original canonical modes $b_i$.) Finally, the initial covariance matrix in the $a_i$ basis [cf.~Eq.~\eqref{eq:Mdef}] reads $M(t=0) = \tilde{U}_0\,M_y(t=0)\,\tilde{U}_0^T$.

\subsubsection{Unitary evolution} \label{sec:UnitaryEvolution}

Once initialized, the system undergoes unitary time evolution with respect to the noisy Hamiltonian $H(t)$.  After time $t$ the initial state becomes
\begin{eqnarray}
    \ket{\psi(t)} &=& U(t) \ket{\psi(0)}
    \nonumber \\
    &\equiv& a_t \ket{0_t} + b_t \ket{1_t} + c_t \ket{v_t},
    \label{psit}
\end{eqnarray}
where $U(t)$ is the time evolution operator (we discuss the second line shortly).  Our microscopic numerical simulations are instead carried out in the Heisenberg picture, for which unitary evolution is encoded in Eq.~\eqref{eq:dMdt}; there noise is included through time dependence in $A(t)$, which is essentially the Hamiltonian expressed in the Majorana basis. Appendix~\ref{app:noise} details our treatment of noise (e.g., the procedure we employ for generating individual noise trajectories for a given realization) and numerical solution of Eq.~\eqref{eq:dMdt}.

We stress that in general $\ket{0_t} \neq U(t) \ket{0_{t = 0}}$ since $\ket{0_t}$ is defined through instantaneous $H(t)$ eigenstates (and similarly for $\ket{1_t}$) \footnote{$\ket{0_t}$ and $\ket{1_t}$ are the two lowest-energy eigenstates of the instantaneous Hamiltonian in the same global parity sector as the evolving state $\ket{\psi(t)}$ (assumed even in this discussion), which may or may not coincide with the parity of the absolute instantaneous ground state.}.  In the second line of Eq.~\eqref{psit} we thus expressed the time-evolved state in terms of qubit states $|0_t\rangle$ and $\ket{1_t}$ at time $t$, \emph{and} a ket $\ket{v_t}$ that signifies dynamically generated excited states.  Generation of weight on the latter via non-zero $c_t$ corresponds to `leakage' of the qubit away from the computational subspace.  Leakage can arise from an odd or even number of fermionic excitations, thus respectively flipping or preserving $P(t)$. Using Eq.~\eqref{psit} we can define a qubit leakage time $T_{\rm leak}$ by writing
\begin{equation}
    \overline{|a_t|^2} + \overline{|b_t|^2} \equiv 1-f(t/T_{\rm leak}), \label{eq:a2pb2}
\end{equation}
where the overline indicates noise averaging.  On the right side,  $f(t/T_{\rm leak})$ is a system-dependent function that vanishes at $t = 0$ and grows---thus shifting weight onto excited states---on a characteristic time scale $T_{\rm leak}$. An alternative leakage metric can be extracted from
\begin{equation}
    \overline{\bra{\psi(t)}P(t)\ket{\psi(t)}} \equiv 1- \tilde f(t/\tilde T_{\rm leak})
    \label{Tleaktilde}
\end{equation}
with $\tilde f$ a different function that grows on a time scale $\tilde T_{\rm leak}$.  The latter measures the weight on excited states containing only odd numbers of fermionic excitations, and thus provides an upper bound on the leakage time defined above, i.e., $T_{\rm leak} \leq \tilde T_{\rm leak}$.  

At this point it is worth commenting on an alternative scheme wherein one defines the qubit in terms of a fixed basis of maximally localized Majorana operators obtained from, say, the time-averaged Hamiltonian $A_0$ or, for a given noise realization, the initial Hamiltonian $A(t=0)$. For an extreme case, suppose that $H(t)$ supports exact instantaneous zero-energy Majorana modes at any time $t$, but that \emph{adiabatic} noise causes the locations of the Majorana modes to vary with time.  The system will then evolve out of the low-energy subspace spanned by the fixed Majorana operators, whereas the qubit subspace should clearly be perfectly preserved---as captured by tracking instantaneous Majorana operators.  In general we expect that employing a fixed basis \emph{overestimates} qubit errors, with the difference being most pronounced when the noise is `slow' and of `large' amplitude.  We will quantify such effects later in Sec.~\ref{sec:fixedvinst}.  

\subsubsection{Readout} \label{sec:readout}

The final step of the protocol involves readout.  We again assume that the nearby quantum dot can perform this task, i.e., measure $X_t = \pm1$, instantaneously and without error.  Under this assumption the relevant physical quantity is the noise-averaged expectation value of $X_t$ with respect to the time-evolved state:
\begin{equation}
      \overline{\bra{\psi(t)} X_t \ket{\psi(t)}} = \cos(\omega_0 t) f(t/T_2),
\end{equation}
which defines the qubit precession frequency $\omega_0$ and dephasing time $T_2$. Compared to Eq.~\eqref{eq:X_decay}, the $t$ subscript on the left side explicitly indicates that the expectation value is (ideally) taken with respect to the instantaneous Majoranas at time $t$.

To calculate the quantum expectation value $\langle\psi(t)|X|\psi(t)\rangle$ for a given noise realization in our microscopic simulations, we basically invert the initialization procedure described above. Specifically, we conjugate $M(t)$ with $\tilde{U}$ (which defines the measurement basis; see below) to arrive at
\begin{equation}
    M_y(t) = \tilde{U}^T\,M(t)\,\tilde{U}, \label{eq:M0tildeOft}
\end{equation}
from which any $\langle i\gamma_i\gamma_j\rangle$ can be read off directly. Measurements involving a product of more than two operators can be evaluated with use of Wick's theorem; an important example is $P(t)$, which (partially) probes leakage out of the computational subspace [recall Eq.~\eqref{Tleaktilde}].

As with initialization, there exists a choice as to precisely what is meant by the Majorana operators $\gamma_{1,2,3,4}$ being measured. Namely, these could be defined as (maximally localized) near-zero-energy modes with respect to, say, the \emph{fixed} time-averaged Hamiltonian, $A_0$, or the \emph{instantaneous} Hamiltonian, $A(t)$ \footnote{An alternative choice similar to the former case could be to use modes derived from the fixed \emph{initial} Hamiltonian $A(t=0) \neq A_0$ for a given noise realization. We take this approach in Sec.~\ref{sec:fixedvinst}.} Operationally, to measure in the basis corresponding to the time-averaged Hamiltonian such that $\gamma_{1,2,3,4} = \gamma^{(0)}_{1,2,3,4}$, we use Eq.~\eqref{eq:M0tildeOft} with $\tilde{U} = \tilde{U}_0$ (the same orthogonal matrix used in the initialization procedure). On the other hand, to measure in the instantaneous basis we need to find a set of maximally localized modes $\gamma_{1,2,3,4} = \gamma_{1,2,3,4}(t)$ at every time $t$, which are encoded in a time-dependent orthogonal matrix $\tilde{U} = \tilde{U}(t)$ to be used in Eq.~\eqref{eq:M0tildeOft}. In our microscopic simulations, we implement both instantaneous and fixed measurement bases. Employing a fixed basis is computationally cheaper as it does not require bringing the Hamiltonian into canonical form and finding maximally localized modes at each time step. Additionally, there is an inherent sign ambiguity in the canonical modes (and thus also maximally localized modes) upon bringing $A(t)$ into canonical form numerically, i.e., the Hamiltonian is invariant upon taking $b_{2k-1}\to-b_{2k-1}$ and $b_{2k}\to-b_{2k}$ for a given $k$ in Eq.~\eqref{eq:Hcanon}. Therefore, measurements in the instantaneous basis such as $\langle i\gamma_1(t)\gamma_3(t)\rangle$ will be plagued by $\gamma_1(t)$ and $\gamma_3(t)$ having arbitrary relative sign for different $t$. For these reasons, we use the fixed basis defined by the time-averaged Hamiltonian for the data presented in Secs.~\ref{NumericalResults}. Note that measuring in the fixed basis corresponding to the time-averaged Hamiltonian plausibly mimics `imperfect' quantum dot readout that is slow on the time scale of the typical noise correlation time. In Sec.~\ref{sec:fixedvinst}, we circumvent the sign ambiguity problem in the instantaneous basis by noise averaging instead $\langle i\gamma_1(t)\gamma_3(t)\rangle^2$ and are thereby able to directly compare the two measurement approaches (and thus qubit encodings).

\section{Analytical estimation of qubit time scales} \label{sec:analytics}

In this section we use standard techniques to derive analytic formulas for the dephasing and leakage times.  These estimates will be compared with time scales extracted from our time-dependent numerical simulations in Sec.~\ref{NumericalResults}.

\subsection{Dephasing time} \label{sec:T2_analytic}

We first restrict attention to the low-energy subspace, neglecting the quasiparticle continuum above the superconducting gap. This approximation is justified provided noise is slow compared to the gap scale.  Leakage of the qubit into excited states will be discussed in the next subsection.  

Within the low-energy subspace, we model the qubit by the minimal two-level Hamiltonian in Eq.\ (\ref{qubitH}). Time dependence in the instantaneous energy splitting $E(t)$ arises due to stochastic temporal variations in parameters in the `parent' microscopic Hamiltonian.  We denote fluctuations in these microscopic parameters about their time-averaged values by $\lambda_i(t)$, which are assumed uncorrelated with one another.  This set can include variations in the electrochemical potential, magnetic field, etc.~in different spatial regions of the device.  For example, if only the global chemical potential varies in time by an amount $\delta\mu(t)$, then we simply have $\lambda_1(t) = \delta \mu(t)$; if instead uncorrelated chemical potential variations arise in the left and right halves of the wires, then we have $\lambda_1(t) = \delta \mu_L(t)$ and $\lambda_2(t) = \delta\mu_R(t)$; and so on. Assuming the fluctuations are weak, we Taylor expand the energy to second order in $\lambda_i$'s, yielding
\begin{equation}
      E(t) \approx E_0 + \sum_i \lambda_i(t)E'_i + \frac{1}{2} \sum_{i,j} \lambda_i(t)\lambda_j(t) E''_{ij},
      \label{E_expansion}
\end{equation}
where we defined short-hand notation
\begin{equation}
    E'_i = \frac{dE}{d\lambda_i}|_{\lambda_i = 0},~~~E''_{ij} = \frac{d^2E}{d\lambda_i d\lambda_j}|_{\lambda_{i,j} = 0}.
\end{equation}

To noise average we will assume that each $\lambda_i$ exhibits Gaussian noise correlations with mean and variance
\begin{align}
\overline{\lambda_i(t)} &= 0,
\label{lambda_ave}
\\
    \overline{\lambda_i(t)\lambda_j(t')} &= \delta_{ij}S_i(t-t').
    \label{noise_corr}
\end{align}
We further take $S_i(t)$ to be Gaussian as a function of time, i.e.,
\begin{equation}
    S_i(t) = D_i^2 e^{-t^2/(2\tau_i)^2},
    \label{noise_power}
\end{equation}
corresponding to a power spectrum
\begin{equation}
    \tilde S_i(\omega) = D_i^2 \sqrt{\frac{4\pi}{\kappa_i^2}}\, e^{-(\omega/\kappa_i)^2}.
    \label{eq:Sw_Gaussian}
\end{equation}
In Eqs.~\eqref{noise_power} and \eqref{eq:Sw_Gaussian}, $D_i$ and $\tau_i \equiv 1/\kappa_i$ respectively denote the fluctuation amplitude and characteristic noise correlation time for fluctuator $\lambda_i$; note that the frequency scale $\kappa_i$ plays the role of a high-frequency cutoff in the power spectrum.
The qubit precession frequency $\omega_0$ then follows from
\begin{equation}
    \hbar \omega_0 = \overline{E(t)} = E_0 + \frac{1}{2} \sum_i D_i^2 E''_{ii}.
    \label{omega0_shifted}
\end{equation}
Notice that noise generically shifts the mean energy splitting $\hbar \omega_0$ away from the noise-free value $E_0$.

We could in principle also adopt the more realistic case of $1/f$ noise, which leads to similar qualitative conclusions (see, e.g., Ref.~\cite{aasen_milestones_2016} for a  discussion in a related context).  However, in the microscopic models discussed below, $1/f$ noise requires the introduction of a high-frequency cutoff which leads to corrections that make quantitative comparisons between theoretical and numerical estimates challenging.

In the two-level system approximation used here, it is straightforward to evaluate the readout for the Ramsey-type protocol described in the preceding section.  For a particular noise realization one obtains 
\begin{align}
      Q(t) \equiv \bra{\psi(t)} X_t \ket{\psi(t)} = \cos\left(\int_0^t dt' E(t')/\hbar \right).
      \label{Pt}
\end{align}
Appendix~\ref{app:averaging} noise averages $Q(t)$; in the long-time limit $t \gg \tau_i$ we find 
\begin{equation}
      \overline{ Q(t \gg \tau_i)} \approx \cos(\omega_0t)e^{-t/T_2}
      \label{eq:T2analytic_func}
\end{equation}
with dephasing time
\begin{equation}
      T_2 = \frac{\hbar^2}{\sqrt{\pi}}\left[\sum_i\tau_i\left(D_i E_i'\right)^2 + \frac{1}{2}\sum_{ij}\frac{\tau_i \tau_j}{\sqrt{\tau_i^2 + 
      \tau_j^2}}\left(D_i E_{ij}''D_j\right)^2 \right]^{-1}.
      \label{T2analytic}
\end{equation}
(Appendix~\ref{app:averaging} also examines the short-time limit $t \ll \tau_i$, which is sometimes relevant for our numerical simulations.) 
Upon varying parameters, e.g., to transition between an ABS and topological qubit, we expect that $T_2$ will vary by far most dramatically through the $E_i'$ and $E_{ij}''$ factors.  These factors quantify the rigidity of the qubit splitting with respect to fluctuations and can be efficiently derived from microscopics through exact diagonalization of \emph{noise-free} static Hamiltonians, in principle containing various bells and whistles.  We will carry out such an analysis in Sec.~\ref{NumericalResults} in parallel with our explicit time-dependent qubit simulations.

\subsection{Leakage time} \label{sec:Tleak_analytic}

We will estimate the qubit leakage time by studying just a single noisy wire in the tetron using Fermi's golden rule.  For analytical tractability, we will further employ a \emph{fixed basis} for Majorana modes in the device; as stressed in Sec.~\ref{sec:UnitaryEvolution} the analytical estimates that follow should thus be viewed as an \emph{upper bound} on the true leakage time that would be obtained by following the instantaneous basis of maximally localized Majorana modes.  

Let us describe the system within the spinful-wire model (extension to the Kitaev-chain description is straightforward).  We write the full microscopic Bogoliubov-de Gennes Hamiltonian as $H=H_0+\delta H(t)$, 
where $H_0$ is given by Eq.~\eqref{eq:nanowire} and $\delta H(t)$ incorporates noise.  
The static part of the Hamiltonian is diagonalized upon decomposing the fermion $\psi_\sigma(x)$ with spin $\sigma$ at position $x$ as
\begin{equation}
  \psi_\sigma(x) = \Phi_{1\sigma}(x) \gamma_1^{(0)} + \Phi_{2\sigma}(x) \gamma_2^{(0)} + \sum_{E>0} \varphi_{E\sigma}(x) f_E.
  \label{psi_decomposition}
\end{equation}
Here $\Phi_{1,2\sigma}$ denote the maximally localized Majorana wave functions associated with the fixed-basis Majorana operators $\gamma_{1,2}^{(0)}$ derived from $H_0$, while $\varphi_{E\sigma}$ denote wave functions for above-gap excitations created by operators $f_E^\dagger$.  After this decomposition $H_0$ becomes
\begin{equation}
    H_0 = \frac{i}{2} \varepsilon_{12} \gamma_1^{(0)} \gamma_2^{(0)} + \sum_{E>0} E f_E^\dagger f_E  
\end{equation}
with $\varepsilon_{12}$ the noise-free splitting of states with $i\gamma_1 \gamma_2 = \pm 1$.  

We write the noise terms as
\begin{equation}
    \delta H(t) = \sum_i \lambda_i(t) \int_0^L dx \psi^\dagger \Lambda_i(x) \psi
\end{equation}
for some matrices $\Lambda_i(x)$ dependent on the nature of the parameter fluctuations $\lambda_i(t)$.  For instance, if $\lambda_1(t) = \delta \mu(t)$ represents a global chemical potential fluctuation, then $\Lambda_1 = -I$.  Note that we neglect noise in the pairing channel for simplicity, though such terms could be readily incorporated if desired.
Rewriting $\delta H(t)$ using Eq.~\eqref{psi_decomposition} yields
\begin{equation}
    \delta H(t) = \sum_{E>0} \{[A_{1E}(t) \gamma_1^{(0)} + A_{2E}(t) \gamma_2^{(0)}]f_E + \mathrm{H.c.}\} + \cdots.
\end{equation}
The ellipsis denotes pieces involving two Majorana operators or two $f_E$ operators, while the coefficients above read
\begin{equation}
    A_{1,2E}(t) = \sum_i \lambda_i(t) \int_0^L dx \Phi_{1,2}^\dagger(x) \Lambda_{i}(x) \varphi_{E}(x).
\end{equation}
Next we introduce the complex fermion $f_0 = (\gamma_1 + i \gamma_2)/2$ and define $A_{\pm,E}(t) = A_{1E}(t) \pm i A_{2E}(t)$ so that
\begin{equation}
     \delta H(t) = \sum_{E>0} \{[A_{-,E}(t) f_0 + A_{+,E}(t) f_0^\dagger]f_E + \mathrm{H.c.}\} + \cdots.
\end{equation}

According to Fermi's golden rule, the noise-averaged leakage rates $\Gamma_\pm$ out of the ground states with $f_0^\dagger f_0=0$ ($\Gamma_-$) and $f_0^\dagger f_0=1$ ($\Gamma_+$) are given by
\begin{equation}
    \Gamma_\pm = \frac{2\pi}{\hbar} \sum_{E>0} \int_{\omega,\omega'} \overline{A_{\pm,E}(\omega)A_{\pm,E}^*(\omega')}\delta(E\mp \varepsilon_{12}-\hbar\omega).
\end{equation}
The noise correlations in Eqs.~\eqref{lambda_ave} through \eqref{eq:Sw_Gaussian} yield
\begin{align}
    \overline{A_{\pm,E}(\omega)A_{\pm,E}^*(\omega')} &=2\pi \delta(\omega-\omega')
    \nonumber \\
    &\times 2 \sqrt{\pi}\sum_j\tau_j D_j^2  e^{-(\omega \tau_j)^2}B_{\pm,j}(E) ,
\end{align}
where
\begin{align}
    B_{\pm,j}(E) = 
     \bigg{|}\int_x[\Phi_1^\dagger(x) \pm i \Phi_2^\dagger(x)]\Lambda_j(x) \varphi_E(x)\bigg{|}^2.
     \label{B}
\end{align}
Feeding this expression into $\Gamma_\pm$ leads to
\begin{equation}
    \Gamma_\pm = 2 \sqrt{\pi}\sum_j\frac{\tau_j D_j^2}{\hbar^2}\sum_{E>0} B_{\pm,j}(E) e^{-[(E\mp \varepsilon_{12}) \tau_j/\hbar]^2} .
\end{equation}
Suppose now that $E_g$ is the bulk gap and $\rho(E)$ is the density of states for the above-gap excitations; assuming the coefficients $B_{\pm,j}$ vary smoothly with energy, we approximate the leakage rates as
\begin{align}
    \Gamma_\pm &\approx 2 \sqrt{\pi}\sum_j\frac{\tau_j D_j^2}{\hbar^2}B_{\pm,j}(E_g)
    \nonumber \\
    &\times 
    \int_{E_g}^\infty dE\rho(E) e^{-[(E\mp \varepsilon_{12}) \tau_j/\hbar]^2} .
\end{align}

Leakage will be most important in the vicinity of the topological phase transition where the bulk gap approaches zero. In this regime, the excitation spectrum can be viewed as arising from gapped-out counterpropagating bulk Majorana fermions with mode velocity $v$ and dispersion 
\begin{equation}
    E(k) = \sqrt{(\hbar vk)^2 + E_g^2}.
    \label{Ek}
\end{equation}
The density of states follows as 
\begin{equation}
    \rho(E) = \frac{L}{\pi \hbar v}\frac{E}{\sqrt{E^2-E_g^2}}
\end{equation}
for $E>E_g$ and vanishes otherwise.  
[We have implicitly assumed periodic boundary conditions in writing Eq.~\eqref{Ek}, though the choice of boundary conditions should not significantly influence $\rho(E)$.]
Assuming $\varepsilon_{12} \ll E_g$, we then obtain
\begin{equation}
    \Gamma_\pm \approx \frac{L}{\hbar v}\sum_j\frac{ D_j^2}{\hbar}B_{\pm,j}(E_g) e^{-(E_g \tau_j/\hbar)^2}.
    \label{GammaFinal}
\end{equation}

The coefficients $B_{\pm,j}(E_g)$ depend on details of the wave functions and noise sources.  We will use scaling arguments to roughly quantify these factors in three regimes:

$(i)$ \emph{Ultra-short-range noise}.~Suppose first that a noise source acts near one of the two Majorana modes, but only on a very local region of size $\xi^{\rm noise}_{j} \ll \xi$ (again, $\xi$ denotes the spatial extent of the Majorana wave function).  In this case $\Lambda_j(x)$ in Eq.~\eqref{B} has appreciable weight only over a distance $\xi^{\rm noise}_j$ so that 
\begin{equation}
    B_{\pm,j}(E_g) \propto \left(\frac{\xi^{\rm noise}_j}{\sqrt{\xi L}}\right)^2 = \left(\frac{\xi^{\rm noise}_j}{\xi}\right)^2 \frac{\xi}{L}.
\end{equation}
The square root in the denominator simply reflects normalization factors in the Majorana and above-gap-excitation wave functions.  Note that $B_{+,j} = B_{-,j}$ here, which implies approximately equal \footnote{We took $\varepsilon_{12}/E_g = 0$ to arrive at Eq.~\eqref{GammaFinal}; corrections from nonzero $\varepsilon_{12}/E_g$ can still give different leakage rates even with short-range-correlated noise.} leakage rates out of the low-energy states with $i\gamma_1 \gamma_2 = \pm 1$.

$(ii)$ \emph{Short-range noise}.~If $\gamma_{1,2}$ are separated by a sufficiently long distance $L_{12}$, one can envision a short-range-noise scenario wherein $\xi < \xi^{\rm noise}_j < L_{12}$.  That is, the noise source acts over the entire window of one of the Majorana modes, but does not influence its partner.  (This scenario is reasonable for bona fide Majorana zero modes separated by the length of the wire, but is less natural for accidental Andreev bound states.)  Here we get
\begin{equation}
      B_{\pm,j}(E_g) \propto \left(\frac{\xi}{\sqrt{\xi L}}\right)^2 = \frac{\xi}{L},
\end{equation}
again yielding equal leakage rates out of the $i\gamma_1 \gamma_2 = \pm 1$ states.

$(iii)$ \emph{Long-range noise}.~Finally, consider a long-range-correlated noise source with $\xi^{\rm noise}_j > L_{12}$.  The integral in Eq.~\eqref{B} now picks up weight from \emph{both} Majorana wave functions, yielding similar scaling to the previous case but with unequal leakage rates for the $i\gamma_1\gamma_2 = \pm 1$ states:
\begin{equation}
      B_{\pm,j}(E_g) \propto (1\pm \eta_j) \frac{\xi}{L}
\end{equation}
for $-1< \eta_j <1$. This regime offers the interesting possibility of a `dark state': when $\eta_j = \pm1$ the transition rate for one of the two low-energy states vanishes.  

The three cases above can be put on equal footing by writing
\begin{equation}
    B_{\pm,j}(E_g) \approx c_j (1\pm \eta_j)\left(\frac{\xi^{\rm noise}_j}{\xi}\right)^{b_j} \frac{\xi}{L},
\end{equation}
the proportionality constant $c_j$, exponent $b_j$, and $\eta_j$ are determined by properties of noise source $j$.  We thereby obtain leakage times
\begin{align}
    T_{\rm leak}^\pm &= 1/\Gamma_\pm \nonumber \\
    &\approx \hbar^2\bigg{[}\sum_i c_i \frac{\xi}{v} D_i^2 (1\pm \eta_i)\left(\frac{\xi^{\rm noise}_i}{\xi}\right)^{b_i}e^{-(E_g \tau_i/\hbar)^2} \bigg{]}^{-1}.
    \label{Tleak_analytic}
\end{align}
Comparison to the dephasing time in Eq.~\eqref{T2analytic} (first term) is instructive.  We see that the time scale $\xi/v$ appearing in $T_{\rm leak}^\pm$ plays the role of $\tau_i$ in $T_2$; similarly, the exponential factor $e^{-(E_g \tau_i/\hbar)^2}$ plays the role of the `energetic rigidity' $E_i'$ characterizing the qubit states.  Upon approaching the topological phase transition---either from the Andreev-bound-state or true Majorana-zero-mode regime---collapse of the bulk gap $E_g$ readily allows leakage to dominate the qubit lifetime.  

\section{Numerical Results} \label{NumericalResults}

We are now in position to present results on both quasi-analytic evaluation of the dephasing and leakage times (using the formulas derived in Secs.~\ref{sec:T2_analytic} and \ref{sec:Tleak_analytic}) as well as full microscopic simulations of noisy tetron dynamics for both the Kitaev and spinful nanowire models.

\subsection{Kitaev tetron} \label{sec:kitaev}

We first consider a tetron qubit built from two parallel Kitaev chains [Eq.~\eqref{eq:Kitaev}] each consisting of $L$ physical sites. The time-averaged chemical potentials of the two wires are offset by a small amount $\mu_\mathrm{offset}$ to break degeneracies which would occur for identical wires; that is, $\mu^{(0)}_{\mathrm{top}/\mathrm{bot}} = \mu \pm \mu_\mathrm{offset}$. We adopt a noise model consisting of two independent fluctuators acting on the respective chemical potentials such that $\mu_{\mathrm{top}/\mathrm{bot}}(t) = \mu^{(0)}_{\mathrm{top}/\mathrm{bot}} + \delta\mu_{\mathrm{top}/\mathrm{bot}}(t)$. Each fluctuator $\delta\mu_{\mathrm{top}/\mathrm{bot}}(t)$ obeys Gaussian noise correlations with a Gaussian noise power spectrum as specified in  Eqs.~\eqref{lambda_ave} through \eqref{eq:Sw_Gaussian}. For simplicity, we take the amplitude of typical fluctuations $D$ and noise correlation time $\tau\equiv1/\kappa$ equal for both fluctuators: $D_{\mathrm{top}/\mathrm{bot}} = \delta\mu_{\mathrm{top}/\mathrm{bot}}^\mathrm{typ} = \delta\mu_\mathrm{typ}$ and $\kappa_{\mathrm{top}/\mathrm{bot}} = \kappa$.

\begin{figure}[t]
\begin{center} 
\includegraphics[width=0.9\columnwidth]{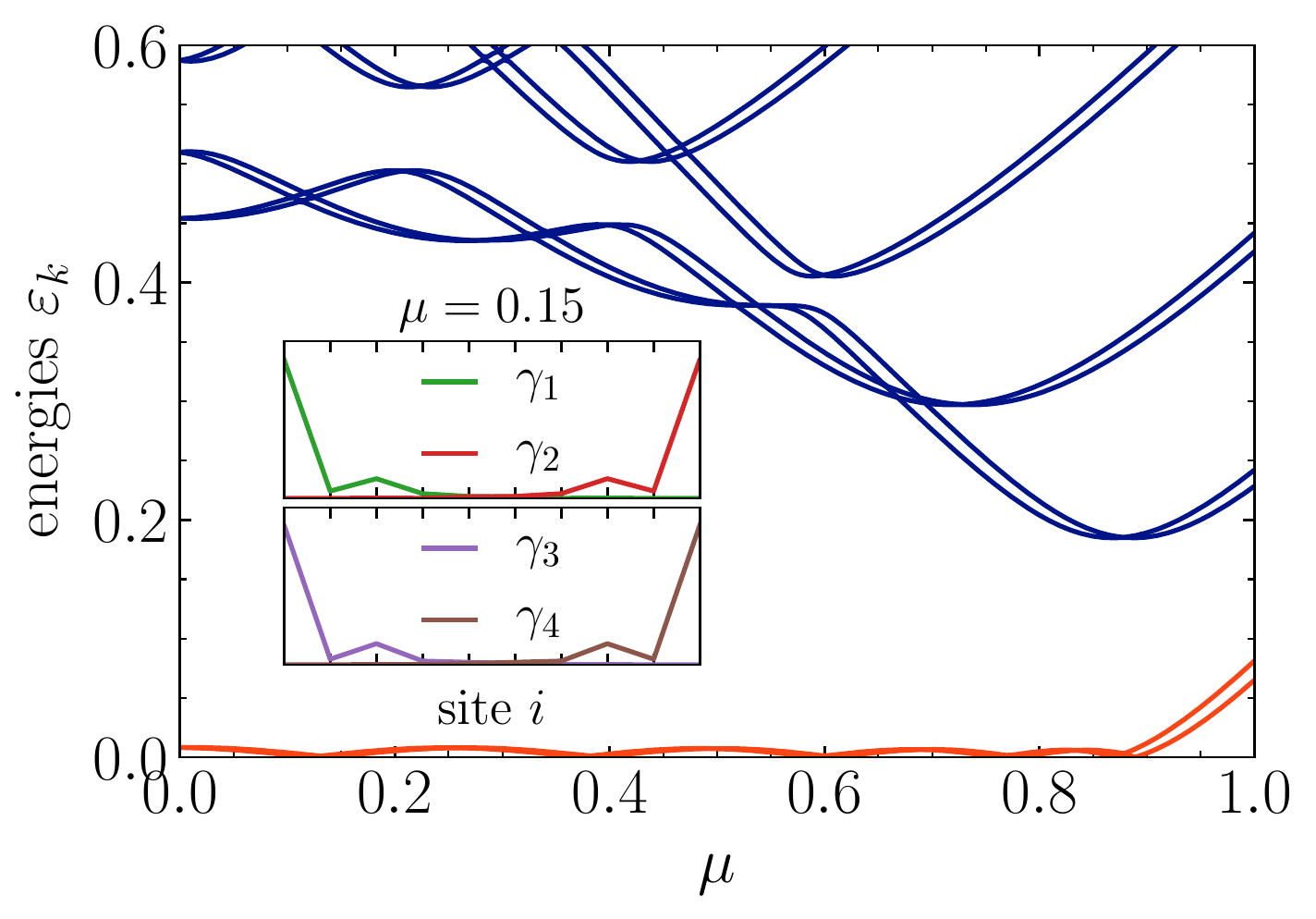}
\end{center}
\caption{Energy spectra of the Kitaev tetron as a function of $\mu$ for the configuration used in our microscopic numerical simulations, cf.~Figs.~\ref{fig:noise_averaging}, \ref{fig:comp_Kitaev}, \ref{fig:meas_basis_mu0.15}, \ref{fig:meas_basis_mu0.8}, and \ref{fig:single_realization}. Here, each wire of the tetron consists of an $L=10$ site Kitaev chain with $J_\mathrm{top/bot}=J=1$ and $\Delta_\mathrm{top/bot}=\Delta=0.4$; a slight chemical potential shift of $\mu_\mathrm{offset}=0.01$ is introduced between the two wires to break any associated degeneracies. The low-energy qubit manifold is highlighted in orange, corresponding to the energies $\varepsilon_{1,2}$ in Eq.~\eqref{eq:Hcanon}. In the inset, we plot the spatial dependence of the wave function square modulus for the maximally localized Majorana modes $\gamma_{1,2,3,4}$ (see Sec.~\ref{sec:qubitdef}) at $\mu=0.15$.
\label{fig:kitaev_setup}}
\end{figure}

\begin{figure}[t]
\begin{center}
\includegraphics[width=0.9\columnwidth]{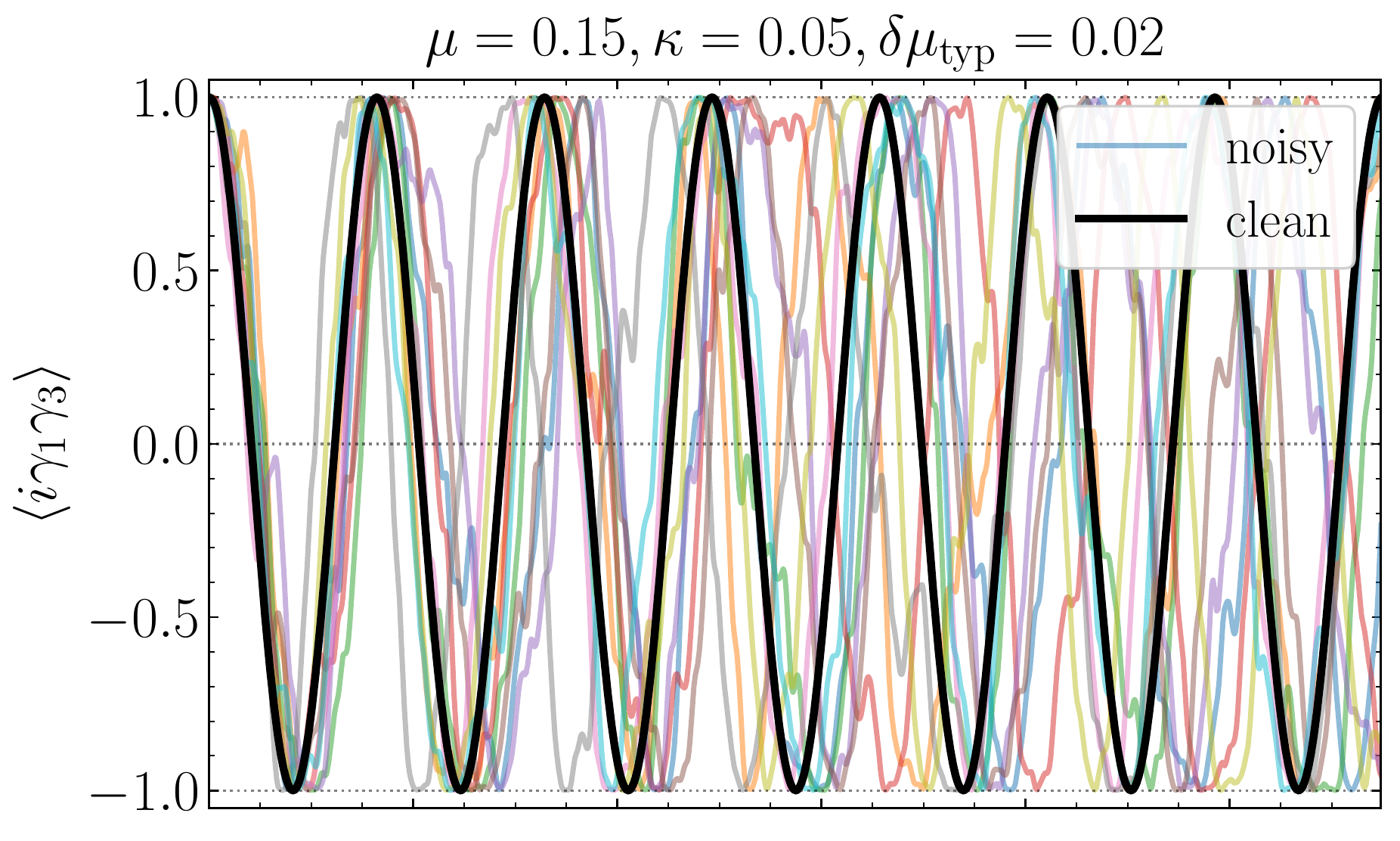}
\includegraphics[width=0.9\columnwidth]{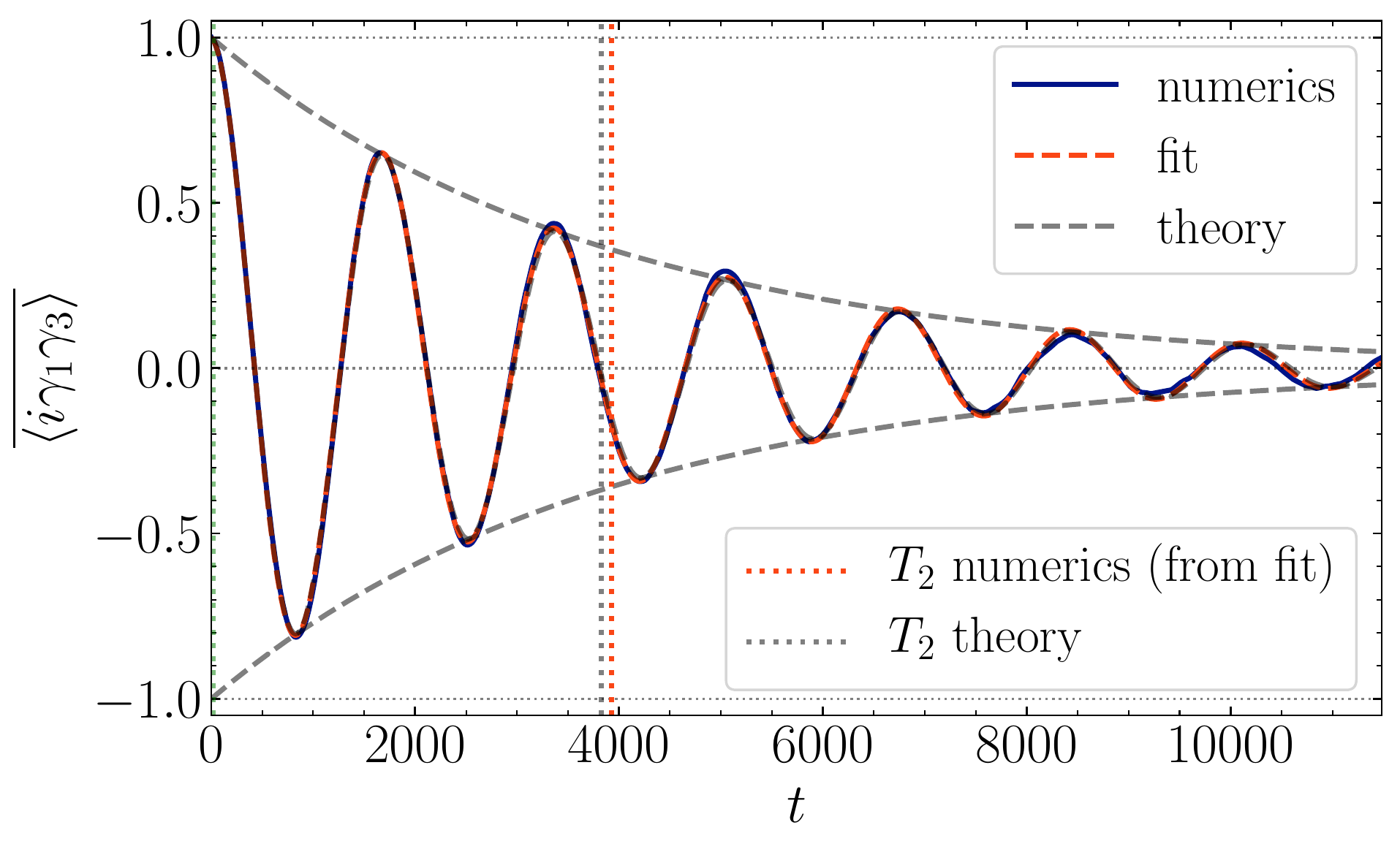}
\end{center}
\caption{Example single noise realizations (top) and noise-averaged data (bottom) for the (quantum-averaged) quantity $\langle i\gamma_1\gamma_3 \rangle$ (with $\gamma_{1,3}=\gamma_{1,3}^{(0)}$) simulated in the Kitaev tetron of Fig.~\ref{fig:kitaev_setup}, here at $\mu=0.15$. The top panel shows the results of ten independent noise realizations (`noisy'; light curves) with $\kappa=0.05$ and $\delta\mu_\mathrm{typ}=0.02$ (one global $\mu$ fluctuator on each wire of the tetron), as well as for a system without noise (`clean'; black curve).
In the bottom panel, we present the noise-averaged signal (`numerics'; solid blue) and a fit (`fit'; dashed orange) guided by the theoretical prediction of Eq.~\eqref{eq:T2analytic_func}. The theoretical prediction itself (and its envelope) obtained using Eqs.~\eqref{omega0_shifted}-\eqref{T2analytic} is also plotted (`theory'; dashed gray). The orange and gray vertical dotted lines indicate the respective $T_2$ times.
\label{fig:noise_averaging}}
\end{figure}

\begin{figure*}
\begin{center} 
\includegraphics[width=0.24\textwidth]{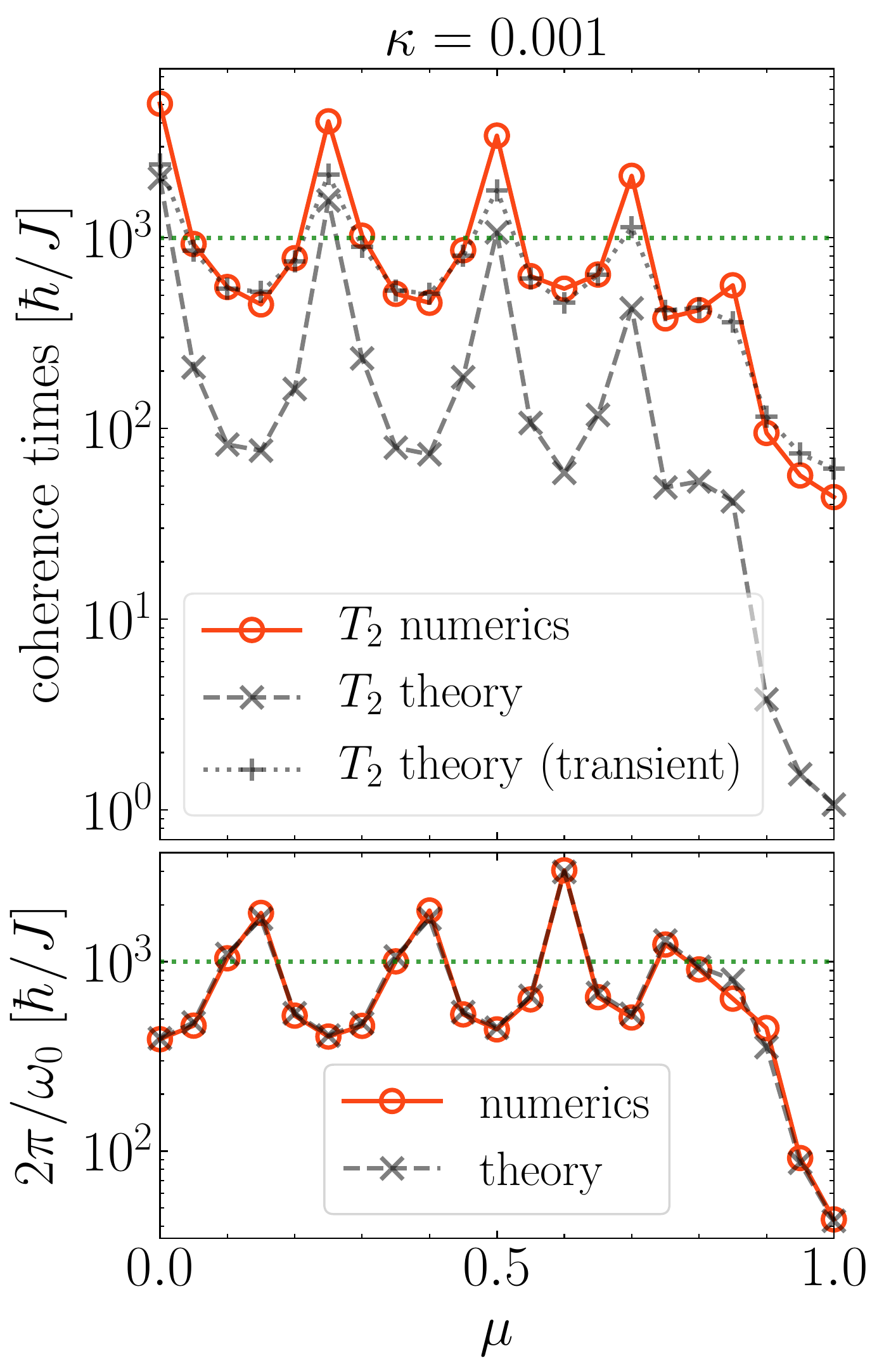}
\includegraphics[width=0.24\textwidth]{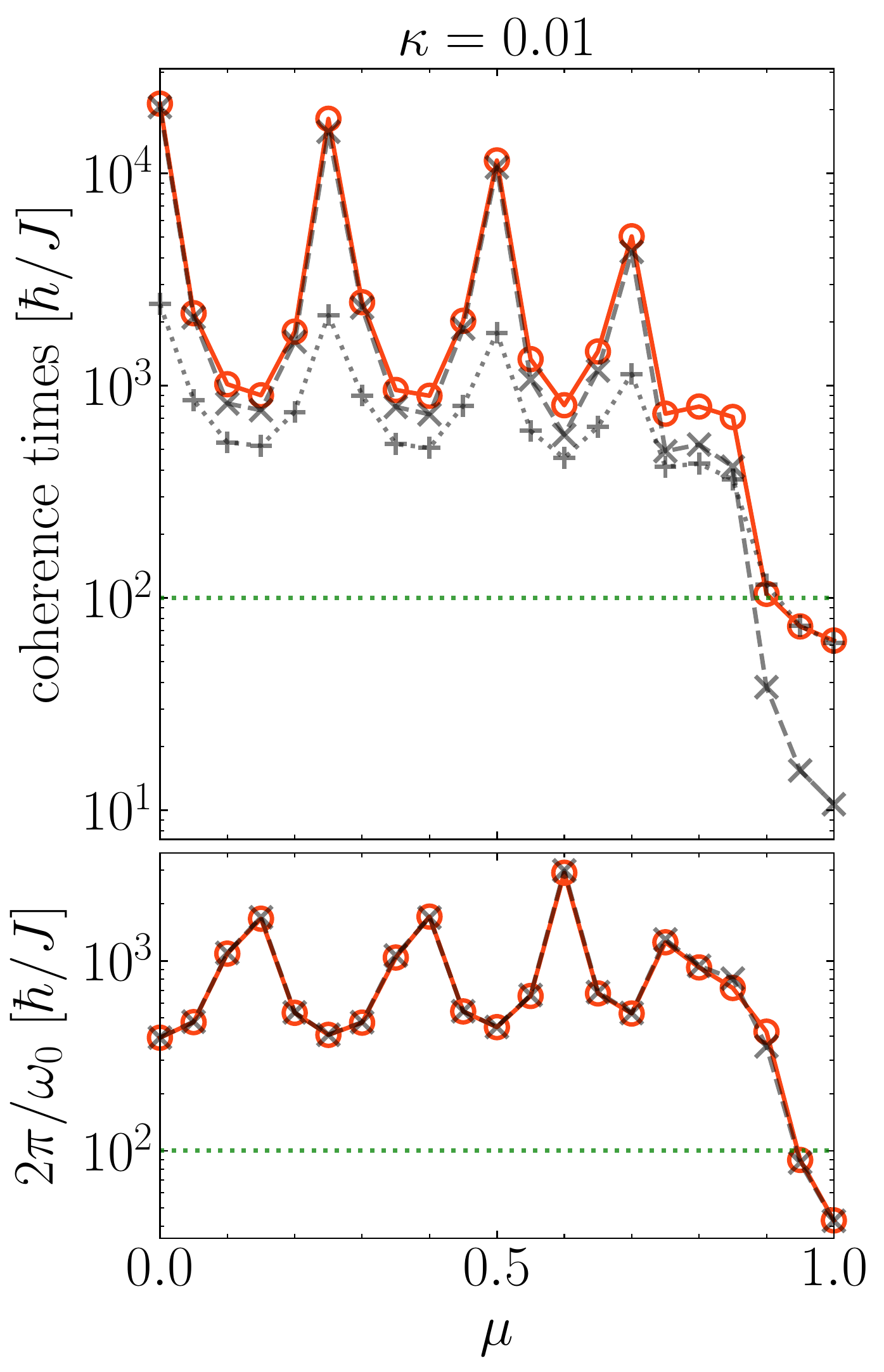}
\includegraphics[width=0.24\textwidth]{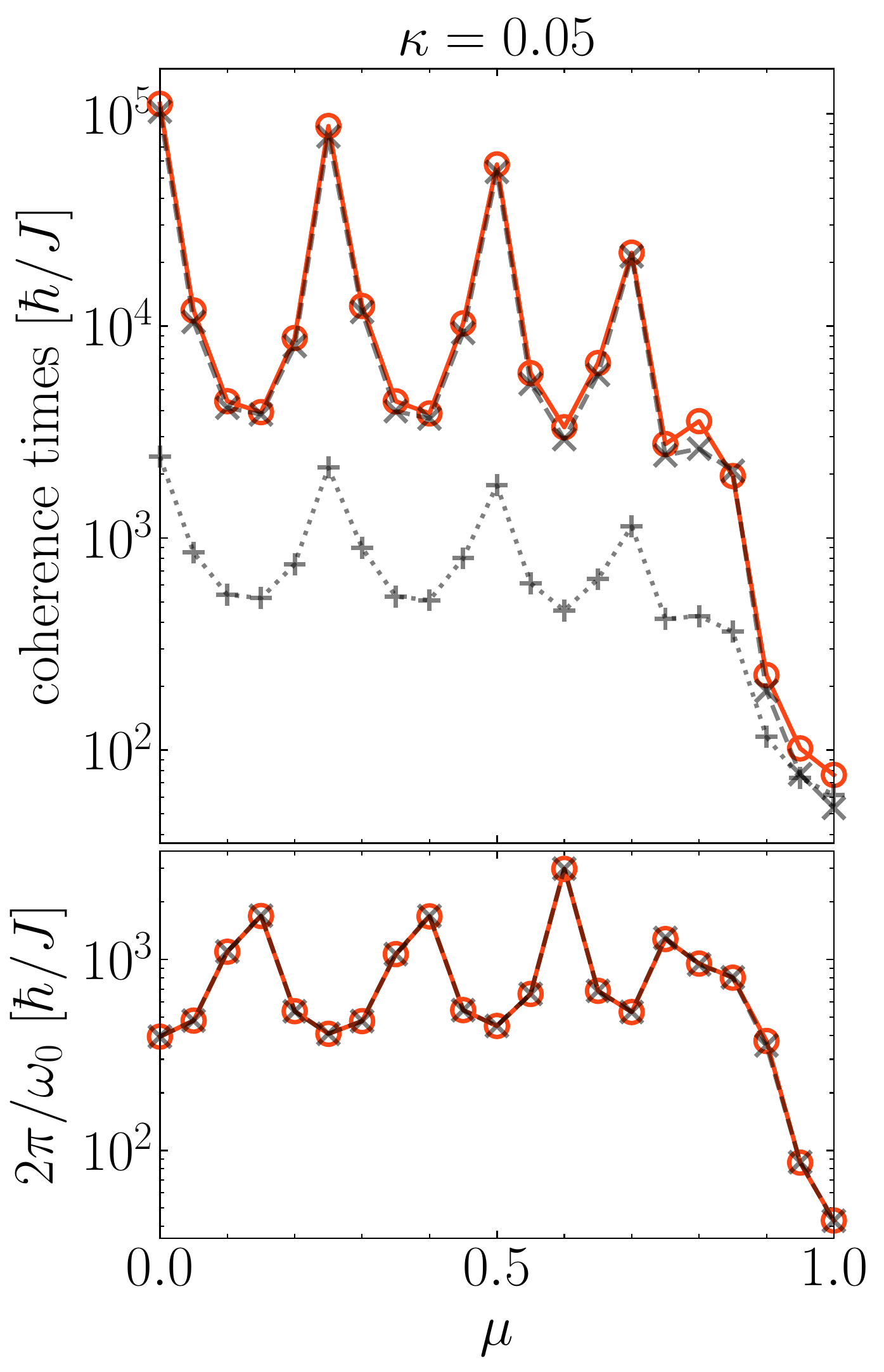}
\includegraphics[width=0.24\textwidth]{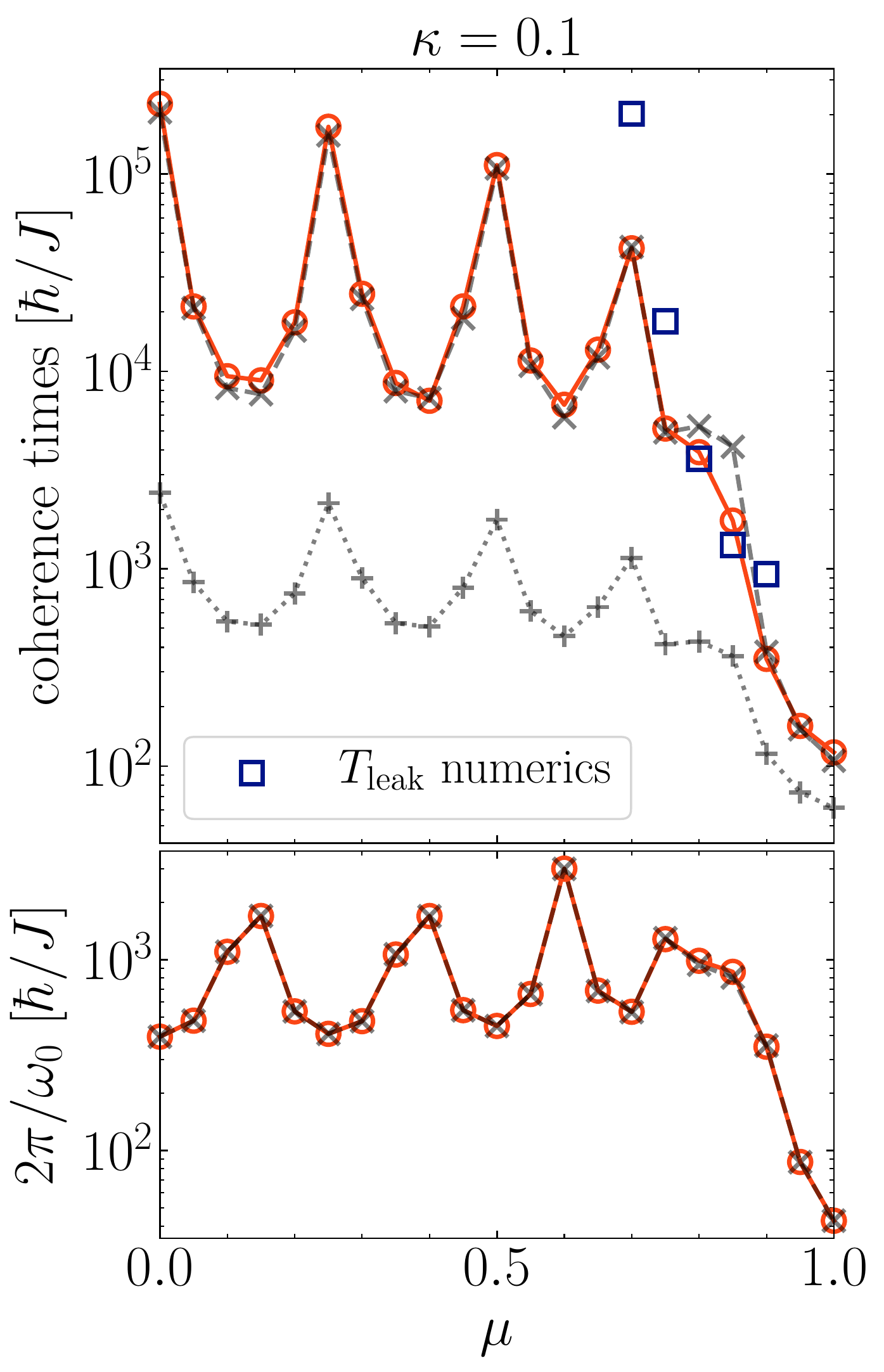}
\end{center}
\caption{Numerically extracted coherence times (top panels) and qubit precession periods (bottom panels) from full microscopic numerical simulations of a noisy Kitaev tetron (see Fig.~\ref{fig:kitaev_setup} and \ref{fig:noise_averaging} for details of the setup). For the dephasing time scales, we plot in the top panels both the values obtained numerically via measurement of $\overline{\langle i\gamma_1\gamma_3\rangle}$ (`$T_2$ numerics') as well as the analytic predictions of Eqs.~\eqref{T2analytic} and \eqref{eq:T2_short-time} [`$T_2$ theory' and `$T_2$ theory (transient)', respectively]; similarly, in the bottom panels, we show both numerically extracted (`numerics') and analytically predicted [Eq.~\eqref{omega0_shifted}] (`theory') precession periods. From left to right, the four panels correspond to increasing $\kappa=0.001,0.01,0.05,0.1$ (roughly the high-frequency cutoff of the noise power spectrum) or, equivalently, decreasing noise correlation times $\tau=1/\kappa$ (the dotted green lines in the two leftmost panels indicate the time scale $\tau$). For all data, we take independent global chemical potential fluctuators on each wire of the tetron, each with typical fluctuation amplitude $\delta\mu_\mathrm{typ}=0.02$. Only at $\kappa=0.1$ and near the phase transition out of the topological phase for $\mu \sim 1$ does the system exhibit detectable leakage times (`$T_\mathrm{leak}$ numerics'); see text for more details. 
\label{fig:comp_Kitaev}}
\end{figure*}

Given such a setup, it is straightforward to numerically evaluate the analytic (perturbative) estimates for the dephasing and leakage times as derived above in Sec.~\ref{sec:analytics}. In addition, we perform full microscopic numerical experiments of the noisy tetron dynamics by generating $N_\mathrm{real}$ independent noise realizations, evolving under the time-dependent Schr\"{o}dinger equation [which in terms of the covariance matrix is given by Eq.~\eqref{eq:dMdt}] for each realization, and finally noise-averaging the resulting measurements. (Details of our noise generation procedure and subsequent solution of the evolution equation can be found in Appendix~\ref{app:noise}.) As spelled out above in Sec.~\ref{sec:protocol}, for our Ramsey-type protocol we initialize the system into a state corresponding to fixed $X_0 = i\gamma^{(0)}_1\gamma^{(0)}_3 = +1$, where the four near-zero-energy Majorana modes $\gamma^{(0)}_{1,2,3,4}$ are defined with respect to the time-averaged Hamiltonian $A_0$. Throughout the evolution we monitor the expectation values $\langle i\gamma_1\gamma_3\rangle$, $\langle i\gamma_1\gamma_3\rangle^2$ (see Sec.~\ref{sec:readout}), and $\langle(i\gamma_1\gamma_2)(i\gamma_3\gamma_4)\rangle$ with $\gamma_{1,2,3,4}$ defined in terms of either the fixed, time-averaged basis used for initialization [i.e., $\gamma_{1,2,3,4} = \gamma^{(0)}_{1,2,3,4}$] or the basis derived from the instantaneous Hamiltonian $A(t)$ [i.e., $\gamma_{1,2,3,4} = \gamma_{1,2,3,4}(t)$].
Finally, we also track the overlap of the evolving wave function onto the low-energy subspace, for both the fixed and instantaneous basis, respectively corresponding to $|a_0|^2 + |b_0|^2$ and $|a_t|^2 + |b_t|^2$ from Eq.~\eqref{psit}.

Within a topological phase, both the qubit precession period $2\pi/\omega_0$ and dephasing time $T_2$ are expected to be exponentially long in the wire length $L$.  Simulating the full microscopic dynamics over these time scales thus becomes prohibitive for modest $L$ even for noninteracting electronic models. Evaluating the relevant time scales using the analytic perturbative predictions, on the other hand, faces no such limitations as this procedure merely requires computing derivatives of the lowest-energy Majorana hybridization energies, which are obtainable with a sparse eigensolver applied to the Hermitian matrix $iA$.

We present data obtained for numerical experiments performed on a tetron built from two Kitaev wires each containing $L = 10$ physical sites, and with varyious $\mu, \kappa$ but all other parameters fixed according to $J_{\mathrm{top}/\mathrm{bot}}=J=1$, $\Delta_{\mathrm{top}/\mathrm{bot}}=\Delta=0.4$, $\mu_\mathrm{offset}=0.01$, and---when noise is present---$\delta\mu_\mathrm{typ}=0.02$.  (With $\hbar=J=1$, energies are in units of $J$, times are in units of $\hbar/J$, and frequencies, e.g., $\kappa$, are in units of $J/\hbar$.)
Figure~\ref{fig:kitaev_setup} shows the energy spectra for this qubit configuration as a function of $\mu$ in the noise-free limit.  Note that for this system size the topological `phase transition' near $\mu = 1$ is significantly smeared by finite-size effects.  Within the topological regime at $|\mu | \lesssim 1$, however, fairly well-formed Majorana zero modes arise. The inset plots the spatial profiles of the near-zero-energy maximally localized Majorana modes $\gamma_{1,2,3,4} = \gamma^{(0)}_{1,2,3,4}$ at $\mu = 0.15$.

Figure~\ref{fig:noise_averaging} contains simulation results for a noisy system with $\mu = 0.15$ and $\kappa = 0.05$.  The top panel shows the evolution of $\langle i\gamma_1\gamma_3\rangle$---defined here in terms of the fixed basis shown in the inset of Fig.~\ref{fig:kitaev_setup}---for ten independent noise realizations.  Also shown for comparison are data for a noiseless run (thick black curve).  The bottom panel shows the results of averaging $N_\mathrm{real} = O(10^3)$ noise realizations \footnote{Error bars are typically on the order of the symbol size or smaller for all data that we present.}, a fit of this noise-averaged numerical data to the functional form $\cos(\omega_0 t) e^{-t/T_2}$, as well as the time dependence predicted \footnote{Throughout, when evaluating the energy splitting $E$ in Eqs.~\eqref{T2analytic} and \eqref{omega0_shifted} for a given microscopic model, for simplicity we take $E = \varepsilon_1 + \varepsilon_2$ [with $\varepsilon_{1,2} \geq 0$, cf.~Eq.~\eqref{eq:Hcanon}] without enforcing a fixed global parity.} by the (small-noise-amplitude) analytic calculations of Sec.~\ref{sec:T2_analytic} [cf.~Eqs.~\eqref{eq:T2analytic_func} and \eqref{T2analytic}]. We see excellent agreement between the numerical data and analytical prediction. These parameters yield negligible leakage (as determined by the relevant measurements discussed above; not shown) on the time scale of the dephasing time.  That is, $T_\mathrm{leak} \gg T_2$.

Finally, in Fig.~\ref{fig:comp_Kitaev}, we present results of a comprehensive study varying $\mu$ at different values of $\kappa=0.001, 0.01, 0.05$, and $0.1$.  The bottom panels plot the qubit precession period $2\pi/\omega_0$ versus $\mu$ as obtained by both the microscopic numerical simulations (`numerics') and analytic prediction (`theory') [see Eq.~\eqref{omega0_shifted} and note that the predicted qubit precession frequency is independent of $\kappa$].  The top panels similarly show the corresponding coherence times $T_2$ and, in the rightmost plot, $T_\mathrm{leak}$. Numerical $T_2$ data corresponds to fits of the noise-averaged, fixed-basis $\langle i\gamma_1\gamma_3\rangle$ data to an oscillatory exponential, precisely as in Fig.~\ref{fig:noise_averaging}. For the analytic $T_2$ predictions, we show the time scales predicted both in the limit $t \gg 1/\kappa$ (`$T_2$ theory') and $t \ll 1/\kappa$ (`$T_2$ theory (transient)'). In the transient case, the decay is expected to follow a Gaussian instead of an exponential [see Eqs.~\eqref{eq:Pt_short-time} and \eqref{eq:T2_short-time} in Appendix \ref{app:averaging}].  As a guide, in the left two panels where the noise is `slow', the horizontal green lines indicate the time scale $\tau = 1/\kappa$.  Coherence times falling below that scale indicate importance of the transient regime.  For consistency, however, we fit the data for all numerical experiments to an exponential form. Nonetheless, when the coherence times fall below $\tau = 1/\kappa$, the extracted time scales from the numerics track the $t \ll 1/\kappa$ prediction reasonably well. To summarize, when the dephasing time measured in the numerics is much larger than $1/\kappa$, it matches the (long time) analytic prediction very well; and when it dips below $1/\kappa$, it follows the (short time, `transient') analytic prediction.

As in the example presented in Fig.~\ref{fig:noise_averaging}, for all of the data points with $\kappa \leq 0.05$ in Fig.~\ref{fig:comp_Kitaev}, we detect negligible leakage out of the low-energy qubit subspace over the time scale of dephasing. Only for $\kappa=0.1$ at $\mu\sim1$ upon exiting the topological phase do we observe appreciable leakage. The quantities $T_\mathrm{leak}$ in the rightmost panel of Fig.~\ref{fig:comp_Kitaev} were extracted by fitting $|a_t|^2 + |b_t|^2$ [see Eqs.~\eqref{psit}-\eqref{eq:a2pb2}] to the (phenomenological) form $Ae^{-t/T_\mathrm{leak}} + (1-A)$; only data points for which $A>0.1$ are plotted. We see from the energy spectra in Fig.~\ref{fig:kitaev_setup} that the dip in leakage times for $\mu\lesssim1$ coincides with the finite-size crossover into the trivial phase on this small $L=10$ site Kitaev tetron, i.e. the point where the gap is minimal. On the other hand, for small $\kappa \leq 0.05$ the minimum finite-size gap exceeds $\kappa$ so that the qubit lifetime is governed by the dephasing time $T_2$ for all $\mu$.

\subsection{Spinful nanowire tetron} \label{sec:spinful}

\begin{figure}[t]
\begin{center} 
\includegraphics[width=0.9\columnwidth]{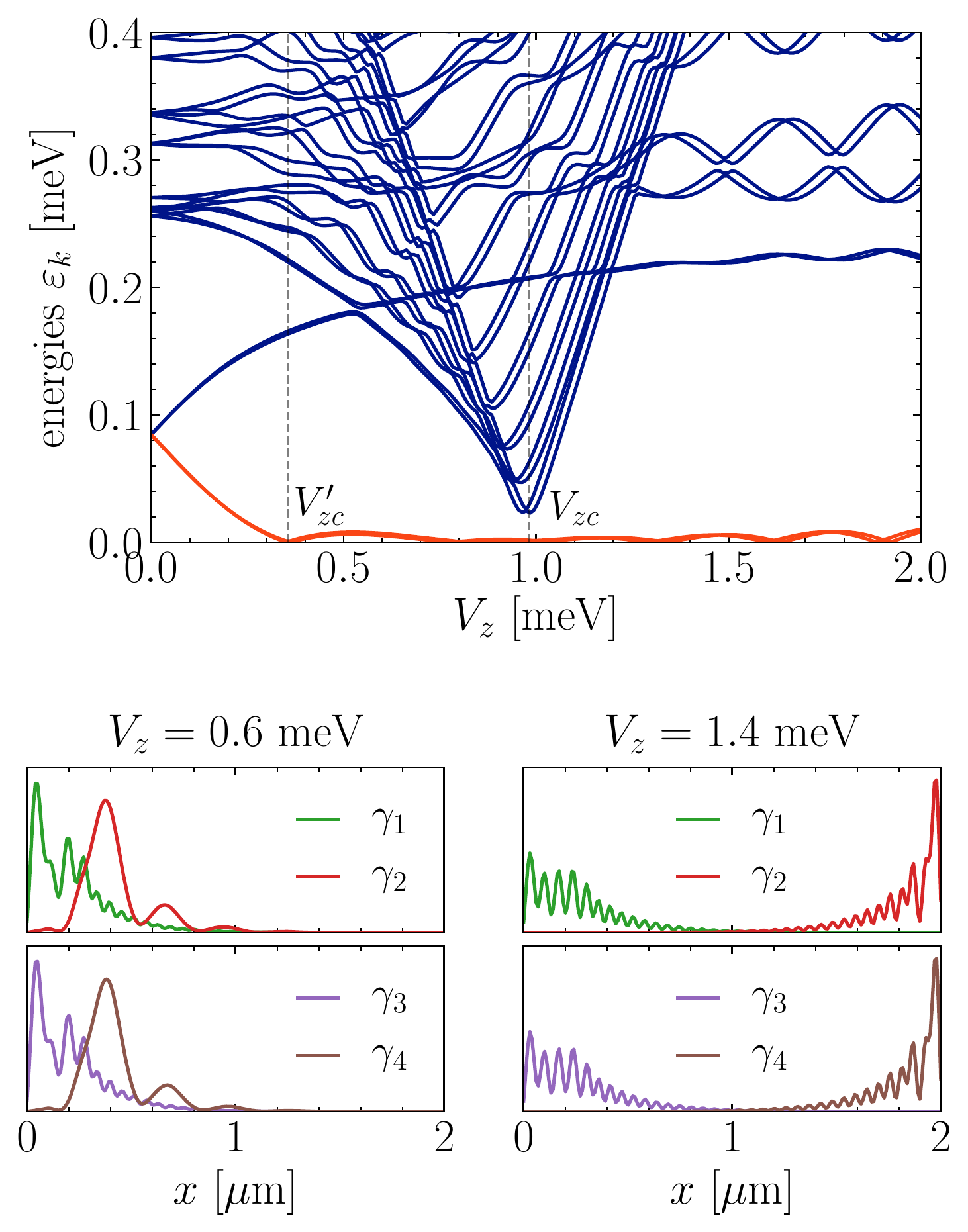}
\end{center}
\caption{Energy spectra (top) and representative near-zero-energy Majorana wave functions (bottom) for the continuum-limit $L=2~\mu\mathrm{m}$ spinful tetron configuration analyzed in Fig.~\ref{fig:Vzscan_summary}. The chosen parameters were inspired by Ref.~\cite{moore_two-terminal_2018} (see text for all details) to give a sizable window of a non-topological ABS regime for $V_{zc}' < V_z < V_{zc}$ before the onset of the topological phase transition at $V_{zc}$. In the bottom panels, we show the square modulus of the wave functions for the maximally localized Majorana modes $\gamma_{1,2,3,4}$ (summing both spin components) at $V_z=0.6~\mathrm{meV}$ (partially-separated ABS qubit) and $V_z=1.4~\mathrm{meV}$ (topological qubit).
\label{fig:spinful_setup_cont}}
\end{figure}

\begin{figure}[t]
\begin{center} 
\includegraphics[width=0.9\columnwidth]{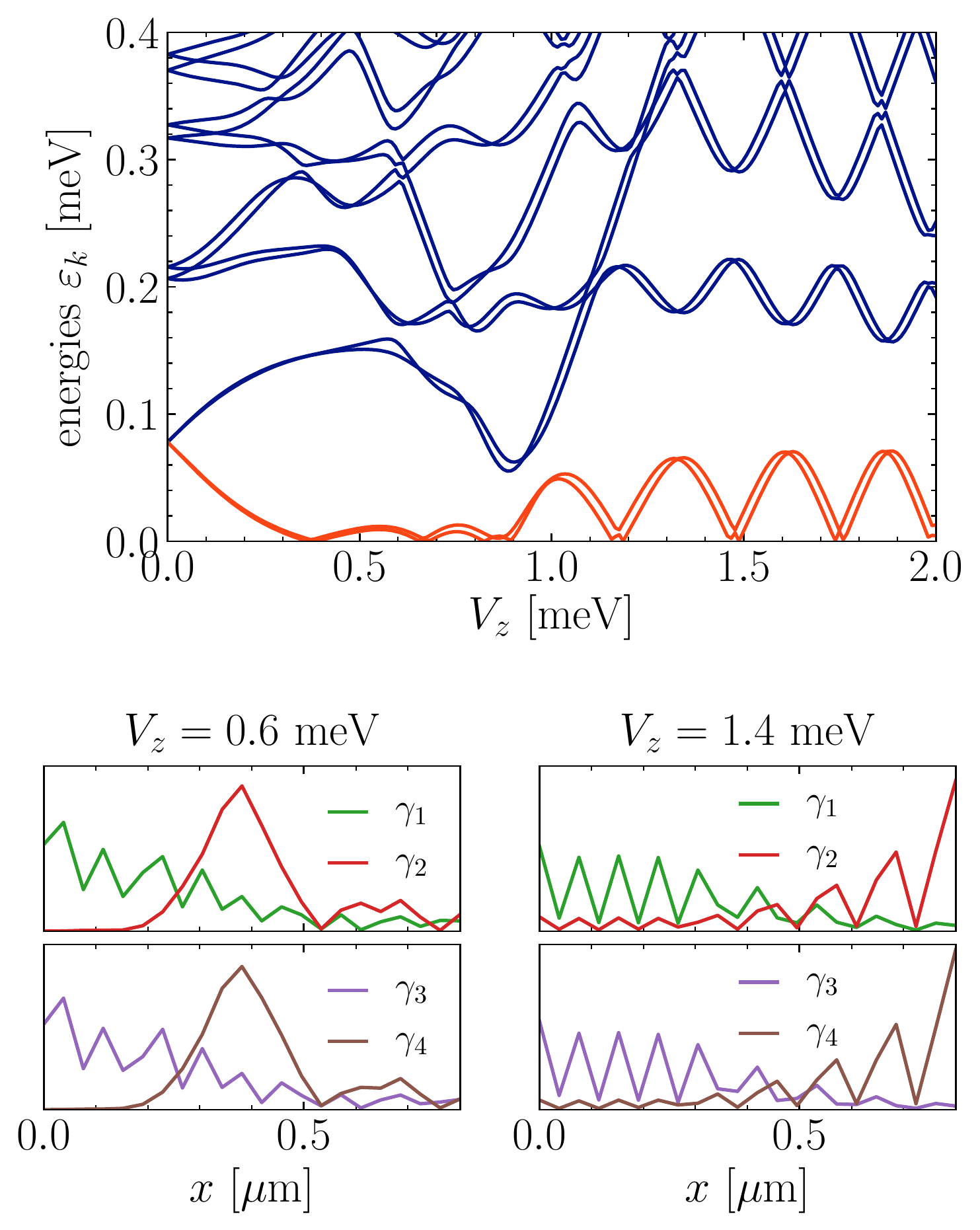}
\end{center}
\caption{Energy spectra (top) and representative near-zero-energy Majorana wave functions (bottom) for the $L=0.8~\mu\mathrm{m}$ spinful tetron configuration studied in Fig.~\ref{fig:comp_spinful}; this data is analogous to Fig.~\ref{fig:spinful_setup_cont} but with some parameter changes---see text for all details. Here we discretize the spinful nanowire model of Eq.~\eqref{eq:nanowire} onto $N_\mathrm{site}=22$ sites per wire to make subsequent full microscopic simulations of the noisy dynamics numerically tractable, while still maintaining several of the qualitative features of the continuum-limit model of Figs.~\ref{fig:Vzscan_summary} and \ref{fig:spinful_setup_cont}.
\label{fig:spinful_setup_disc}}
\end{figure}

We now turn to tetrons assembled from two spinful nanowires governed by the Hamiltonian defined in Eq.~\eqref{eq:nanowire}. Guided by the data presented in Ref.~\cite{moore_two-terminal_2018}, we focus on a set of parameters that yields  low-energy edge ABSs over a sizable window of Zeeman fields $V_{zc}' < V_z < V_{zc}$ before entering a topological regime with bona fide Majorana zero modes at larger fields $V_z > V_{zc}$. Above in Fig.~\ref{fig:Vzscan_summary}, we presented data of qubit coherence times and precession periods for such a system as a function of $V_z$ evaluated numerically according to the analytical estimates of Sec.~\ref{sec:analytics}. The corresponding (time-averaged) system parameters are as follows: 
$\Delta(x)=\frac{\Delta_0}{2}\left[\tanh\left(\frac{x-x_0}{\ell_\Delta}\right)+1\right]$ with $\Delta_0=0.25~\mathrm{meV}$, $\ell_\Delta=0.03~\mu\mathrm{m}$, and $x_0=0.3~\mu\mathrm{m}$; $V(x)=\frac{V_0}{2}\left[-\tanh\left(\frac{x-x_0}{\ell_V}\right)+1\right]$ with $V_0=3.8\Delta_0$ and $\ell_V=0.03~\mu\mathrm{m}$; $m=0.03m_e$; $\mu=V_0=3.8\Delta_0$ (with top/bottom wire offset $\mu_\mathrm{offset}=0.01~\mathrm{meV}$); and $\alpha=500~\mathrm{meV}\,\mathrm{\AA}$. For our discretization, we take a lattice spacing $a\approx0.01~\mu\mathrm{m}$ corresponding to $N_\mathrm{sites}=100,200,300,400$ sites (per wire) for the sequence of sizes $L=1,2,3,4~\mu\mathrm{m}$. In Fig.~\ref{fig:spinful_setup_cont}, we show additional data for the $L=2~\mu\mathrm{m}$ system.  The top panel plots the energy spectrum versus $V_z$ while the bottom panels illustrate the spatial profiles of the maximally localized near-zero-energy Majorana modes $\gamma_{1,2,3,4}$ at $V_z = 0.6~\mathrm{meV}$ (non-topological ABS qubit) and $1.4~\mathrm{meV}$ (topological qubit). The chosen noise configuration consists of four independent fluctuators: one global, spatially constant chemical potential ($\mu$) and Zeeman field ($V_z$) fluctuator per wire each with Gaussian correlated noise with inverse correlation times $\kappa=0.05~\mathrm{meV}/\hbar$ and typical fluctuation amplitudes $\delta\mu_\mathrm{typ}=\delta V_{z,\mathrm{typ}}=0.01~\mathrm{meV}$.

\begin{figure*}[t]
\begin{center} 
\includegraphics[width=0.24\textwidth]{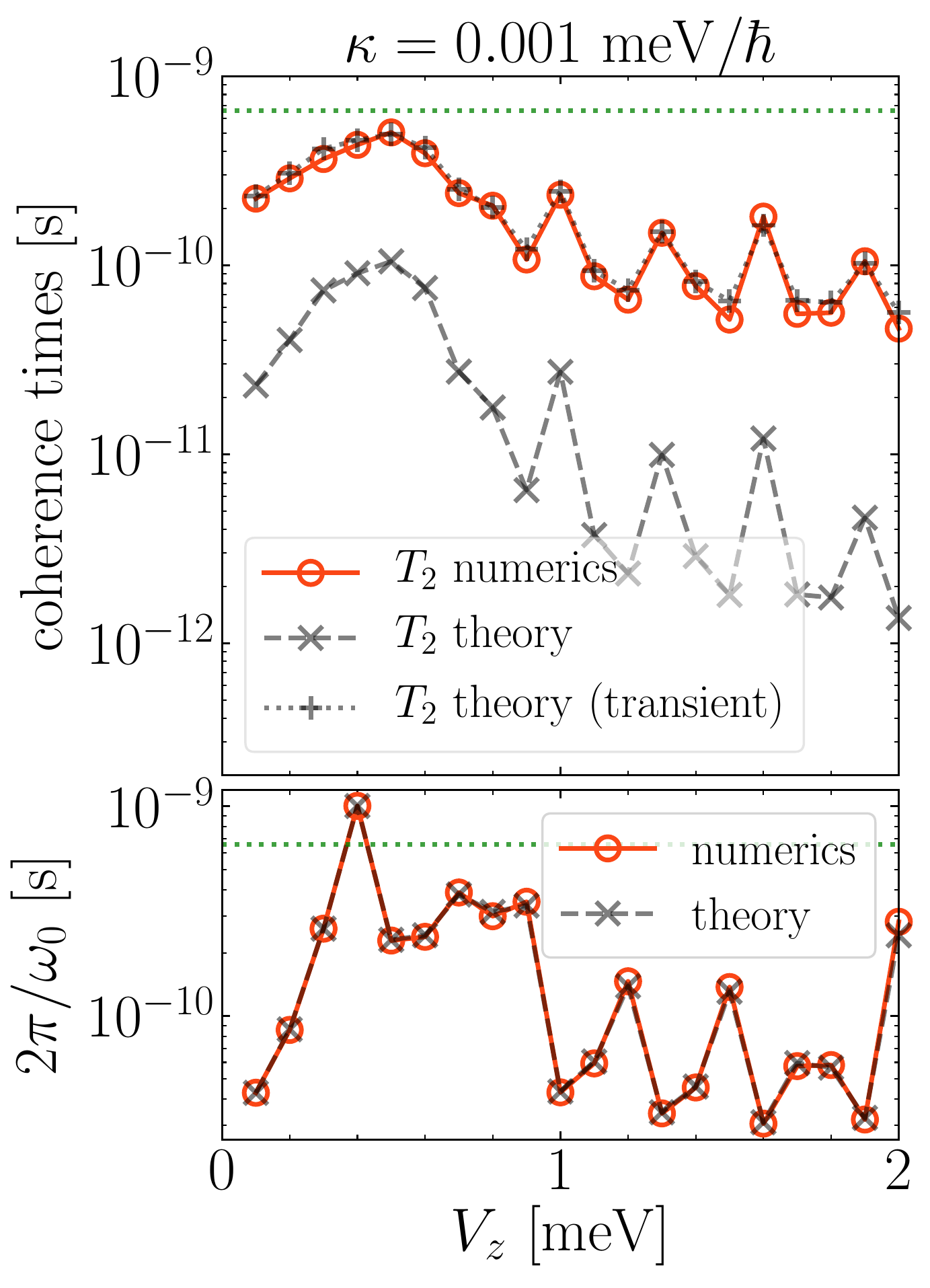}
\includegraphics[width=0.24\textwidth]{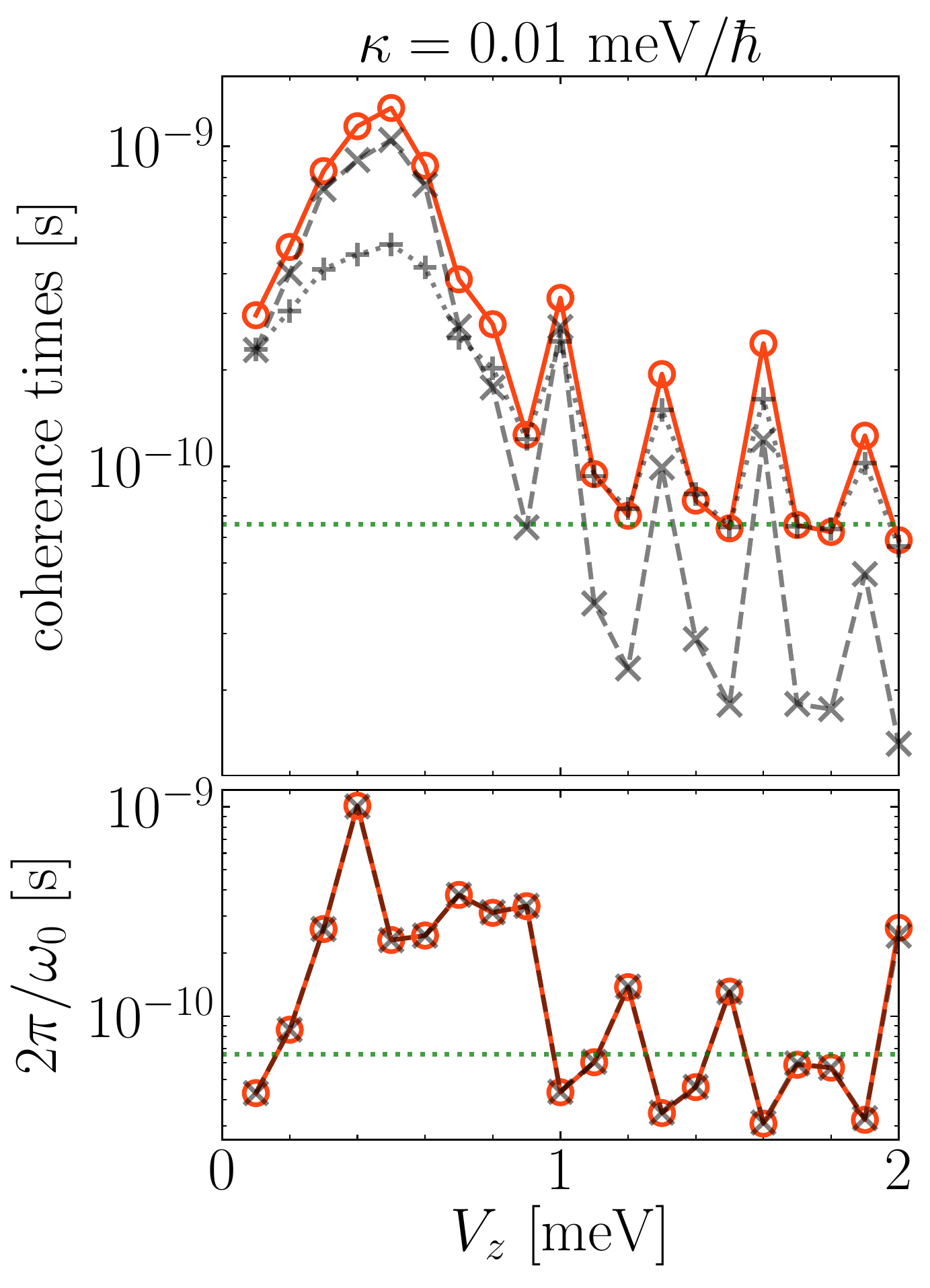}
\includegraphics[width=0.24\textwidth]{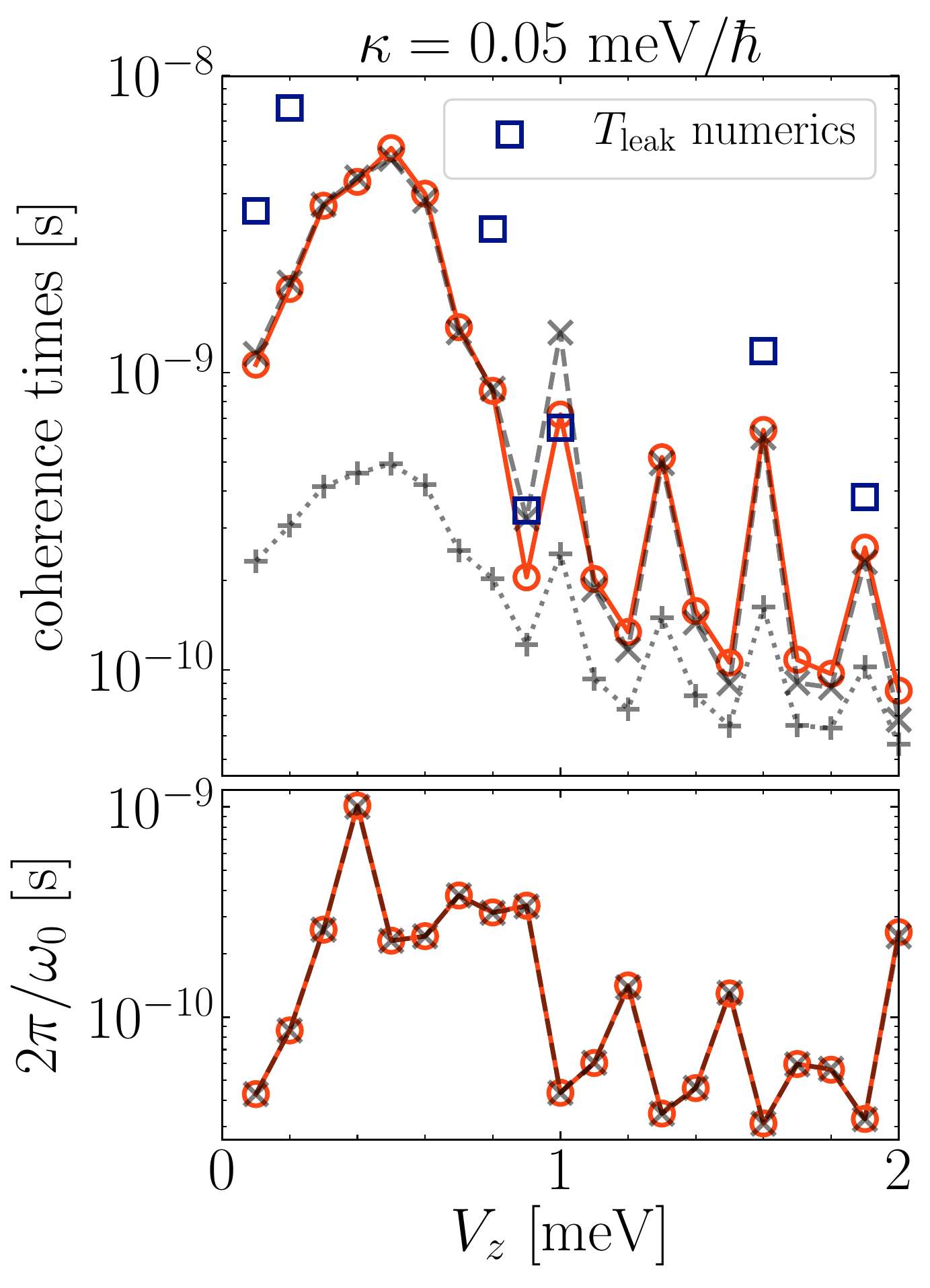}
\includegraphics[width=0.24\textwidth]{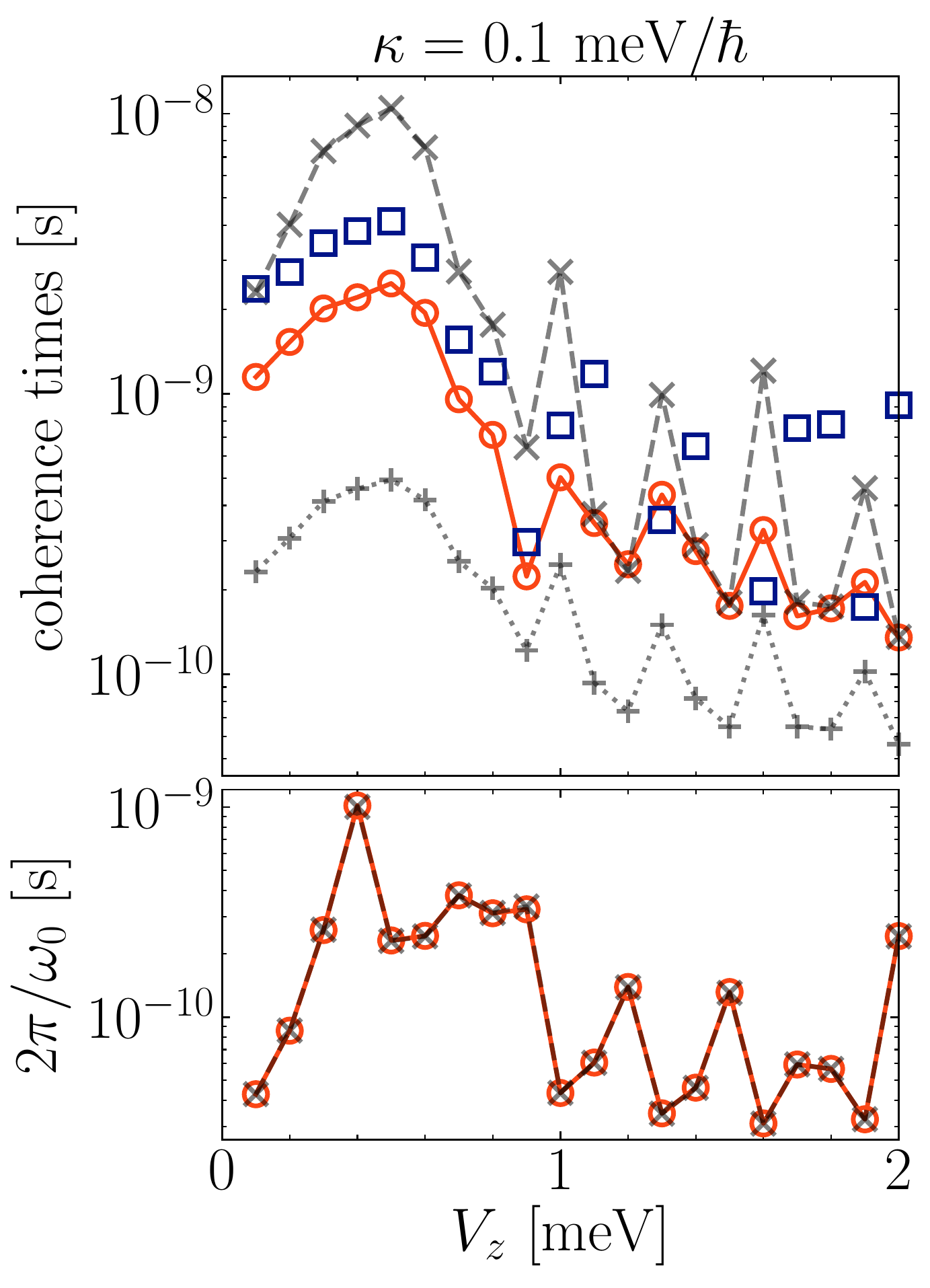}
\end{center}
\caption{Numerically extracted coherence times (top panels) and qubit precession periods (bottom panels) from full microscopic numerical simulations of a noisy spinful nanowire tetron (see text and Fig.~\ref{fig:spinful_setup_disc} for details of the setup). This data is analogous to the Kitaev tetron data of Fig.~\ref{fig:kitaev_setup}; and all formatting, conventions, and means of data analysis are identical. Again, the presence of numerically obtained $T_\mathrm{leak}$ points at $\kappa = 0.05,0.1~\mathrm{meV}/\hbar$ indicate a detectable leakage time via measurement of $\overline{|a_t|^2} + \overline{|b_t|^2}$. Importantly, when this time scale drops below the analytically predicted $T_2$ time, the numerically extracted $T_2$ time (obtained via measurement of $\overline{\langle i\gamma_1\gamma_3\rangle}$) itself gets cut off and drops below the analytic value.
\label{fig:comp_spinful}}
\end{figure*}

Given these parameters, the $T_2$ and $\omega_0$ values plotted in Fig.~\ref{fig:Vzscan_summary} were obtained by numerically evaluating Eqs.~\eqref{T2analytic} and \eqref{omega0_shifted} via sparse diagonalization of the Hermitian matrix $iA$ and computing the requisite energy derivatives. (Only the leading-order terms were included in the evaluations for the more conceptual Fig.~\ref{fig:Vzscan_summary}, while for Figs.~\ref{fig:comp_Kitaev} and \ref{fig:comp_spinful} the second-order terms were also calculated and are in fact necessary to obtain quantitative agreement with the microscopic numerics.) For $T_\mathrm{leak}$ we take a simplified approach which focuses entirely on the predicted exponential dependence: $T_\mathrm{leak} \sim e^{{(E_g/\hbar\kappa)}^2}$. Specifically, we evaluate Eq.~\eqref{Tleak_analytic} at fixed $c_i = 1$, $\eta_i = 0$, $b_i = 0$, $v = \alpha/\hbar~(= 500~\mathrm{meV}\,\mathrm{\AA}/\hbar)$, and $\xi = 200~\mathrm{nm}$ (recall $D_i = \delta\mu_\mathrm{typ} = \delta V_{z,\mathrm{typ}} = 0.01~\mathrm{meV}$ and $\kappa_i = \kappa = 0.05~\mathrm{meV}/\hbar$). While the analysis of Sec.~\ref{sec:Tleak_analytic} considered only a single nanowire, it can be safely applied to the noisy tetron considered here as the wires are taken to be decoupled and the spatial profiles of all noise sources have support on only one wire or the other (two global fluctuators per wire). In evaluating Eq.~\eqref{Tleak_analytic}, we thus consider only two fluctuators, and for the excitation gap $E_g$, we take an average of the excitation gaps [$\varepsilon_{3,4}$ from Eq.~\eqref{eq:Hcanon}] for the top and bottom wires (recall the finite $\mu_\mathrm{offset}$). This procedure ignores the complicated dependencies of the (qualitatively less important) subexponential factors on the wave function amplitudes, noise profile details, etc., but it faithfully captures the dominant exponential dependence and thus suffices for our purposes here.

As highlighted in Sec.~\ref{sec:executive}, Fig.~\ref{fig:Vzscan_summary} demonstrates several important points: First, within the ABS regime ($V_{zc}' < V_z < V_{zc}$) the qubit coherence time and precession frequency exhibit weak, nonuniversal dependence on wire length $L$.  Second, the qubit lifetime drops sharply near the topological phase transition ($V_z \sim V_{zc}$) due to noise-induced leakage out of the qubit subspace, leading to a leakage-limited qubit.  Third, deep in the topological phase the system forms an exponentially protected, dephasing-limited topological qubit (though for still longer wires the topological qubit will eventually become leakage limited as the bulk excitation gap is roughly independent of $L$).  Fourth, within the topological phase the dephasing time and qubit precession period oscillate out of phase and eventually decrease upon increasing $V_z$, reflecting an enlargment of the coherence length $\xi$ at `large' $V_z$.

Performing full microscopic simulations of such noisy spinful tetrons containing hundreds of lattice sites in a manner that parallels the Kitaev simulations presented in Fig.~\ref{fig:comp_Kitaev} is numerically intractable. We can, however, construct a small-$L$  `toy' system far from the continuum limit of Eq.~\eqref{eq:nanowire} that is amenable to simulations and  exhibits qualitatively similar features to the systems considered in Fig.~\ref{fig:Vzscan_summary}. For this system, we choose a set of parameters [including functional forms of the external potentials $V(x)$ and $\Delta(x)$] identical to those used in Figs.~\ref{fig:Vzscan_summary} and \ref{fig:spinful_setup_cont} with the following exceptions: now $\alpha=400~\mathrm{meV}\,\mathrm{\AA}$, $\ell_V=0.08~\mu\mathrm{m}$, and $L=0.8~\mu\mathrm{m}$ discretized into $N_\mathrm{sites}=22$ (per wire) such that the `lattice constant' $a = \frac{L}{N_\mathrm{sites}-1} \approx 0.038~\mu\mathrm{m}$. In Fig.~\ref{fig:spinful_setup_disc}, we show the corresponding energy spectra as a function of external Zeeman field $V_z$ as well as the wave function amplitudes for the maximally localized Majorana modes $\gamma_{1,2,3,4}$ at $V_z = 0.6~\mathrm{meV}$ (ABS qubit) and $1.4~\mathrm{meV}$ (`topological' qubit). For the noise setup, we again take one global $\mu$ and $V_z$ fluctuator on each wire of the tetron with typical fluctuation amplitudes $\delta\mu_\mathrm{typ}=\delta V_{z,\mathrm{typ}}=0.01~\mathrm{meV}$ and identical inverse correlation times $\kappa$. 

Using these parameters, we show in Fig.~\ref{fig:comp_spinful} data for a spinful tetron analogous to the Kitaev tetron results of Fig.~\ref{fig:comp_Kitaev}. Here, as in Fig.~\ref{fig:Vzscan_summary}, we sweep over $V_z$; the four panels correspond to $\kappa=0.001,0.01,0.05,0.1~\mathrm{meV}/\hbar$. The extraction of the numerical (fitted) time scales (`numerics') and evaluation of the analytical time scales (`theory') were carried out in exactly the same manner as in Fig.~\ref{fig:comp_Kitaev}. Again, in the two leftmost panels at small $\kappa=0.001,0.01~\mathrm{meV}/\hbar$ we (1) see the importance of the transient regime when comparing numerical $T_2$ times to the theoretical predictions and (2) observe no detectable leakage out of the ground state manifold on the simulated time scales for all $V_z$. As seen in Fig.~\ref{fig:comp_spinful}, the crossover from the ABS regime to topological regime occurs near $V_z\approx0.9~\mathrm{meV}$ where the excitation gap is minimal. In this vicinity, for $\kappa=0.05~\mathrm{meV}/\hbar$, we see that the numerically obtained leakage times become on the order of or smaller than the theoretically predicted dephasing times. In turn the \emph{numerical} dephasing times drop below the analytic prediction: the noise-averaged signal $\overline{\langle i\gamma_1\gamma_3\rangle}$ is now damped by both dephasing effects (cf.~Fig.~\ref{fig:noise_averaging}) as well as leakage effects in which the amplitude of oscillations for \emph{individual noise realizations} also decay in time (see also Fig.~\ref{fig:meas_basis_mu0.8}). While our theoretical analysis in Sec.~\ref{sec:analytics} treated the effects of dephasing and leakage independently, our full microscopic simulations as presented here are able to capture both effects simultaneously and faithfully. (The same physics is also occurring near the crossover out of the topological phase in the rightmost panel of Fig.~\ref{fig:comp_Kitaev}.)

Finally, at $\kappa=0.1~\mathrm{meV}/\hbar$ in the rightmost panel of Fig.~\ref{fig:comp_spinful}, we observe an intricate interplay between dephasing and leakage. Here, $\kappa$ is sufficiently large to cutoff the numerically extracted $T_2$ time below the pure dephasing limit even in the ABS regime for $V_z\lesssim0.9~\mathrm{meV}$. In the vicinity of the crossover we see a sharp reduction in the leakage time due to gap closing, thereby limiting the qubit lifetime as discussed above in the context of Fig.~\ref{fig:Vzscan_summary}. (Note that for the particular qubit simulated in Fig.~\ref{fig:comp_spinful}, due to e.g.~the extremely short wire length, even pure dephasing effects cause a reduction in coherence times in the `topological' regime relative to the ABS one; there is thus a concomitant drop near the corresponding crossover which is unrelated to the leakage-induced qubit lifetime reduction due to gap closing.) Within the `topological' phase, for points corresponding to a minimal excitation gap [for example at $V_z=1.3,1.6,1.9~\mathrm{meV}$ (cf.~Fig.~\ref{fig:spinful_setup_cont})], $\kappa=0.1~\mathrm{meV}/\hbar$ is again sufficiently large to cause leakage to largely dictate the qubit coherence time. (In principle, leakage-limited lifetimes deep within the topological phase could also occur for sufficiently fast noise in the continuum limit analytical study of Fig.~\ref{fig:Vzscan_summary}; however, studying these effects quantitatively in that framework would require a less crude evaluation of the analytical $T_\mathrm{leak}$ prediction.) On the other hand, for other values of $V_z$ with excitation gaps in excess of $\hbar\kappa=0.1~\mathrm{meV}$, the qubit lifetime is limited by dephasing, and the numerically extracted time scales match the analytic $T_2$ prediction impressively well.

\subsection{Estimation of coherence times for realistic nanowire models} \label{sec:realistic}

\begin{figure}
    \centering
    \includegraphics[width=\columnwidth]{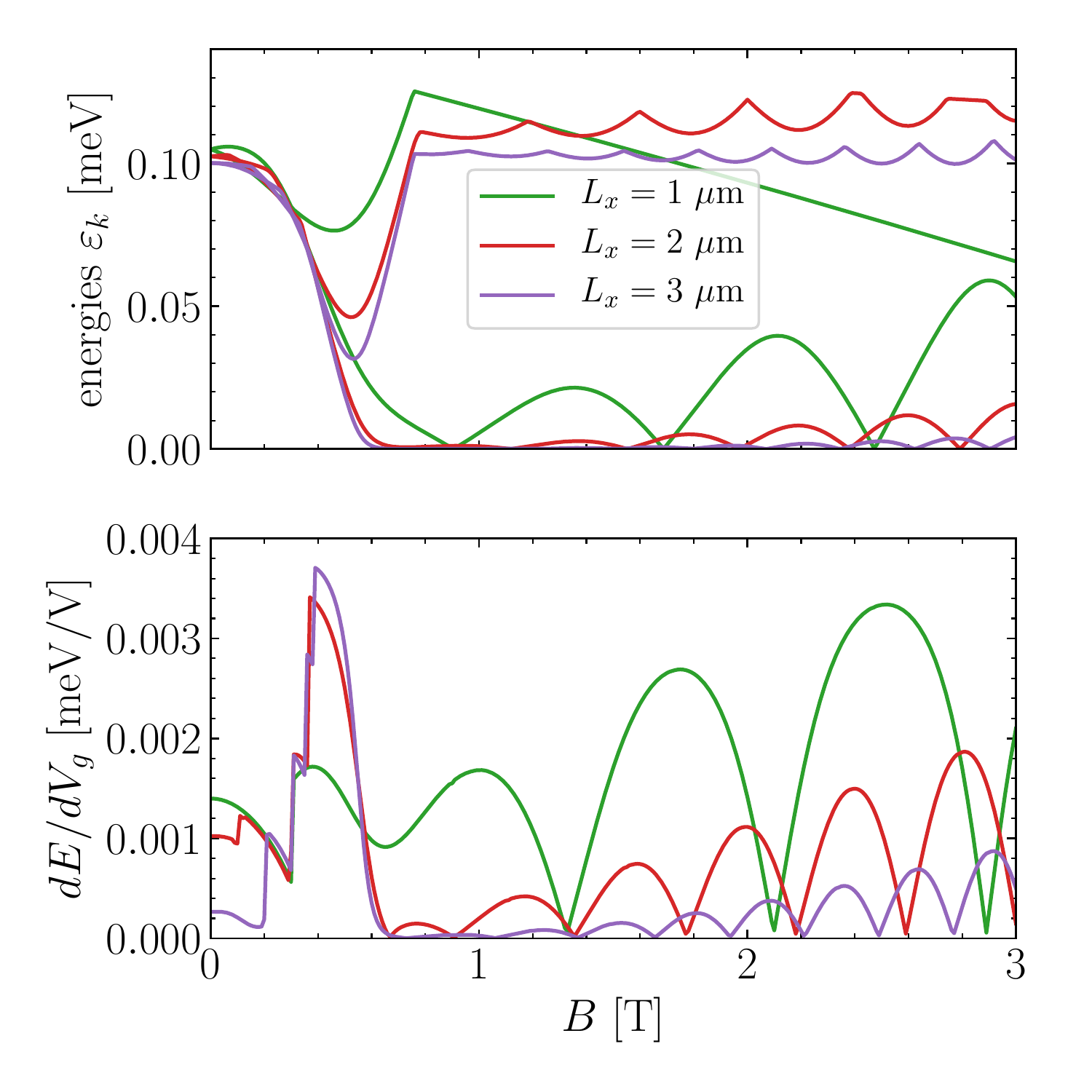}
    \caption{Dephasing analysis of `slab geometry' nanowire. The upper panel shows the low-energy spectrum for a single wire, while the lower panel shows the derivative of the qubit splitting $E$ with respect to gate voltage $V_g$. Given some knowledge about the noise environment, this quantity can be directly related to the dephasing time $T_2$ via Eq.~\eqref{T2analytic}. For a detailed list of parameters used in these simulations, see main text. We find that also for this more realistic, multi-band model of a nanowire, the qualitative features such as exponential dependence of dephasing time on the length of the wire as well as the correlation between splitting amplitude and dephasing time are preserved.}
    \label{fig:slab}
\end{figure}

A key advantage of the approach outlined in Sec.~\ref{sec:T2_analytic} to estimate dephasing times is that it relies purely on the derivative of the Majorana splitting with respect to the fluctuating variables. Such simple spectral properties can be obtained using sparse diagonalization techniques and are thus easy to evaluate also for models where a computation of the full dynamics would be prohibitively expensive. To further illustrate this point, we perform such an estimate for a more sophisticated model of a nanowire that include a realistic electrostatic potential and several subbands. We closely follow the approach of Ref.~\cite{antipov2018}, which considers a rectangular wire of dimensions $L_x \times L_y \times L_z$ proximitized on the top by a superconductor of thickness $d_z$; here, $z$ is the direction transverse to the superconductor and $x$ is the long direction of the wire, i.e., $L_x \gg L_y, L_z$. We simplify the model in two key ways: First, we ignore the effect of subbands in the $y$ direction.  And second, instead of treating the superconductor explicitly, we directly induce superconductivity in the semiconductor through a mean-field pairing term. We perform simulations with fixed $L_z = 60{\rm\ nm}$ and variable $L_x$. 

The first-quantized normal-state Hamiltonian of the system can be written as
\begin{align} \label{eq:H_N}
H_{\rm N} =& -\frac{1}{2m^*} \left( \partial_x^2 + \partial_z^2 \right) \!-\! \alpha \hat k_x \sigma_y \\
& \!+\!\phi(z) \!+\! \frac{\mu_B g B}{2} \sigma_x, \nonumber
\end{align}
with $m^*$ the effective mass, $\alpha$ the strength of spin-orbit coupling, $\phi(z)$ the electrostatic potential, $g$ the $g$-factor of the semiconducting material, $B$ the external magnetic field, and $\sigma_\alpha$ the Pauli matrices acting on spin. We use $m^* = 0.026$, $g=-15$, $\alpha = 0.05{\rm\ eV\ nm}$.
Using a Nambu-space notation, where $\tau_\alpha$ are Pauli matrices acting in particle-hole space, we can write the Hamiltonian of the superconducting system as
\begin{equation}
    H = H_{\rm N} \tau_z - \Delta \sigma_y \otimes \tau_y,
\end{equation}
where $\Delta$ denotes the strength of superconducting pairing, which we set to $\Delta = 0.1{\rm\ meV}$. The electrostatic potential $\phi(z)$ is obtained from a self-consistent Thomas-Fermi calculation, where the boundary condition near the superconductor is set to an assumed band offset of $300{\rm\ meV}$, while the boundary condition at the other end is tuned via an electrostatic gate with an applied voltage $V_g$. We use a gate voltage of $V_g = -0.288{\rm V}$, which tunes the system close to the bottom of a band and thus favors the formation of a topological phase.

The spectrum is evaluated by employing a finite-difference approximation, with a regular discretization of $a = 2{\rm\ nm}$, and using a shift-and-invert eigensolver. Results are shown in Fig.~\ref{fig:slab}.  Here, we evaluate the derivative of the energy splitting $\varepsilon$ with respect to the applied gate voltage $V_g$. Using Eq.~\eqref{T2analytic}, this estimate can be combined with an estimate for the noise correlation time as well as typical fluctuation amplitude to obtain an estimate for the $T_2$ dephasing time. We note that this calculation can in principle be extended to more realistic models of the system, including explicit treatment of the superconductor and even full three-dimensional models of the device. However, major uncertainy enters the estimates through the properties characterizing the noise, which can only be extracted from experiments.

\section{Effects of fixed versus instantaneous measurement basis} \label{sec:fixedvinst}

In this section, we investigate in more quantitative detail the effects of encoding and measuring the qubit in the basis of maximally localized Majorana modes associated with the \emph{instantaneous} Hamiltonian, as opposed to using a \emph{fixed} basis throughout the time evolution. In our numerical simulations presented in Sec.~\ref{NumericalResults}, we used the fixed basis of maximally localized Majorana modes $\gamma_{1,2,3,4}=\gamma_{1,2,3,4}^{(0)}$ corresponding to the \emph{time-averaged} Hamiltonian specified by $A_0$ to initialize the qubit.  Furthermore, for technical reasons described in Sec.~\ref{sec:readout}, we often employed this same basis for readout of quantities such as $\langle i\gamma_1\gamma_3\rangle$. For a given noise realization, the initial Hamiltonian will differ from the time-averaged one, i.e., $A(t=0) \neq A_0$, and hence this procedure introduces a sort of `quench' in the dynamics at $t=0$, the effects of which become more pronounced for large typical fluctuation amplitudes $D_i$. For the small, seemingly reasonable $D_i$ chosen above this effect is negligible.  We expect, however, the difference in measurement basis choice to manifest itself most clearly when $D_i$ becomes large. Here, it is worth reiterating an important point made in Sec.~\ref{sec:UnitaryEvolution}: Fluctuations, especially those of large amplitude, may significantly alter the maximally localized Majorana modes (e.g., their locations), but in the purely adiabatic limit the qubit subspace is still preserved provided that one tracks and measures in the instantaneous basis. On the other hand, measuring in a fixed basis throughout the evolution will naturally underestimate qubit coherence times, as the noisy dynamics will generically induce a `leakage' out of the fixed basis over time, even if the qubit subspace is perfectly preserved by the instantaneous basis.

Initializing using the time-averaged basis $\gamma_{1,2,3,4}^{(0)}$ is clearly problematic for addressing these matters at large $D_i$; e.g., even at $t=0$ the system may have measurable leakage out of the qubit subspace derived from the instantaneous basis. For this section, we thus employ a different initialization procedure from that used in Sec.~\ref{NumericalResults}: For each noise realization, we find a maximally localized set of near-zero-energy Majorana modes $\gamma_{1,2,3,4}(t=0)$ derived from $A(t=0)$; we subsequently set $i\gamma_1(t=0)\gamma_3(t=0)=+1$ and $i\gamma_2(t=0)\gamma_4(t=0)=\pm1$, with the latter chosen such that the global fermion parity is always fixed to $+1$. We then compare the results of two different measurement bases: (1) the initial basis defined by the $\gamma_{1,2,3,4}(t=0)$ (which is fixed for each noise realization) and (2) the instantaneous basis defined by $\gamma_{1,2,3,4}(t)$. [For small enough $D_i$, we have checked that using the former initial/fixed measurement basis produces results which are numerically indistinguishable from the initialization/readout procedure used in Sec.~\ref{NumericalResults} based on the $\gamma_{1,2,3,4}^{(0)}$.]

\begin{figure}[t]
\begin{center}
\includegraphics[width=0.49\columnwidth]{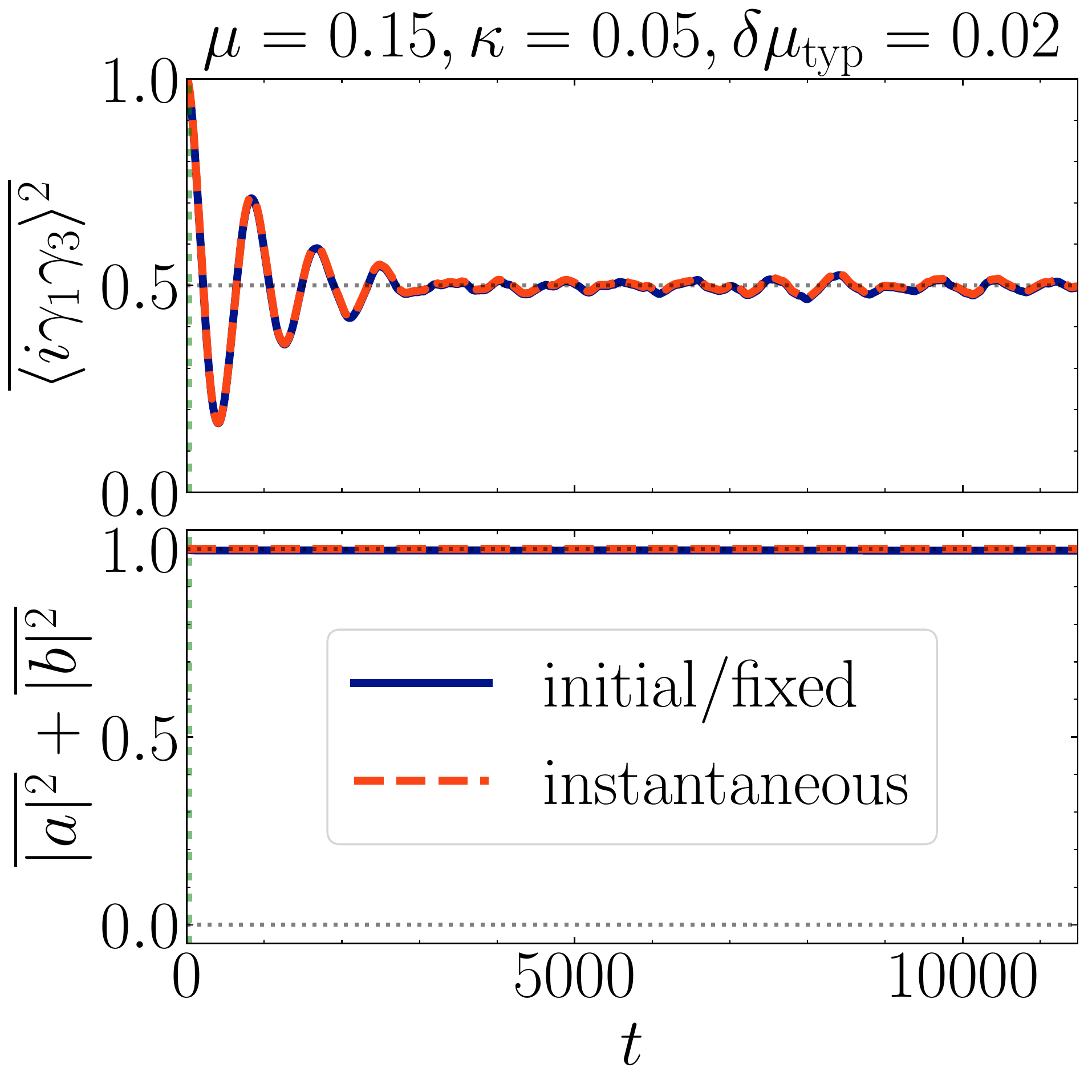}
\includegraphics[width=0.49\columnwidth]{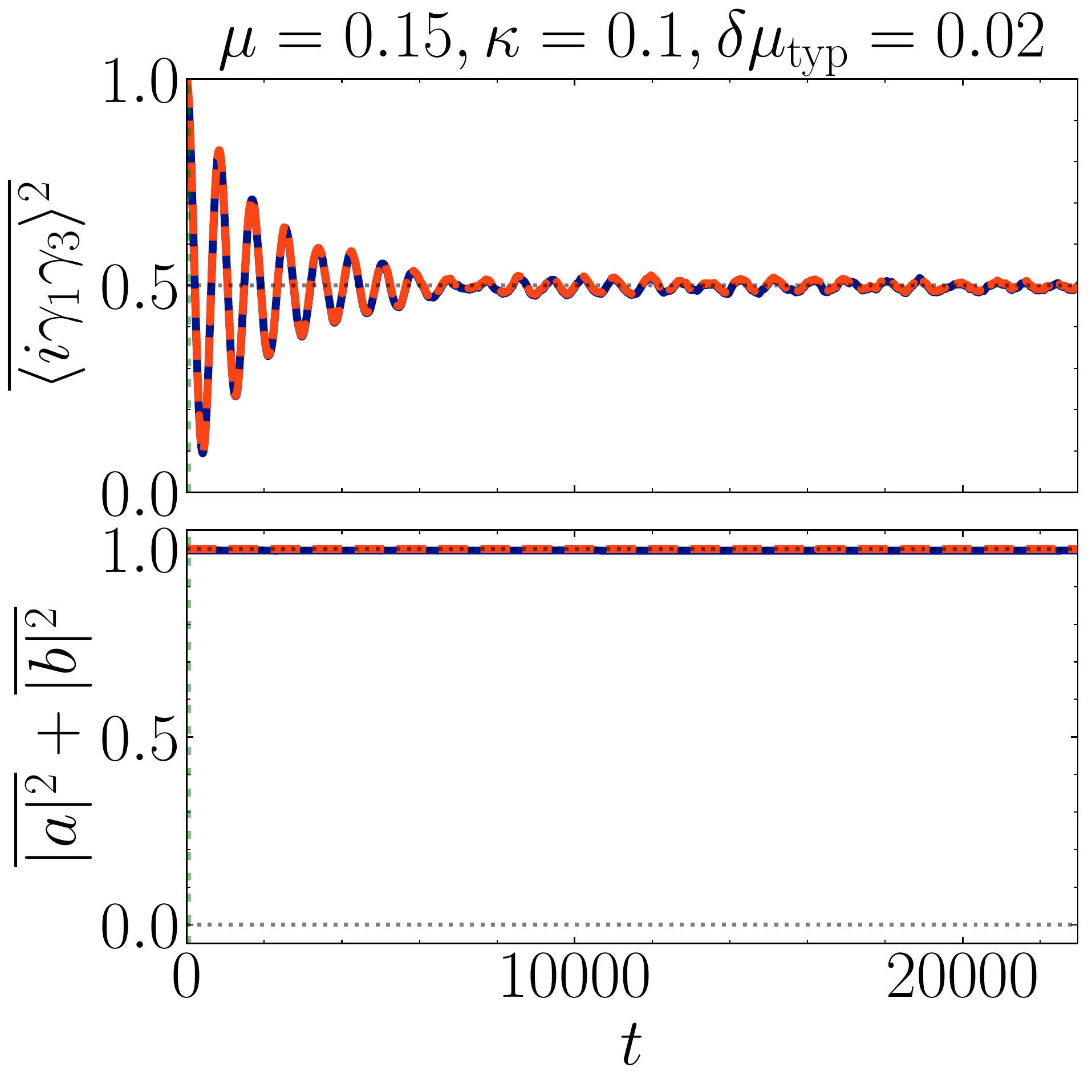}
\end{center}
\begin{center}
\includegraphics[width=0.49\columnwidth]{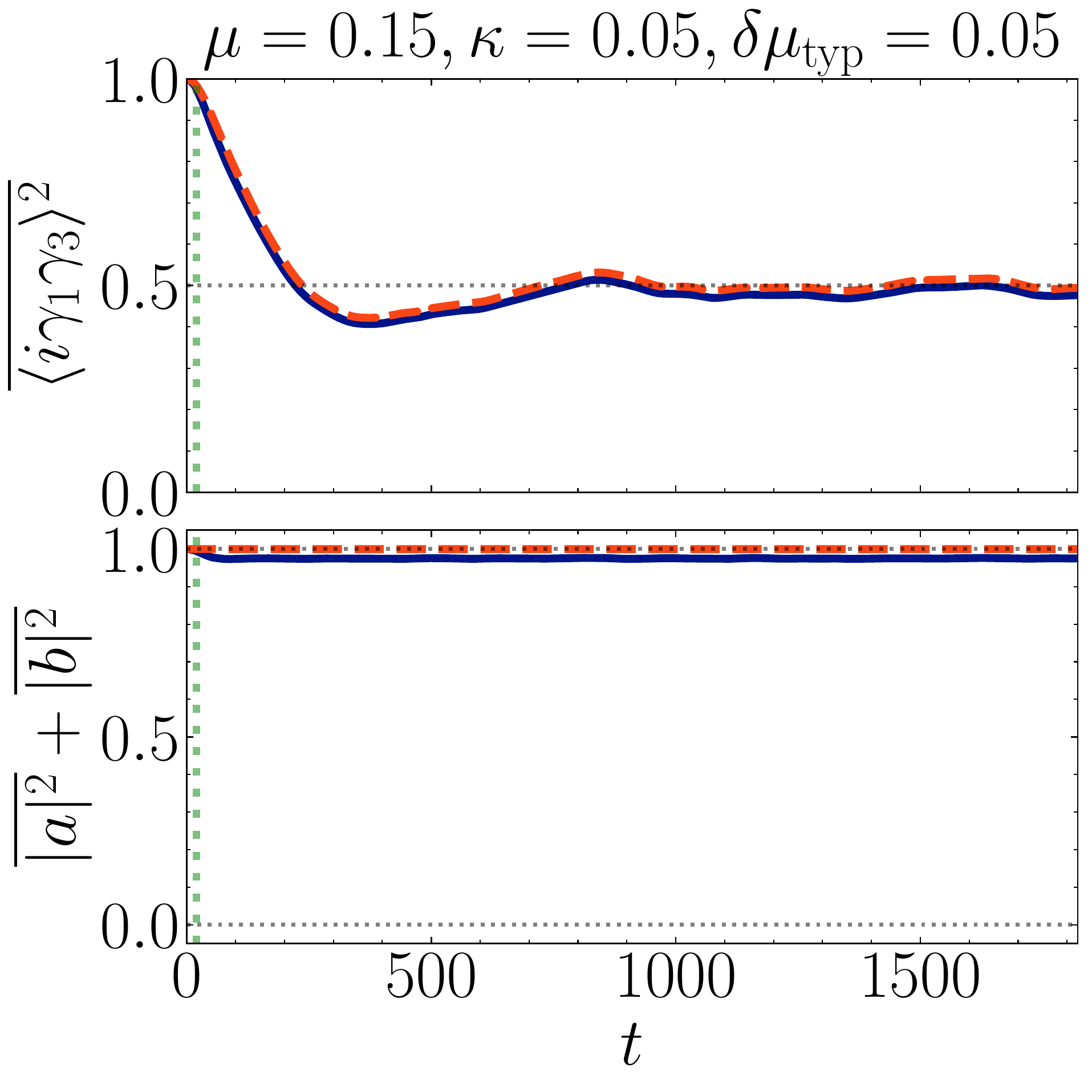}
\includegraphics[width=0.49\columnwidth]{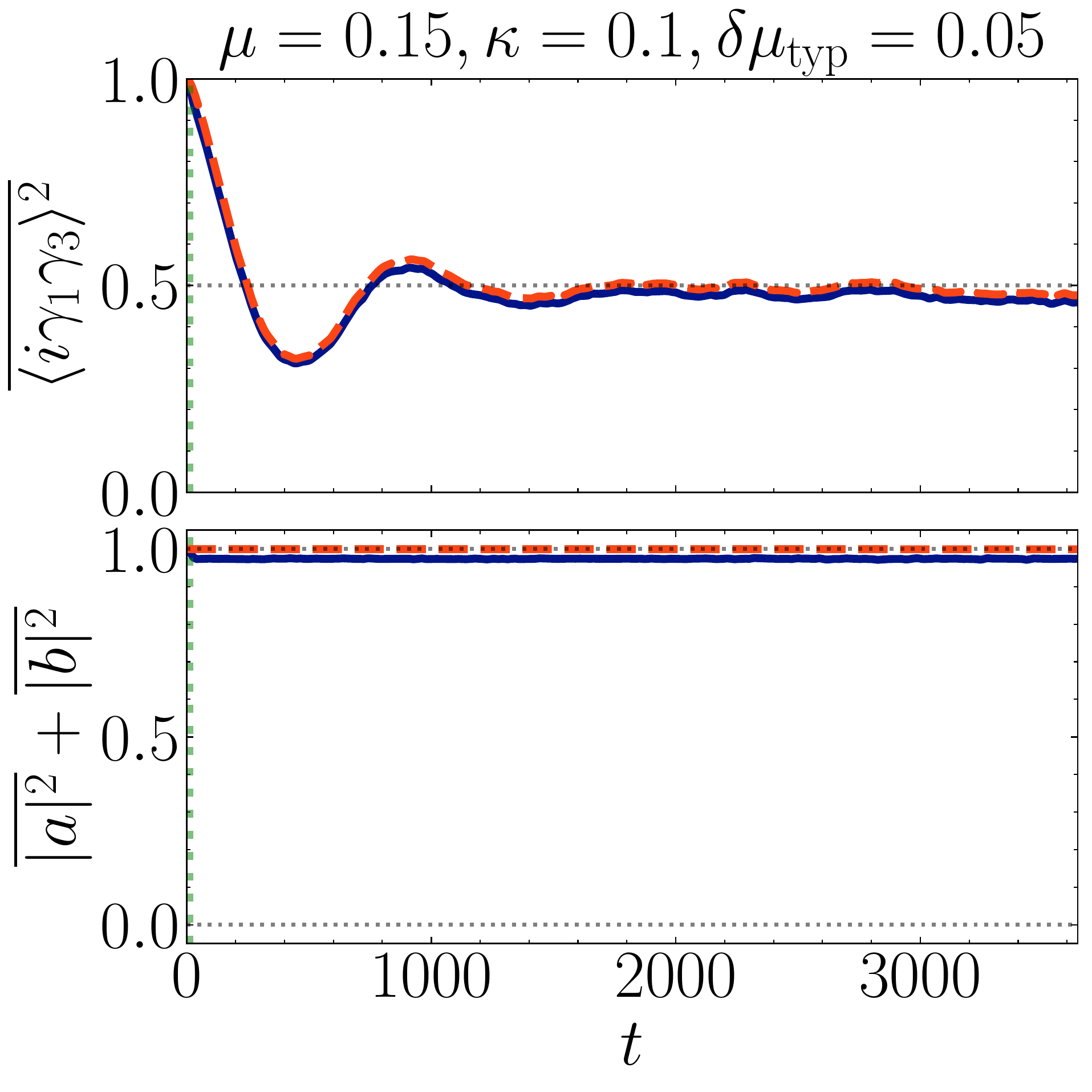}
\end{center}
\vspace{-0.2in}
\begin{center}
\includegraphics[width=0.49\columnwidth]{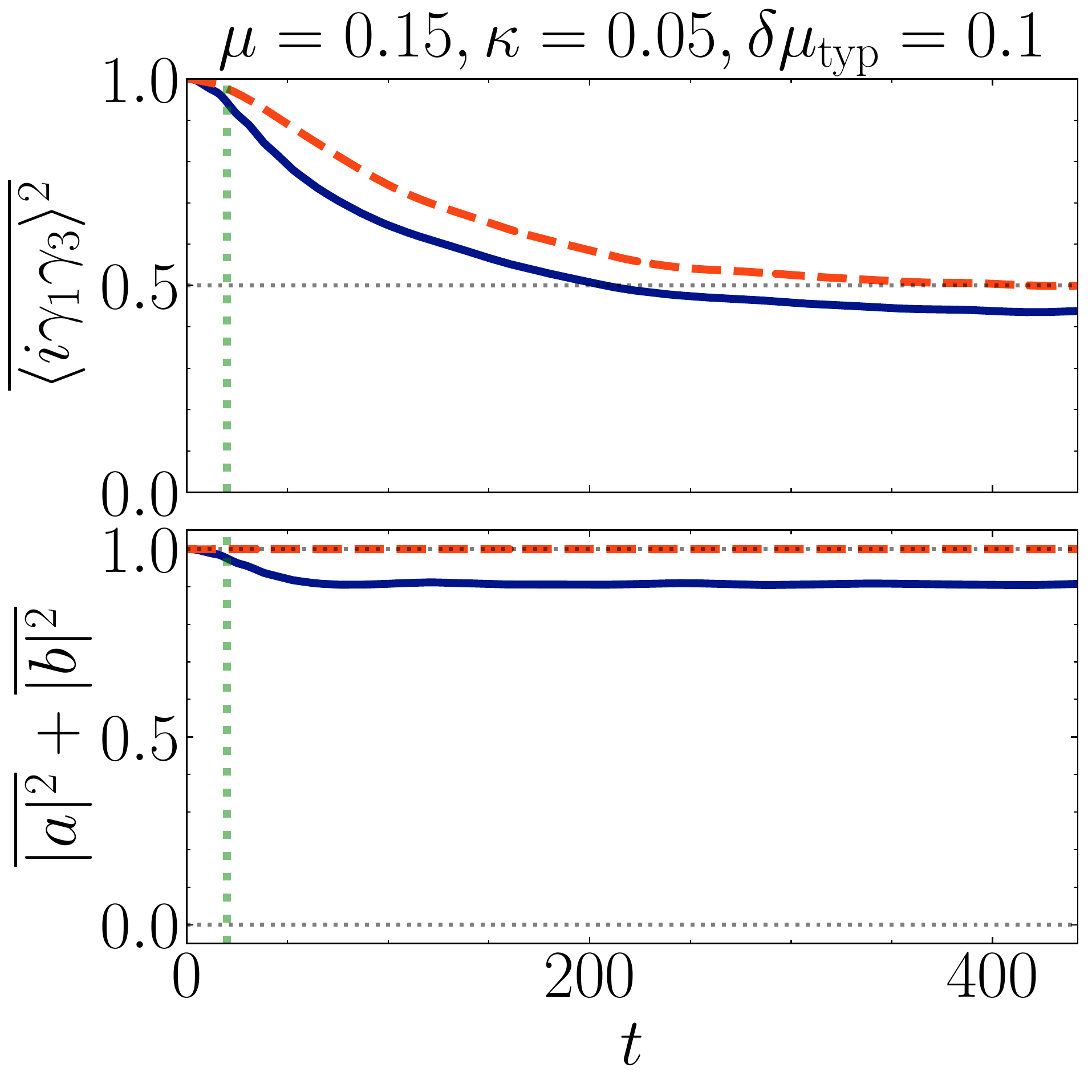}
\includegraphics[width=0.49\columnwidth]{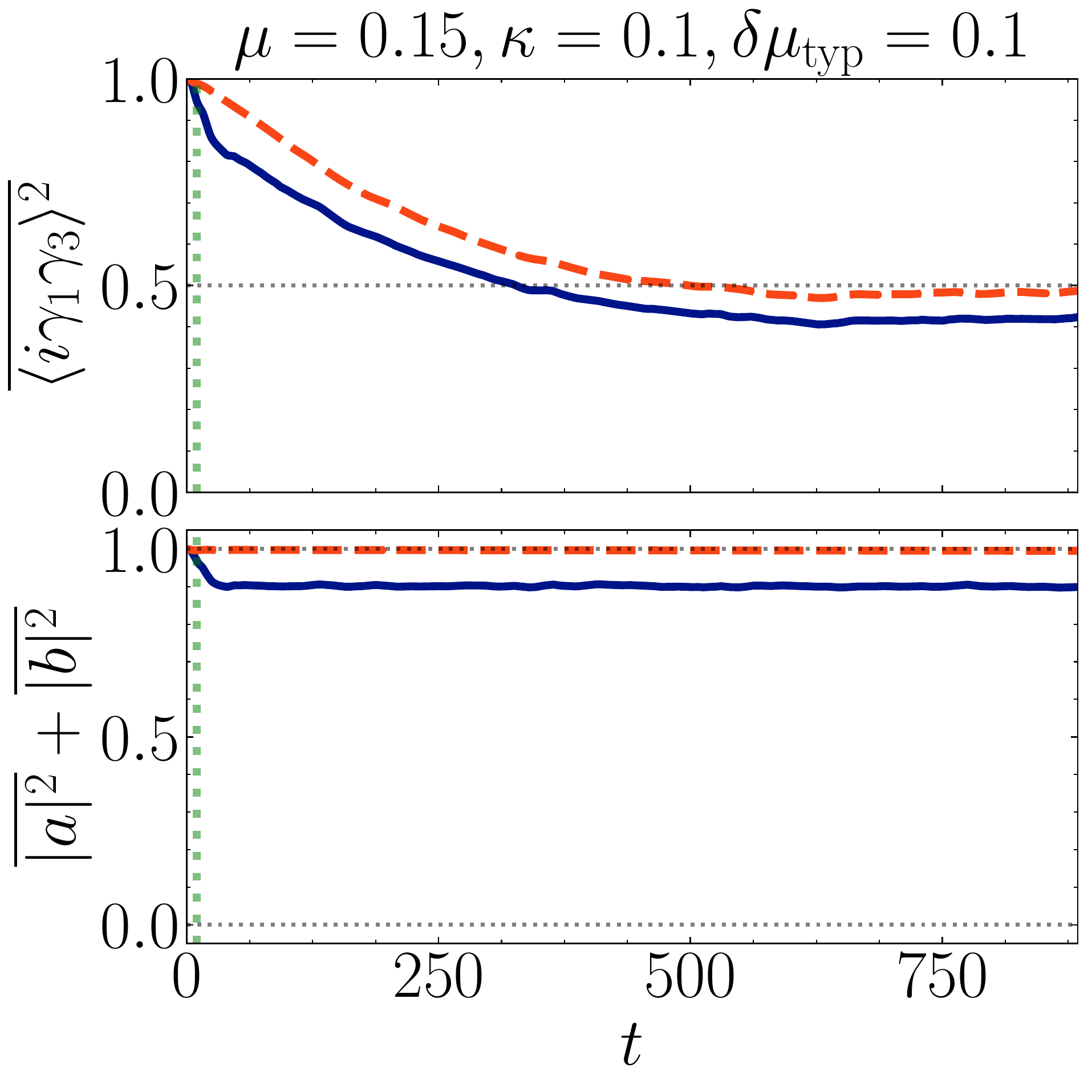}
\end{center}
\caption{Measurements of $\overline{\langle i\gamma_1\gamma_3\rangle^2}$ and $\overline{|a|^2} + \overline{|b|^2}$ in both the fixed/initial (solid blue) and instantaneous (dashed orange) basis for a Kitaev tetron at $\mu=0.15$, $\kappa=0.05,0.1$ (increasing left to right), and $\delta\mu_\mathrm{typ}=0.02,0.05,0.1$ (increasing top to bottom); see text for details of other chosen parameters. Here we initialize the system using a procedure based on the \emph{initial} Hamiltonian encoded by $A(t=0)$. For large enough $\delta\mu_\mathrm{typ}$, a noticeable difference in the choice of measurement basis is observed on a time scale on the order of $\tau=1/\kappa$ (vertical green dotted lines).
\label{fig:meas_basis_mu0.15}}
\end{figure}

\begin{figure}[t]
\begin{center}
\includegraphics[width=0.49\columnwidth]{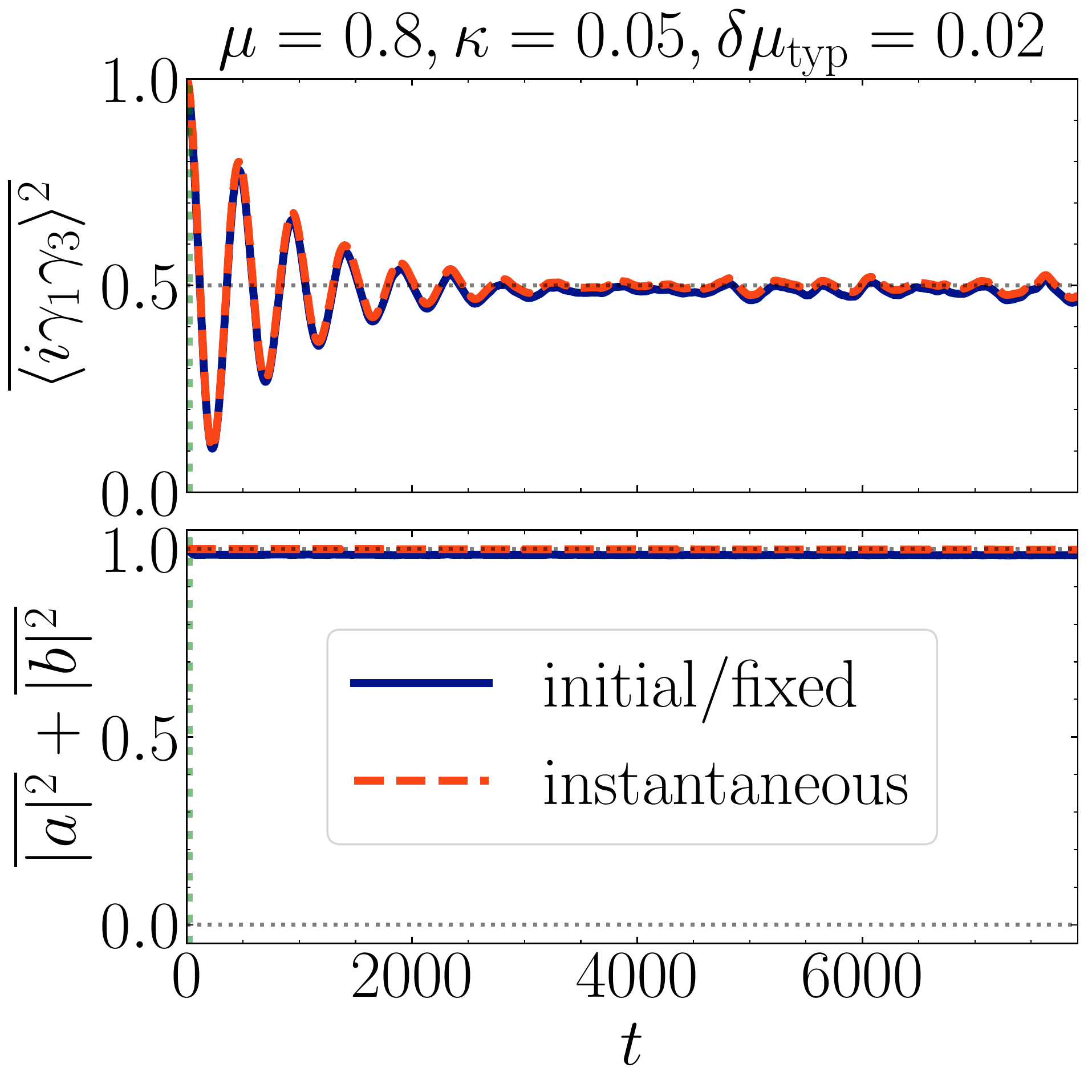}
\includegraphics[width=0.49\columnwidth]{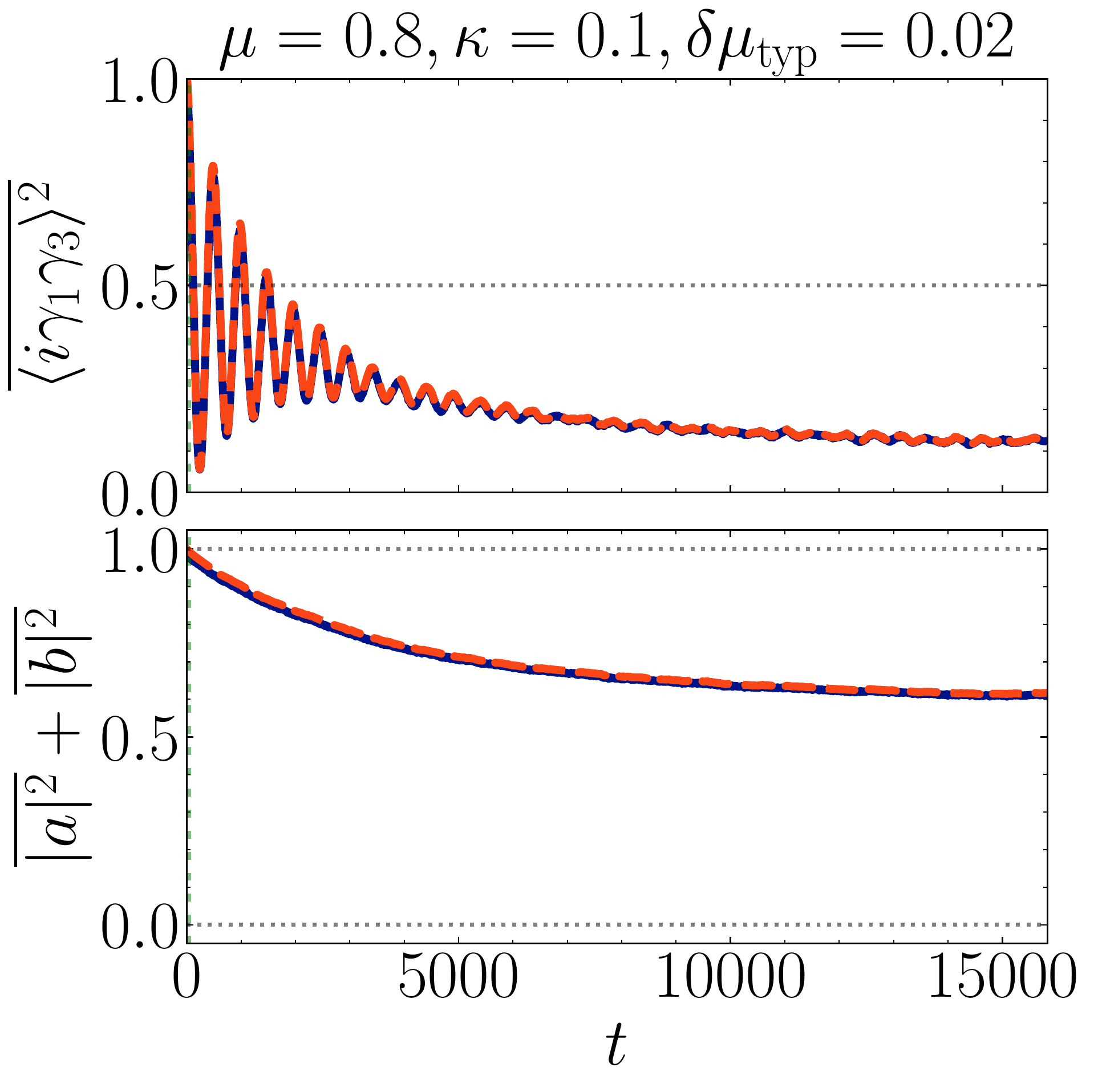}
\end{center}
\vspace{-0.2in}
\begin{center}
\includegraphics[width=0.49\columnwidth]{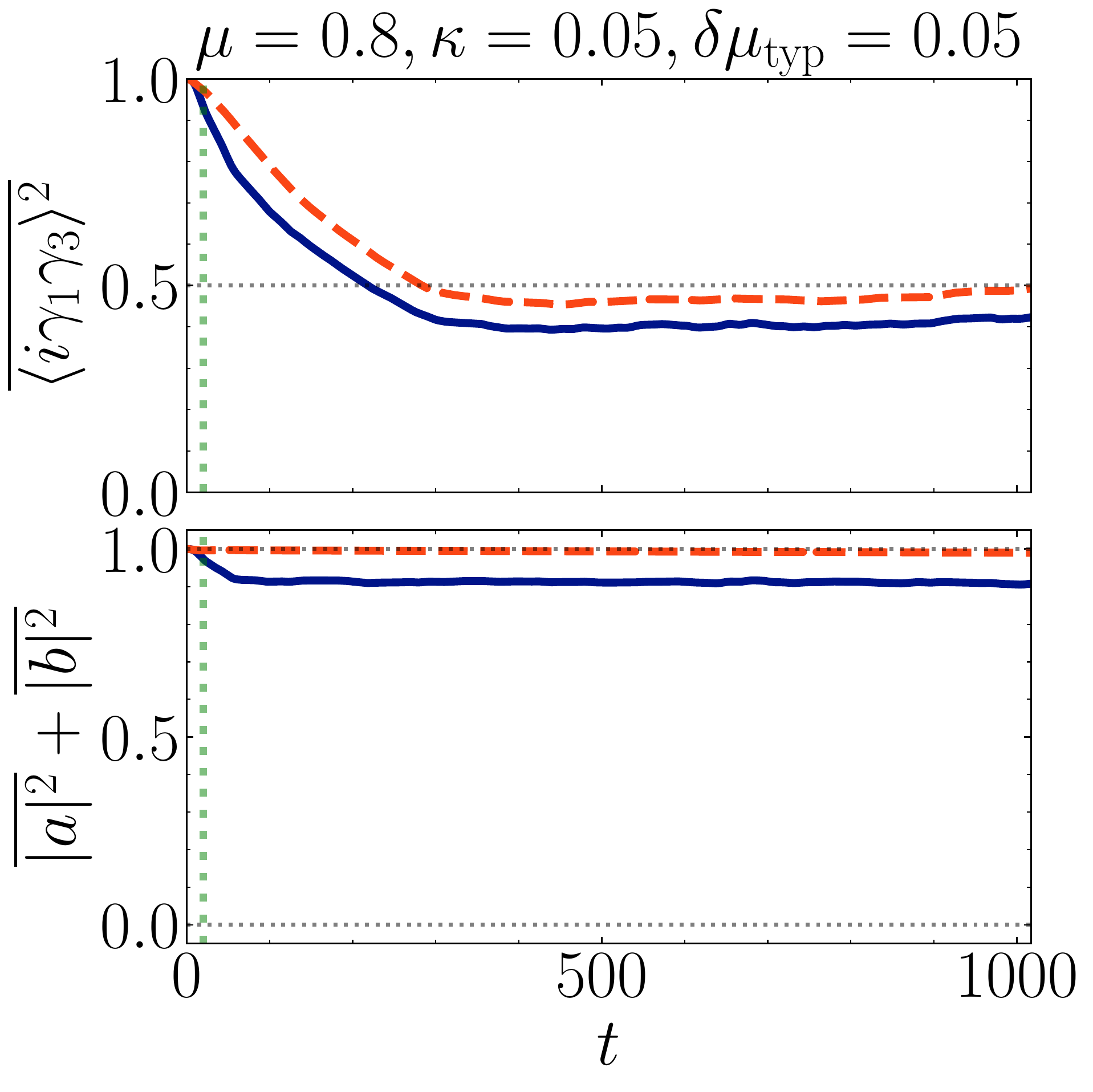}
\includegraphics[width=0.49\columnwidth]{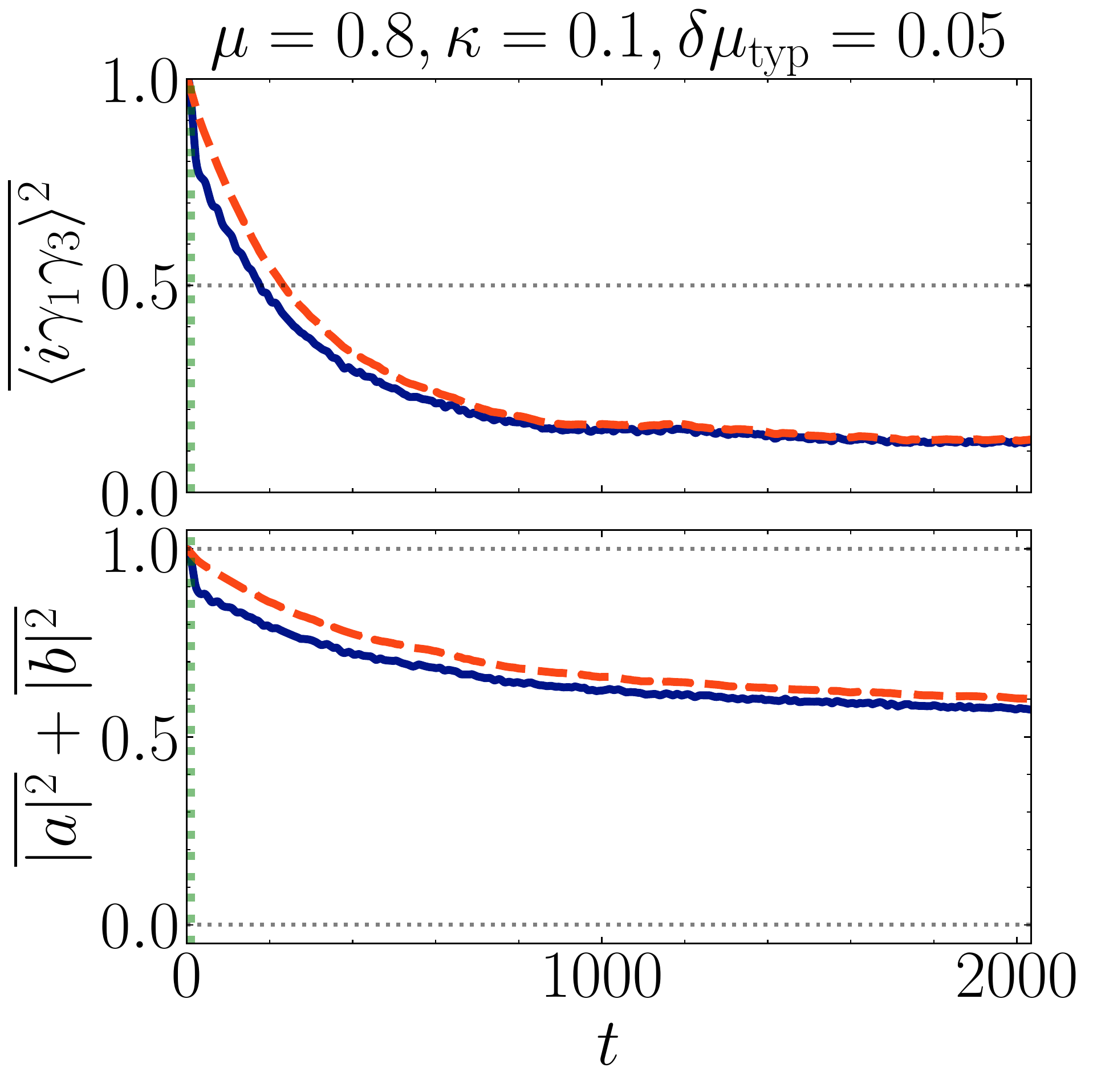}
\end{center}
\vspace{-0.2in}
\begin{center}
\includegraphics[width=0.49\columnwidth]{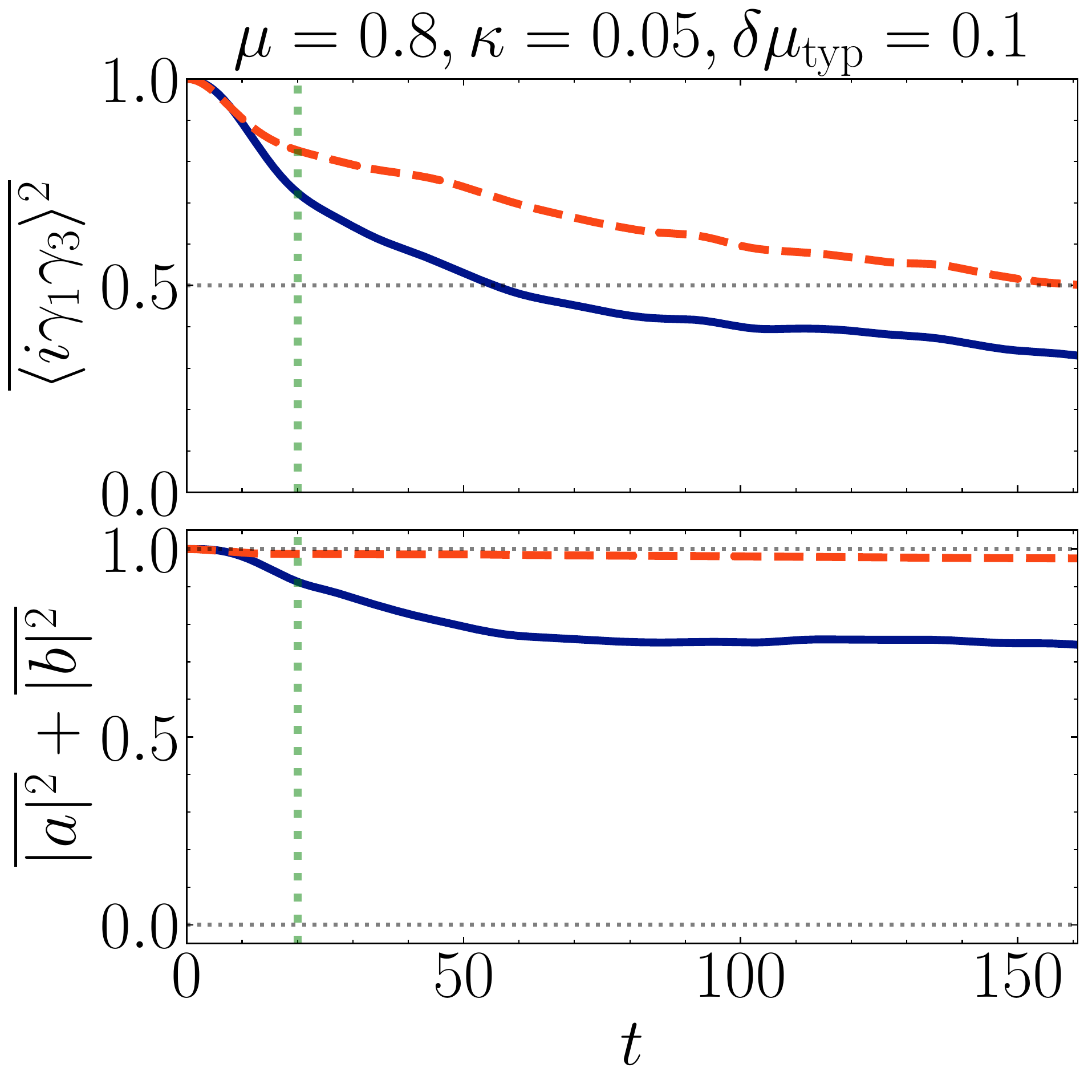}
\includegraphics[width=0.49\columnwidth]{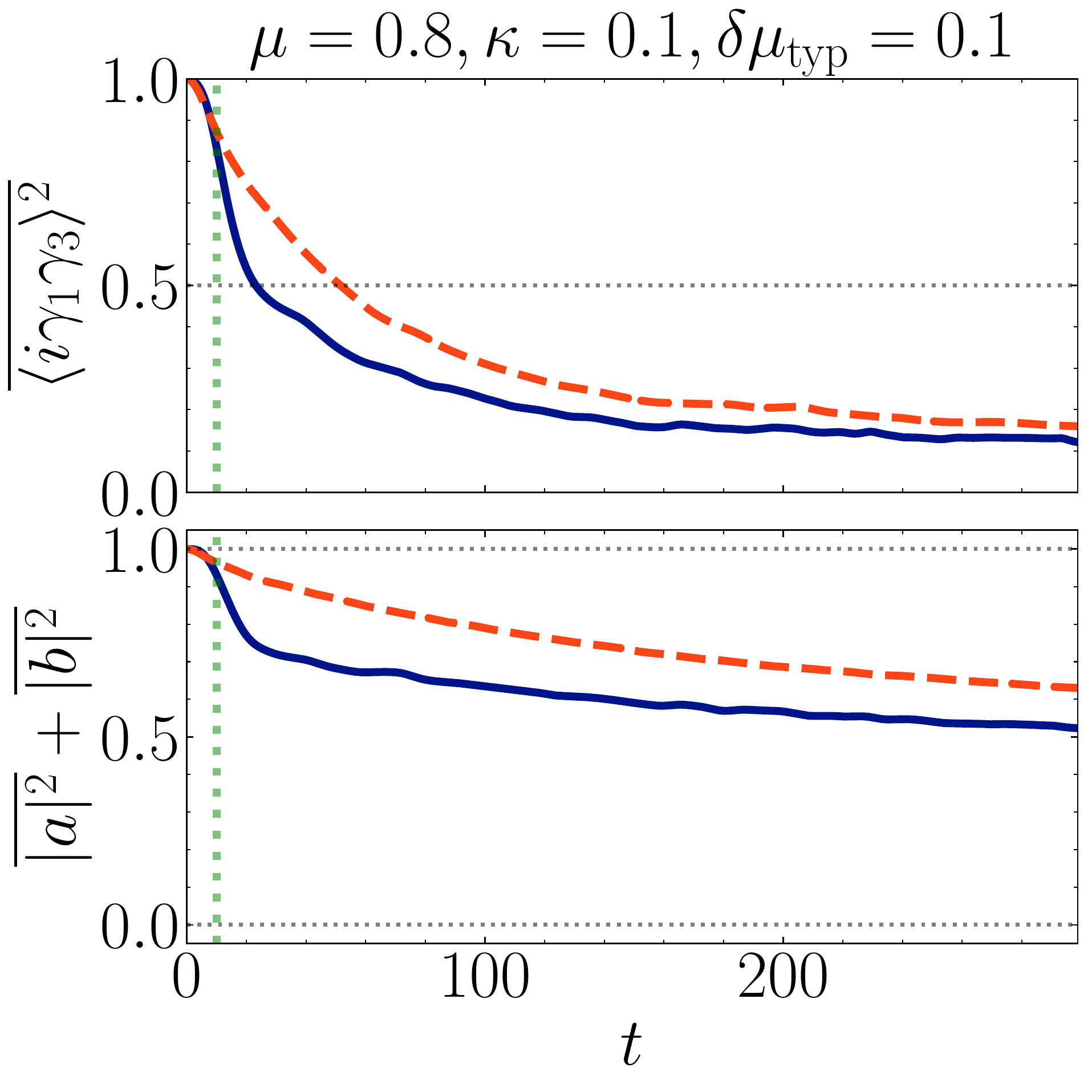}
\end{center}
\caption{Data analogous to that presented in Fig.~\ref{fig:meas_basis_mu0.15} but here taken at $\mu=0.8$. Now leakage out of the instantaneous low-energy qubit subspace is observed at $\kappa=0.1$ (as detected by $\overline{|a_t|^2} + \overline{|b_t|^2}$). For all other data at $\mu=0.15$ in Fig.~\ref{fig:meas_basis_mu0.15} and $\kappa=0.05$ here, leakage out of the instantanous qubit subspace is negligible. Measuring in the fixed basis, on the other hand, generically leads to an effective (topologically unprotected) `leakage' out of the fixed-basis manifold, which can become significant for large $\delta\mu_\mathrm{typ}$ thereby underestimating qubit coherence times.
\label{fig:meas_basis_mu0.8}}
\end{figure}

Figures~\ref{fig:meas_basis_mu0.15} and \ref{fig:meas_basis_mu0.8} present examples of noise-averaged time dynamics of a Kitaev tetron comparing the results of the two different measurement bases. We focus on the following two quantities: $\overline{\langle i\gamma_1\gamma_3\rangle^2}$ and $\overline{|a|^2} + \overline{|b|^2}$, where for the initial/fixed measurement basis $\gamma_i = \gamma_i(t=0)$ and $|a|^2 + |b|^2 = |a_{t=0}|^2 + |b_{t=0}|^2$, while for the instantaneous basis $\gamma_i = \gamma_i(t)$ and $|a|^2 + |b|^2 = |a_{t}|^2 + |b_{t}|^2$. (We consider $\langle i\gamma_1\gamma_3\rangle^2$ instead of $\langle i\gamma_1\gamma_3\rangle$ as the latter would be polluted by the sign ambiguity described in Sec.~\ref{sec:readout} in the case of the instantaneous basis; see also Fig.~\ref{fig:single_realization}.) The noise model is the same as that considered in Sec.~\ref{sec:kitaev}: one global $\mu$ fluctuator on each wire of the tetron (for a total of two fluctuators), each characterized by typical amplitude $\delta\mu_\mathrm{typ}$ and inverse correlation time $\kappa$.  We perform simulations with $\delta\mu_\mathrm{typ}=0.02,0.05,0.1$ and $\kappa=0.05,0.1$. Figure~\ref{fig:meas_basis_mu0.15} corresponds to $\mu=0.15$ which is deep in the topological phase, while  Fig.~\ref{fig:meas_basis_mu0.8} corresponds to $\mu=0.8$ near the crossover into the trivial phase (see Fig.~\ref{fig:kitaev_setup}). The top panels show the time dependence of $\overline{\langle i\gamma_1\gamma_3\rangle^2}$ , while the bottom panels show $|a|^2 + |b|^2$; solid blue (dashed orange) curves represent measurements in the initial/fixed (instantaneous) basis. In the case of pure dephasing, $\overline{\langle i\gamma_1\gamma_3\rangle^2}$ is  expected to take the following form (for times $t\gg1/\kappa$): $\frac12\left[\cos(2\omega_0 t)e^{-4t/T_2} + 1\right]$, i.e., relative to $\overline{\langle i\gamma_1\gamma_3\rangle}$ [cf.~Eq.~\eqref{eq:T2analytic_func}] it oscillates at twice the frequency and approaches $\frac12$ (instead of 0) at four times the decay rate.

For relatively small $\delta\mu_\mathrm{typ}=0.02$, we see in the first row of data presented in Figs.~\ref{fig:meas_basis_mu0.15} and \ref{fig:meas_basis_mu0.8} that the chosen measurement basis negligibly impacts the results \footnote{The residual oscillations in $\overline{\langle i\gamma_1\gamma_3\rangle^2}$ at long times are a convergence artifact associated with noise averaging the square.}.  This conclusion holds both in the case of (nearly) pure dephasing dynamics ($\mu=0.15$ at $\kappa=0.05,0.1$ and $\mu=0.8$ at $\kappa=0.05$) and dynamics involving both dephasing and leakage ($\mu=0.8$ at $\kappa=0.1$). Note that in the latter case, $\overline{\langle i\gamma_1\gamma_3\rangle^2}$ approaches a value less than $\frac12$ as the oscillations die out---leakage causes the traces for individual noise realizations to themselves decay in oscillation amplitude over time (in contrast to Fig.~\ref{fig:noise_averaging}).

As we increase $\delta\mu_\mathrm{typ}$ in the middle ($\delta\mu_\mathrm{typ}=0.05$) and bottom ($\delta\mu_\mathrm{typ}=0.1$) rows of Figs.~\ref{fig:meas_basis_mu0.15} and \ref{fig:meas_basis_mu0.8}, we see that the difference in measurement basis becomes more pronounced. In addition, at fixed $\delta\mu_\mathrm{typ}$ this difference is most severe for slow noise; this point is particularly apparent in the bottom row of Fig.~\ref{fig:meas_basis_mu0.8} comparing $\kappa=0.05$ to $\kappa=0.1$. Furthermore, for both considered measurements, we note that the difference becomes manifest only after a time scale on the order of the noise correlation time $\tau=1/\kappa$ (indicated by vertical green dotted lines). Ultimately, we can clearly see from this data that the relevant coherence time diagnostics decay more quickly when evaluated in a fixed basis \footnote{Although for the qubit and noise model considered here, it is difficult to find a regime where clear `Ramsey oscillations' of $\braket{i\gamma_1\gamma_3}$ persist \emph{and} the basis choice gives rise to a clear difference.}. These results are fully consistent with the qualitative picture sketched above in Sec.~\ref{sec:protocol}.

\section{Discussion} \label{sec:discussion}

In this paper we explored a Ramsey-type protocol that probes qubit dynamics in proximitized nanowire devices---which can support either a  topological Majorana-based qubit or a trivial ABS qubit depending on parameters.  These two scenarios are challenging to distinguish in local probes, e.g., transport, yet display vastly different noise sensitivity as quantified by the qubit dephasing time revealed by our protocol. The required measurements are more challenging than transport but yield correspondingly more detailed information including the qubit lifetime, time-domain detection of the qubit splitting, and the presence of a topological phase transition.  In our study we employed both analytical estimates of dephasing and leakage times as well as explicit numerical simulations of the noisy qubit dynamics.  An appealing feature of the analytical estimates is that they can be readily evaluated (modulo uncertainties in noise details that require experimental input) even in state-of-the-art microscopic models.  Some proof-of-concept simulations were presented in Sec.~\ref{sec:realistic}, and it would be valuable to further quantify the `rigidity' of the qubit splitting to fluctuations in future modeling efforts.  Another feature highlighted by our study is the distinction between fixed-basis and instantaneous Majorana modes in a noisy environment.  We argued on general grounds, and confirmed in our explicit simulations, that examining the former underestimates the true qubit lifetime, with the effect becoming increasingly prominent as the noise becomes slower and of larger amplitude.

In actual experiments, additional imperfections could obfuscate some of the features of the dephasing time discussed in this manuscript. For example, due to long-range inhomogeneities (on a scale longer than the superconducting coherence length) in the electrostatic potential and other system parameters, a nanowire may not undergo a topological phase transition simultaneously at all positions. Instead, the critical field may vary in space, so that some regions can enter the topological phase earlier than others. In such cases, the minimal gap when sweeping the magnetic field does not scale inversely with the system size, but instead inversely with the size of the largest contiguous region undergoing the phase transition. However, the analysis of Sec.~\ref{sec:Tleak_analytic} shows that the relevant regions would be those whose critical states significantly overlap with the Majorana wavefunctions---i.e., regions proximate to the ends of the wire. Thus, the leakage time near the phase crossover regime will be determined by the properties of those critical regions.

Moreover, throughout this paper (e.g., in Fig.~\ref{fig:Vzscan_summary}) we have ignored magnetic-field-induced suppression of the bulk superconducting gap. Such effects are clearly important in present-day experiments \cite{deng_majorana_2016,zhang_quantized_2018} and may ultimately limit the feasibility of our proposed fixed-length study in those devices.  Nevertheless, we are hopeful that future devices will harbor a larger window of field strengths over which one can scrutinize trends in splittings and coherence times.

In practice, the qubit frequency and the dephasing time, which are both tuned exponentially via the wire length and the coherence length, need to fall into an appropriate window for the effects discussed in this paper to be observable. If the wires are too long, the lifetime of the qubit may become limited by error processes not included here, for example quasi-particle poisoning, which render the effects we discuss unobservable. Estimates for the quasi-particle poisoning times vary widely, but can easily exceed one microsecond~\cite{albrecht2017transport}. Conversely, if the wires are too short, both the qubit precession period and its lifetime may be shorter than time-domain experiments can resolve. This time scale is mostly limited by how quickly the coupling between the qubit and the measurement dot can be tuned, which likely limits the resolvable time scales to about one nanosecond. Given these constraints, the ideal regime where the effects discussed in this paper can be observed still encompasses several orders of magnitude in time scale, and thus a significant window of wire lengths. As future direction, it would be interesting to develop protocols that can probe an even broader range of time scales.

Finally, we focused entirely on the minimal `tetron' qubit design for simplicity.  A natural future direction is to extend our study to hexon devices that allow much more flexibility, including additional readout channels, measurement-based braiding, etc.  Quantifying the role of noise in such higher-level applications would be an important and illuminating exercise. All in all, it is a remarkable feature of hardware-based topological qubits that we can directly simulate---in the time domain and with a single simulation---the effects of device-level noise on topologically protected quantum information.

\acknowledgements

The authors thank Andrey Antipov, William Cole, Torsten Karzig, Enrico Rossi, and Mike Zaletel for useful discussions. This work was supported by CRC 183 of Deutsche Forschungsgemeinschaft (F.v.O.); QuantERA project TOPOQUANT (F.v.O.); sabbatical support from IQIM, an NSF physics frontier center funded in part by the Moore Foundation (F.v.O.); the Army Research Office under Grant Award W911NF17-1-0323 (J.A.); the NSF through grant DMR-1723367 (J.A.); the Caltech Institute for Quantum Information and Matter, an NSF Physics Frontiers Center with support of the Gordon and Betty Moore Foundation through Grant GBMF1250 (R.V.M. and J.A.); the Walter Burke Institute for Theoretical Physics at Caltech (R.V.M. and J.A.); and the Gordon and Betty Moore Foundation’s EPiQS Initiative, Grant GBMF8682 (J.A.). Part of this work was performed at the Aspen Center for Physics, which is supported by National Science Foundation grant PHY-1607611 (R.V.M.).

\appendix

\section{Details of noisy time-dependent simulations and noise generation} \label{app:noise}

\begin{figure*}[t]
\begin{center}
\includegraphics[width=0.45\textwidth]{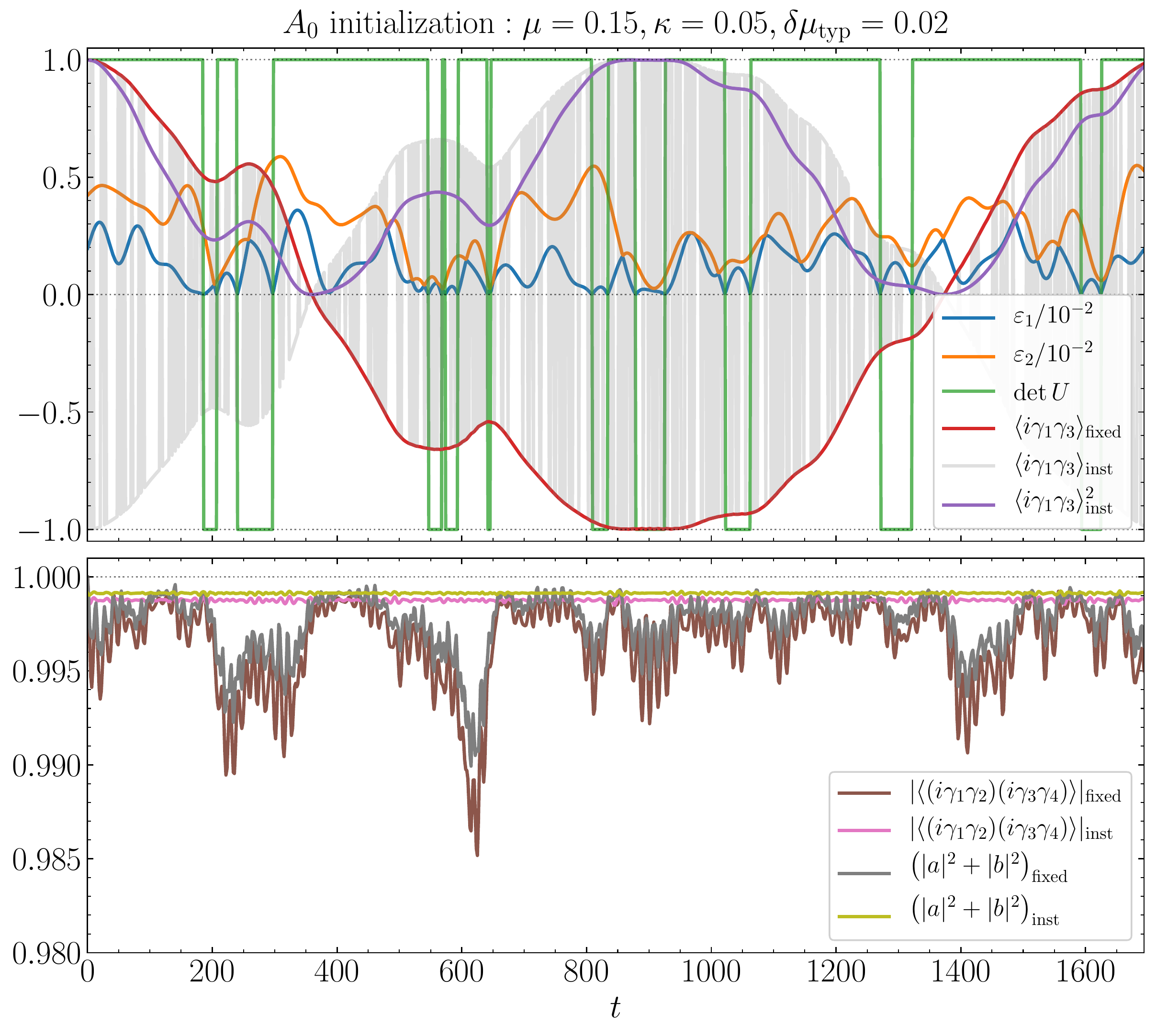}
\hspace{0.25in}
\includegraphics[width=0.45\textwidth]{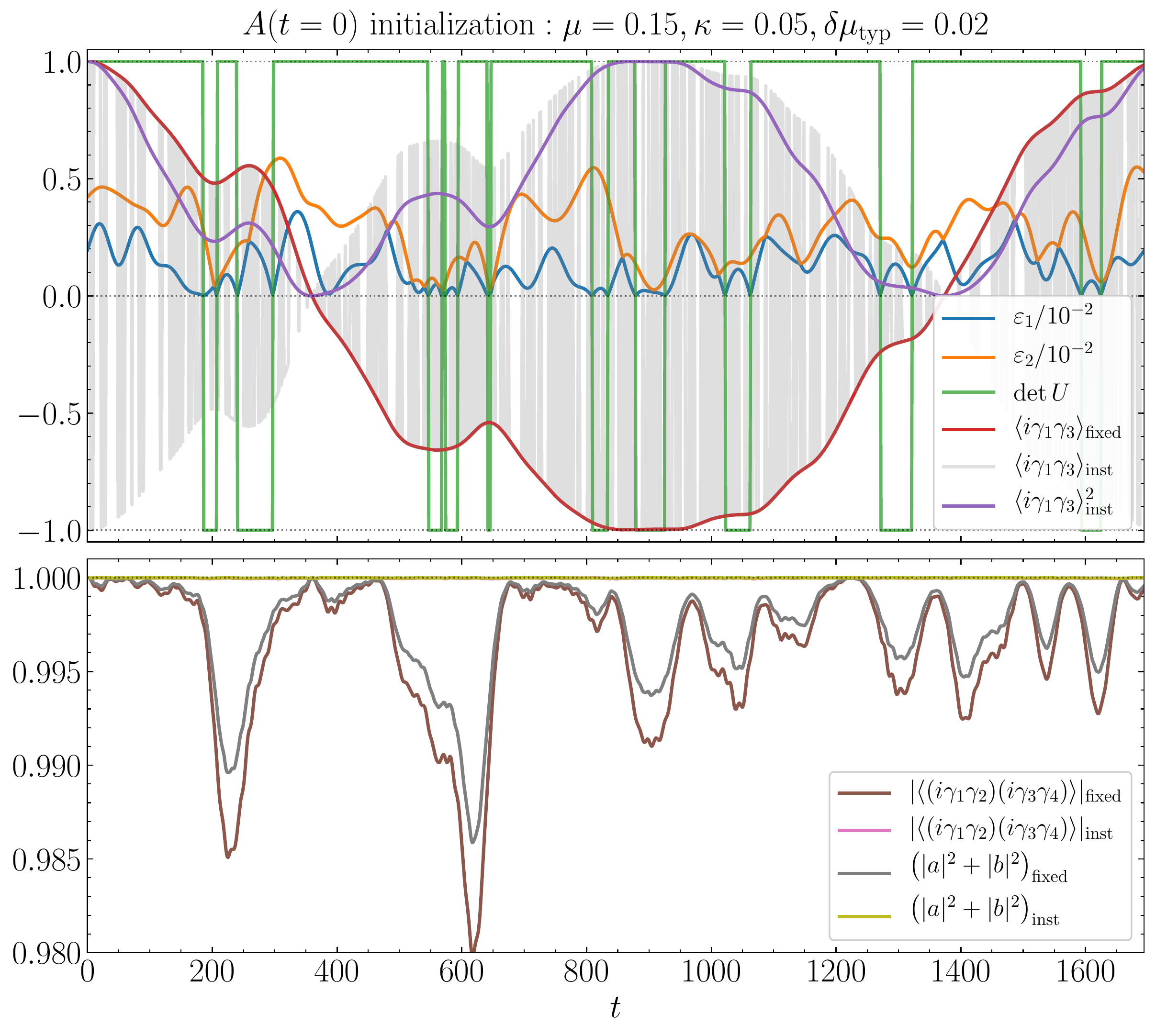}
\end{center}
\caption{Example time dependence of quantum-averaged quantities for a single noise realization our Ramsey-type protocol in the Kitaev tetron. Here, $\mu=0.15$, $\kappa=0.05$, and $\delta\mu_\mathrm{typ}=0.02$ with all other parameters and details identical to those in Secs.~\ref{sec:kitaev} and \ref{sec:fixedvinst} (see Figs.~\ref{fig:kitaev_setup}, \ref{fig:noise_averaging}, \ref{fig:comp_Kitaev}, and \ref{fig:meas_basis_mu0.15}). In the left (right) panel, we show data obtained with the initialization procedure based on $A_0$ [$A(t=0)$] as in Sec.~\ref{sec:kitaev} (Sec.~\ref{sec:fixedvinst}). The top panels include the dephasing diagnostics $\langle i\gamma_1\gamma_3\rangle$ and $\langle i\gamma_1\gamma_3\rangle^2$, as well as the (instantaneous) quantities $\varepsilon_{1,2}$ and $\det U$ from Eqs.~\eqref{eq:B}-\eqref{eq:Hcanon}; the bottom panels show the leakage diagnostics $\langle(i\gamma_1\gamma_2)(i\gamma_3\gamma_4)\rangle$ and $|a|^2+|b|^2$. Subscripts in the legends indicate measurement in either the corresponding fixed or instantaneous (`inst') basis.
\label{fig:single_realization}}
\end{figure*}

We perform time evolution following the Gaussian covariance matrix formalism described in Sec.~\ref{sec:MajoranaReform}. (See also Refs.~\cite{kraus2010generalized,bravyi_complexity_2017}; for a self-contained overview, see Ref.~\cite{bauer_dynamics_2018}.) Our numerical implementation is based on the \text{DifferentialEquations.jl} package in Julia~\cite{rackauckas_differentialequations.jl_2017,DifferentialEquations.jl}.

To generate a noise realization of some fluctuating quantity $X(t)$ with a given noise power spectrum or correlation function, we use techniques based on fast Fourier transformation. This approach scales superlinear in the desired total time of the simulation ($T_\mathrm{sim}$ below). While for many types of noise (e.g., $1/\omega$ and $1/\omega^2$) more efficient ways of generating noise trajectories are known, in our context we never found this step to be a computational bottleneck. In addition, this approach has the advantage of being very general.

Consider a random variable $X(t)$ with $\overline{X(t)} = 0$ and $\overline{X^2(t)} = (\delta X_{\rm typ})^2$, and correlation function $\overline{X(t) X(t')} = S(|t - t'|)$. At each time, the variable is drawn from a Gaussian distribution, $P[X(t)] \sim e^{-X^2/2(\delta X_{\rm typ})^2}$. We can define the noise in frequency space $X(\omega)$ as well as the noise power spectrum $S(\omega) \propto \overline{|X(\omega)|^2}$ 
as 
\begin{align}
X(\omega) &= \int_{-\infty}^{\infty} dt\ e^{-i \omega t} X(t), \\
S(\omega) &= \int_{-\infty}^{\infty} dt\ e^{-i \omega t} S(t).
\end{align}

For a simulation up to some finite time $T_{\rm sim}$, we generate a discrete noise trajectory up to some time $T_{\rm noise} \gg T_{\rm sim}$ with a timestep $\Delta t = T_{\rm noise}/N$, i.e., the trajectory is defined on timeslices $t_n = n \Delta t$, $n = 0, \ldots, N-1$. It is important that $\Delta t \ll \omega_{\rm max}$, where $\omega_{\rm max}$ is the highest relevant frequency occurring in the noise. In the example of Gaussian noise characterized by a high-frequency cutoff $\kappa$, it is natural to set $\omega_{\rm max}$ to some sufficiently large multiple of $\kappa$.

To numerically generate a single noise trajectory, we first create a white noise trajectory $Y(t_n)$ by drawing a sample on each time point from an independent and identical Gaussian ensemble of variance 1. We then perform a discrete Fourier transformation,
\begin{align}
Y(\omega_k) &= \sum_{n=0}^{N-1} Y(t_n) e^{-i \omega_k t_n}, &\omega_k &= k \frac{2\pi}{T_{\rm noise}}.
\end{align}
We can then obtain our desired noise trajectory in Fourier space by taking
\begin{align}
X(\omega_k) &= \left[ \frac{S(\omega_k)}{\Delta t} \right]^{1/2} Y(\omega_k)
\end{align}
and finally performing an inverse Fourier transformation to obtain $X(t_n)$:
\begin{align}
X(t_n) = \frac{1}{N} \sum_{k=0}^{N-1} X(\omega_k) e^{i \omega_k t_n}.
\end{align}
In all these steps, it is convenient to choose $N$ even so that the Fourier transformation is real. Furthermore, it is in many cases important to be able to sample $X(t)$ for arbitrary $t$, for example when integrating the Schr\"odinger equation with an adaptive time step. To this end, we perform a linear interpolation between timesteps. If $\Delta t$ is chosen sufficiently small, this will not incur significant numerical error.

In Fig.~\ref{fig:single_realization}, we show the time dependence of all quantum-averaged quantities considered in this work for a single noise realization of our Ramsey-type protocol over one qubit precession period for the Kitaev tetron system of Secs.~\ref{sec:kitaev} and \ref{sec:fixedvinst}. We also show data necessary to derive the instantaneous qubit splitting \cite{Kitaev01_PhysU_44_131}: the two lowest-lying instantaneous energies $\varepsilon_{1,2}$ and the determinant of the instantaneous orthogonal transformation $U$ [see Eqs.~\eqref{eq:B}-\eqref{eq:Hcanon}]. The left (right) panel corresponds to the initialization procedure based on the time-averaged (initial) Hamiltonian $A_0$ [$A(t=0)$] used in Sec.~\ref{sec:kitaev} (Sec.~\ref{sec:fixedvinst}). Quantities with the subscript `fixed' denote measurement in a fixed basis throughout the evolution [e.g., $\gamma_i = \gamma_i^{(0)}$ in the left panel and $\gamma_i = \gamma_i(t=0)$ in the right panel], while quantities with the subscript `inst' denote measurement in the instantaneous basis derived from $A(t)$ [e.g., $\gamma_i = \gamma_i(t)$]. The data shown for $\langle i\gamma_1\gamma_3 \rangle_\mathrm{inst} = \langle i\gamma_1(t)\gamma_3(t) \rangle$ is polluted by the numerical sign ambiguity described in Sec.~\ref{sec:readout}; we show it here over a single realization to illustrate this point. Comparing the envelope of $\langle i\gamma_1\gamma_3 \rangle_\mathrm{inst}$ to $\langle i\gamma_1\gamma_3 \rangle_\mathrm{fixed}$ demonstrates that for these parameters the choice of instantaneous versus fixed measurement basis does not have any impact on the results down to the level of each individual realization. Furthermore, comparing the dephasing diagnostics data in the top panels of Fig.~\ref{fig:single_realization} reveals---again for these chosen parameters---that the choice of initialization procedure makes negligible difference.  For the leakage diagnostics in the bottom panel, the measurements are very close to unity, and we thus show an appropriately zoomed in view. Note the observable high-frequency components in these measurements due to the small `quench' at $t=0$ when using the $A_0$-based initialization procedure.

To calculate noise-averaged quantities, we run $N_\mathrm{real} = O(10^3)$ independent such individual realizations. While this approach is relatively numerically demanding (compared, for example, to the master-equation based approach of Ref.~\cite{hu_majorana_2015}), it has multiple advantages: For one, our techniques are completely general in terms of what noise models can be simulated and what physical quantities can be measured / noise averaged (in contrast to the methods of Ref.~\cite{hu_majorana_2015} which are more limiting in these respects); furthermore, our simulation strategy very closely mimics the actual experimental procedure we propose (although in experiment noise and quantum averaging cannot be distinguished).

\section{Noise-averaging analysis}
\label{app:averaging}

Here we provide details on the noise-averaging of Eq.~\eqref{Pt} assuming the Gaussian noise correlations specified in Eqs.~\eqref{lambda_ave} through \eqref{eq:Sw_Gaussian}.  We will specifically evaluate $\overline{\exp\left(i\int_0^tdt'E(t')\right)}$ with $E(t')$ given by the harmonic approximation in Eq.~\eqref{E_expansion} and $\hbar = 1$ for notational simplicity; the noise average of $Q(t)$ follows straightforwardly from this quantity.  It is convenient to discretize time (for intermediate stages of the calculation) and write
\begin{align}
    \overline{e^{i\int_0^tdt'E(t')}} &= \frac{1}{Z}\int \mathcal{D}\lambda_i(t) e^{i \Delta t\sum_{t' = 0}^t E(t') }
    \nonumber \\
    &\times e^{-\frac{1}{2}\sum_{t' t''}\sum_i S_i^{-1}(t'-t'')\lambda_i(t')\lambda_i(t'')}.
\end{align}
Here $\Delta t$ is the time interval used for discretization, the second line is the weighting factor that gives the desired noise correlations, and $Z$ is a normalization defined as
\begin{equation}
      Z = \int \mathcal{D}\lambda_i(t) e^{-\frac{1}{2}\sum_{t' t''}\sum_i S_i^{-1}(t'-t'')\lambda_i(t')\lambda_i(t'')}.
\end{equation}
Next we introduce
\begin{align}
    M_{ij}(t',t'') = M^0_{ij}(t',t'') + \delta M_{ij}(t',t''),
\end{align}
where 
\begin{align}
    M^0_{ij}(t',t'') &= \delta_{ij}S^{-1}_i(t'-t'')
    \\
    \delta M_{ij}(t',t'') &= 
    \begin{cases}
        -i\Delta t \delta_{t',t''} E_{ij}'',~~~ 0< t' < t \\
        0, ~~~~~~~~\text{otherwise},
    \end{cases}
\end{align}
and also
\begin{align}
    v_i(t') = \begin{cases}
        i\Delta t E_i',~~~ 0< t' < t \\
        0, ~~~~~~~~\text{otherwise}.
    \end{cases}
\end{align}
These definitions allow us to write
\begin{align}
    \overline{e^{i\int_0^tdt'E(t')}} &= \frac{e^{i E_0 t}}{Z}\int \mathcal{D}\lambda_i(t) e^{\sum_i \sum_{t'} v_i(t') \lambda_i(t')}
    \nonumber \\
    &\times e^{-\frac{1}{2} \sum_{ij} \sum_{t't''} \lambda_i(t')M_{ij}(t',t'')\lambda_j(t'')}.
\end{align}
Note that the $t',t''$ sums are unrestricted above.  We can now perform the Gaussian integration to obtain
\begin{align}
    \overline{e^{i\int_0^tdt'E(t')}} &= e^{i E_0 t}e^{-\frac{1}{2} \ln\left(\frac{\det M}{\det M^0}\right)}
    \nonumber \\
    &\times e^{\frac{1}{2} \sum_{ij} \sum_{t't''}v_i(t')M^{-1}_{ij}(t',t'') v_j(t'')}.
    \label{GI}
\end{align}

To proceed we first expand the log term in Eq.~\eqref{GI} to second order in $\delta M$:
\begin{align}
    \ln\left(\frac{\det M}{\det M^0}\right) &= \ln \det[I + (M^0)^{-1} \delta M]
    \nonumber \\
    &\approx {\rm Tr}[(M^0)^{-1} \delta M] -\frac{1}{2} {\rm Tr}[(M^0)^{-1} \delta M]^2
    \nonumber \\
    &= -i t\sum_i D_i^2 E_{ii}'' 
    \nonumber \\
    &+ \frac{1}{2} \sum_{ij}(E_{ij}'')^2 \int_0^t dt' dt'' S_i(t'-t'')S_j(t''-t').
\end{align}
The third line above encodes the leading shift in the qubit precession frequency [cf.~Eq.~\eqref{omega0_shifted}].  At long times, $t \gg \tau_i$, the integrals in the fourth line become
\begin{align}
    \int_0^t dt' dt'' S_i(t'-t'')S_j(t''-t') 
    \approx t\frac{2 \sqrt{\pi}\tau_i \tau_j}{\sqrt{\tau_i^2 + 
      \tau_j^2}}(D_iD_j)^2.
      \label{int1}
\end{align}
In the second line of Eq.~\eqref{GI} we simply replace $M^{-1} \approx (M^0)^{-1}$; the next-leading correction provides a higher-order shift to the qubit precession frequency compared to that captured above.  This approximation yields
\begin{align}
    &\sum_{ij} \sum_{t't''}v_i(t')M^{-1}_{ij}(t',t'') v_j(t'')
    \nonumber \\
    &\approx - \sum_i (E_i')^2 \int_0^t dt' dt'' S_i(t'-t'') \approx t 2\sqrt{\pi} \sum_i \tau_i(D_i E_i')^2,
    \label{int2}
\end{align}
where on the far right side we again assumed the long-time limit. 

Putting everything together, we find
\begin{align}
    \overline{e^{i\int_0^tdt'E(t')}} &=e^{i\left(E_0 + \frac{1}{2}\sum_i D_i^2 E_{ii}''\right)t}
    \nonumber \\
    &\times e^{-\sqrt{\pi}\left[ \sum_i\tau_i\left(D_i E_i'\right)^2 + \frac{1}{2}\sum_{ij}\frac{\tau_i \tau_j}{\sqrt{\tau_i^2 + 
      \tau_j^2}}\left(D_i E_{ij}''D_j\right)^2\right]t}
\end{align}
at $t \gg \tau_i$.
Upong restoring explicit $\hbar$'s, this result indeed recovers the qubit precession frequency and dephasing time quoted in Eqs.~\eqref{omega0_shifted} and \eqref{T2analytic}.  

The short-time limit, $t \ll \tau_i$, can be easily treated as well.  Here we can simply write
\begin{align}
    \int_0^t dt' dt'' S_i(t'-t'')S_j(t''-t') &\approx t^2 S_i(0)S_j(0) = (D_i D_j t)^2
    \\
    \int_0^t dt' dt'' S_i(t'-t'') &\approx t^2 S_i(0)= (D_i t)^2
\end{align}
in Eqs.~\eqref{int1} and \eqref{int2}, so that
\begin{align}
    \overline{e^{i\int_0^tdt'E(t')}} &=e^{i\left(E_0 + \frac{1}{2}\sum_i D_i^2 E_{ii}''\right)t}
    \nonumber \\
    &\times e^{-\frac{1}{2}\left[ \sum_i\left(D_i E_i'\right)^2 + \frac{1}{2}\sum_{ij}\left(D_i E_{ij}''D_j\right)^2\right]t^2}.
\end{align}
The noise-averaged $Q(t)$ is then
\begin{equation}
    \overline{Q(t \ll \tau_i)} \approx \cos(\omega_0 t)e^{-(t/T_2^{\rm short~time})^2}
    \label{eq:Pt_short-time}
\end{equation}
with (restoring $\hbar$'s)
\begin{equation}
    T_2^{\rm short~time} = 2\hbar^2\left[ \sum_i\left(D_i E_i'\right)^2 + \frac{1}{2}\sum_{ij}\left(D_i E_{ij}''D_j\right)^2\right]^{-1}.
    \label{eq:T2_short-time}
\end{equation}

\bibliography{MajoranaNoise}

\begin{thebibliography}{82}%
\makeatletter
\providecommand \@ifxundefined [1]{%
 \@ifx{#1\undefined}
}%
\providecommand \@ifnum [1]{%
 \ifnum #1\expandafter \@firstoftwo
 \else \expandafter \@secondoftwo
 \fi
}%
\providecommand \@ifx [1]{%
 \ifx #1\expandafter \@firstoftwo
 \else \expandafter \@secondoftwo
 \fi
}%
\providecommand \natexlab [1]{#1}%
\providecommand \enquote  [1]{``#1''}%
\providecommand \bibnamefont  [1]{#1}%
\providecommand \bibfnamefont [1]{#1}%
\providecommand \citenamefont [1]{#1}%
\providecommand \href@noop [0]{\@secondoftwo}%
\providecommand \href [0]{\begingroup \@sanitize@url \@href}%
\providecommand \@href[1]{\@@startlink{#1}\@@href}%
\providecommand \@@href[1]{\endgroup#1\@@endlink}%
\providecommand \@sanitize@url [0]{\catcode `\\12\catcode `\$12\catcode
  `\&12\catcode `\#12\catcode `\^12\catcode `\_12\catcode `\%12\relax}%
\providecommand \@@startlink[1]{}%
\providecommand \@@endlink[0]{}%
\providecommand \url  [0]{\begingroup\@sanitize@url \@url }%
\providecommand \@url [1]{\endgroup\@href {#1}{\urlprefix }}%
\providecommand \urlprefix  [0]{URL }%
\providecommand \Eprint [0]{\href }%
\providecommand \doibase [0]{http://dx.doi.org/}%
\providecommand \selectlanguage [0]{\@gobble}%
\providecommand \bibinfo  [0]{\@secondoftwo}%
\providecommand \bibfield  [0]{\@secondoftwo}%
\providecommand \translation [1]{[#1]}%
\providecommand \BibitemOpen [0]{}%
\providecommand \bibitemStop [0]{}%
\providecommand \bibitemNoStop [0]{.\EOS\space}%
\providecommand \EOS [0]{\spacefactor3000\relax}%
\providecommand \BibitemShut  [1]{\csname bibitem#1\endcsname}%
\let\auto@bib@innerbib\@empty
\bibitem [{\citenamefont {{Kitaev}}(2001)}]{Kitaev01_PhysU_44_131}%
  \BibitemOpen
  \bibfield  {author} {\bibinfo {author} {\bibfnamefont {A.~Y.}\ \bibnamefont
  {{Kitaev}}},\ }\href {\doibase 10.1070/1063-7869/44/10S/S29} {\bibfield
  {journal} {\bibinfo  {journal} {Physics Uspekhi}\ }\textbf {\bibinfo {volume}
  {44}},\ \bibinfo {pages} {131} (\bibinfo {year} {2001})},\ \Eprint
  {http://arxiv.org/abs/cond-mat/0010440} {cond-mat/0010440} \BibitemShut
  {NoStop}%
\bibitem [{\citenamefont {Sarma}\ \emph {et~al.}(2015)\citenamefont {Sarma},
  \citenamefont {Freedman},\ and\ \citenamefont {Nayak}}]{sarma_majorana_2015}%
  \BibitemOpen
  \bibfield  {author} {\bibinfo {author} {\bibfnamefont {S.~D.}\ \bibnamefont
  {Sarma}}, \bibinfo {author} {\bibfnamefont {M.}~\bibnamefont {Freedman}}, \
  and\ \bibinfo {author} {\bibfnamefont {C.}~\bibnamefont {Nayak}},\ }\href
  {https://www.nature.com/articles/npjqi20151} {\bibfield  {journal} {\bibinfo
  {journal} {npj Quantum Inf.}\ }\textbf {\bibinfo {volume} {1}},\ \bibinfo
  {pages} {15001} (\bibinfo {year} {2015})}\BibitemShut {NoStop}%
\bibitem [{\citenamefont {Mourik}\ \emph {et~al.}(2012)\citenamefont {Mourik},
  \citenamefont {Zuo}, \citenamefont {Frolov}, \citenamefont {Plissard},
  \citenamefont {Bakkers},\ and\ \citenamefont
  {Kouwenhoven}}]{mourik_signatures_2012}%
  \BibitemOpen
  \bibfield  {author} {\bibinfo {author} {\bibfnamefont {V.}~\bibnamefont
  {Mourik}}, \bibinfo {author} {\bibfnamefont {K.}~\bibnamefont {Zuo}},
  \bibinfo {author} {\bibfnamefont {S.~M.}\ \bibnamefont {Frolov}}, \bibinfo
  {author} {\bibfnamefont {S.~R.}\ \bibnamefont {Plissard}}, \bibinfo {author}
  {\bibfnamefont {E.~P. A.~M.}\ \bibnamefont {Bakkers}}, \ and\ \bibinfo
  {author} {\bibfnamefont {L.~P.}\ \bibnamefont {Kouwenhoven}},\ }\href
  {http://www.sciencemag.org/cgi/doi/10.1126/science.1222360} {\bibfield
  {journal} {\bibinfo  {journal} {Science}\ }\textbf {\bibinfo {volume}
  {336}},\ \bibinfo {pages} {1003} (\bibinfo {year} {2012})}\BibitemShut
  {NoStop}%
\bibitem [{\citenamefont {Deng}\ \emph {et~al.}(2012)\citenamefont {Deng},
  \citenamefont {Yu}, \citenamefont {Huang}, \citenamefont {Larsson},
  \citenamefont {Caroff},\ and\ \citenamefont {Xu}}]{deng_anomalous_2012}%
  \BibitemOpen
  \bibfield  {author} {\bibinfo {author} {\bibfnamefont {M.~T.}\ \bibnamefont
  {Deng}}, \bibinfo {author} {\bibfnamefont {C.~L.}\ \bibnamefont {Yu}},
  \bibinfo {author} {\bibfnamefont {G.~Y.}\ \bibnamefont {Huang}}, \bibinfo
  {author} {\bibfnamefont {M.}~\bibnamefont {Larsson}}, \bibinfo {author}
  {\bibfnamefont {P.}~\bibnamefont {Caroff}}, \ and\ \bibinfo {author}
  {\bibfnamefont {H.~Q.}\ \bibnamefont {Xu}},\ }\href
  {http://pubs.acs.org/doi/10.1021/nl303758w} {\bibfield  {journal} {\bibinfo
  {journal} {Nano Lett.}\ }\textbf {\bibinfo {volume} {12}},\ \bibinfo {pages}
  {6414} (\bibinfo {year} {2012})}\BibitemShut {NoStop}%
\bibitem [{\citenamefont {Das}\ \emph {et~al.}(2012)\citenamefont {Das},
  \citenamefont {Ronen}, \citenamefont {Most}, \citenamefont {Oreg},
  \citenamefont {Heiblum},\ and\ \citenamefont
  {Shtrikman}}]{das_zero-bias_2012}%
  \BibitemOpen
  \bibfield  {author} {\bibinfo {author} {\bibfnamefont {A.}~\bibnamefont
  {Das}}, \bibinfo {author} {\bibfnamefont {Y.}~\bibnamefont {Ronen}}, \bibinfo
  {author} {\bibfnamefont {Y.}~\bibnamefont {Most}}, \bibinfo {author}
  {\bibfnamefont {Y.}~\bibnamefont {Oreg}}, \bibinfo {author} {\bibfnamefont
  {M.}~\bibnamefont {Heiblum}}, \ and\ \bibinfo {author} {\bibfnamefont
  {H.}~\bibnamefont {Shtrikman}},\ }\href
  {http://www.nature.com/doifinder/10.1038/nphys2479} {\bibfield  {journal}
  {\bibinfo  {journal} {Nat. Phys.}\ }\textbf {\bibinfo {volume} {8}},\
  \bibinfo {pages} {887} (\bibinfo {year} {2012})}\BibitemShut {NoStop}%
\bibitem [{\citenamefont {Churchill}\ \emph {et~al.}(2013)\citenamefont
  {Churchill}, \citenamefont {Fatemi}, \citenamefont {Grove-Rasmussen},
  \citenamefont {Deng}, \citenamefont {Caroff}, \citenamefont {Xu},\ and\
  \citenamefont {Marcus}}]{churchill_superconductor-nanowire_2013}%
  \BibitemOpen
  \bibfield  {author} {\bibinfo {author} {\bibfnamefont {H.~O.~H.}\
  \bibnamefont {Churchill}}, \bibinfo {author} {\bibfnamefont {V.}~\bibnamefont
  {Fatemi}}, \bibinfo {author} {\bibfnamefont {K.}~\bibnamefont
  {Grove-Rasmussen}}, \bibinfo {author} {\bibfnamefont {M.~T.}\ \bibnamefont
  {Deng}}, \bibinfo {author} {\bibfnamefont {P.}~\bibnamefont {Caroff}},
  \bibinfo {author} {\bibfnamefont {H.~Q.}\ \bibnamefont {Xu}}, \ and\ \bibinfo
  {author} {\bibfnamefont {C.~M.}\ \bibnamefont {Marcus}},\ }\href
  {https://link.aps.org/doi/10.1103/PhysRevB.87.241401} {\bibfield  {journal}
  {\bibinfo  {journal} {Phys. Rev. B}\ }\textbf {\bibinfo {volume} {87}},\
  \bibinfo {pages} {241401} (\bibinfo {year} {2013})}\BibitemShut {NoStop}%
\bibitem [{\citenamefont {Lee}\ \emph {et~al.}(2013)\citenamefont {Lee},
  \citenamefont {Jiang}, \citenamefont {Houzet}, \citenamefont {Aguado},
  \citenamefont {Lieber},\ and\ \citenamefont
  {De~Franceschi}}]{lee_spin-resolved_2013}%
  \BibitemOpen
  \bibfield  {author} {\bibinfo {author} {\bibfnamefont {E.~J.~H.}\
  \bibnamefont {Lee}}, \bibinfo {author} {\bibfnamefont {X.}~\bibnamefont
  {Jiang}}, \bibinfo {author} {\bibfnamefont {M.}~\bibnamefont {Houzet}},
  \bibinfo {author} {\bibfnamefont {R.}~\bibnamefont {Aguado}}, \bibinfo
  {author} {\bibfnamefont {C.~M.}\ \bibnamefont {Lieber}}, \ and\ \bibinfo
  {author} {\bibfnamefont {S.}~\bibnamefont {De~Franceschi}},\ }\href
  {http://www.nature.com/doifinder/10.1038/nnano.2013.267} {\bibfield
  {journal} {\bibinfo  {journal} {Nat. Nanotechnol.}\ }\textbf {\bibinfo
  {volume} {9}},\ \bibinfo {pages} {79} (\bibinfo {year} {2013})}\BibitemShut
  {NoStop}%
\bibitem [{\citenamefont {Finck}\ \emph {et~al.}(2013)\citenamefont {Finck},
  \citenamefont {Van~Harlingen}, \citenamefont {Mohseni}, \citenamefont
  {Jung},\ and\ \citenamefont {Li}}]{finck_anomalous_2013}%
  \BibitemOpen
  \bibfield  {author} {\bibinfo {author} {\bibfnamefont {A.~D.~K.}\
  \bibnamefont {Finck}}, \bibinfo {author} {\bibfnamefont {D.~J.}\ \bibnamefont
  {Van~Harlingen}}, \bibinfo {author} {\bibfnamefont {P.~K.}\ \bibnamefont
  {Mohseni}}, \bibinfo {author} {\bibfnamefont {K.}~\bibnamefont {Jung}}, \
  and\ \bibinfo {author} {\bibfnamefont {X.}~\bibnamefont {Li}},\ }\href
  {https://link.aps.org/doi/10.1103/PhysRevLett.110.126406} {\bibfield
  {journal} {\bibinfo  {journal} {Phys. Rev. Lett.}\ }\textbf {\bibinfo
  {volume} {110}},\ \bibinfo {pages} {126406} (\bibinfo {year}
  {2013})}\BibitemShut {NoStop}%
\bibitem [{\citenamefont {Albrecht}\ \emph {et~al.}(2016)\citenamefont
  {Albrecht}, \citenamefont {Higginbotham}, \citenamefont {Madsen},
  \citenamefont {Kuemmeth}, \citenamefont {Jespersen}, \citenamefont
  {Nyg\aa{}rd}, \citenamefont {Krogstrup},\ and\ \citenamefont
  {Marcus}}]{albrecht_exponential_2016}%
  \BibitemOpen
  \bibfield  {author} {\bibinfo {author} {\bibfnamefont {S.~M.}\ \bibnamefont
  {Albrecht}}, \bibinfo {author} {\bibfnamefont {A.~P.}\ \bibnamefont
  {Higginbotham}}, \bibinfo {author} {\bibfnamefont {M.}~\bibnamefont
  {Madsen}}, \bibinfo {author} {\bibfnamefont {F.}~\bibnamefont {Kuemmeth}},
  \bibinfo {author} {\bibfnamefont {T.~S.}\ \bibnamefont {Jespersen}}, \bibinfo
  {author} {\bibfnamefont {J.}~\bibnamefont {Nyg\aa{}rd}}, \bibinfo {author}
  {\bibfnamefont {P.}~\bibnamefont {Krogstrup}}, \ and\ \bibinfo {author}
  {\bibfnamefont {C.~M.}\ \bibnamefont {Marcus}},\ }\href
  {http://www.nature.com/articles/nature17162} {\bibfield  {journal} {\bibinfo
  {journal} {Nature}\ }\textbf {\bibinfo {volume} {531}},\ \bibinfo {pages}
  {206} (\bibinfo {year} {2016})}\BibitemShut {NoStop}%
\bibitem [{\citenamefont {Deng}\ \emph {et~al.}(2016)\citenamefont {Deng},
  \citenamefont {Vaitiek\ifmmode~\dot{e}\else \.{e}\fi{}nas}, \citenamefont
  {Hansen}, \citenamefont {Danon}, \citenamefont {Leijnse}, \citenamefont
  {Flensberg}, \citenamefont {Nyg\aa{}rd}, \citenamefont {Krogstrup},\ and\
  \citenamefont {Marcus}}]{deng_majorana_2016}%
  \BibitemOpen
  \bibfield  {author} {\bibinfo {author} {\bibfnamefont {M.~T.}\ \bibnamefont
  {Deng}}, \bibinfo {author} {\bibfnamefont {S.}~\bibnamefont
  {Vaitiek\ifmmode~\dot{e}\else \.{e}\fi{}nas}}, \bibinfo {author}
  {\bibfnamefont {E.~B.}\ \bibnamefont {Hansen}}, \bibinfo {author}
  {\bibfnamefont {J.}~\bibnamefont {Danon}}, \bibinfo {author} {\bibfnamefont
  {M.}~\bibnamefont {Leijnse}}, \bibinfo {author} {\bibfnamefont
  {K.}~\bibnamefont {Flensberg}}, \bibinfo {author} {\bibfnamefont
  {J.}~\bibnamefont {Nyg\aa{}rd}}, \bibinfo {author} {\bibfnamefont
  {P.}~\bibnamefont {Krogstrup}}, \ and\ \bibinfo {author} {\bibfnamefont
  {C.~M.}\ \bibnamefont {Marcus}},\ }\href
  {http://science.sciencemag.org/content/354/6319/1557.abstract} {\bibfield
  {journal} {\bibinfo  {journal} {Science}\ }\textbf {\bibinfo {volume}
  {354}},\ \bibinfo {pages} {1557} (\bibinfo {year} {2016})}\BibitemShut
  {NoStop}%
\bibitem [{\citenamefont {Zhang}\ \emph {et~al.}(2018)\citenamefont {Zhang},
  \citenamefont {Liu}, \citenamefont {Gazibegovic}, \citenamefont {Xu},
  \citenamefont {Logan}, \citenamefont {Wang}, \citenamefont {van Loo},
  \citenamefont {Bommer}, \citenamefont {de~Moor}, \citenamefont {Car},
  \citenamefont {Op~het Veld}, \citenamefont {van Veldhoven}, \citenamefont
  {Koelling}, \citenamefont {Verheijen}, \citenamefont {Pendharkar},
  \citenamefont {Pennachio}, \citenamefont {Shojaei}, \citenamefont {Lee},
  \citenamefont {Palmstr\o{}m}, \citenamefont {Bakkers}, \citenamefont
  {Sarma},\ and\ \citenamefont {Kouwenhoven}}]{zhang_quantized_2018}%
  \BibitemOpen
  \bibfield  {author} {\bibinfo {author} {\bibfnamefont {H.}~\bibnamefont
  {Zhang}}, \bibinfo {author} {\bibfnamefont {C.-X.}\ \bibnamefont {Liu}},
  \bibinfo {author} {\bibfnamefont {S.}~\bibnamefont {Gazibegovic}}, \bibinfo
  {author} {\bibfnamefont {D.}~\bibnamefont {Xu}}, \bibinfo {author}
  {\bibfnamefont {J.~A.}\ \bibnamefont {Logan}}, \bibinfo {author}
  {\bibfnamefont {G.}~\bibnamefont {Wang}}, \bibinfo {author} {\bibfnamefont
  {N.}~\bibnamefont {van Loo}}, \bibinfo {author} {\bibfnamefont {J.~D.~S.}\
  \bibnamefont {Bommer}}, \bibinfo {author} {\bibfnamefont {M.~W.~A.}\
  \bibnamefont {de~Moor}}, \bibinfo {author} {\bibfnamefont {D.}~\bibnamefont
  {Car}}, \bibinfo {author} {\bibfnamefont {R.~L.~M.}\ \bibnamefont {Op~het
  Veld}}, \bibinfo {author} {\bibfnamefont {P.~J.}\ \bibnamefont {van
  Veldhoven}}, \bibinfo {author} {\bibfnamefont {S.}~\bibnamefont {Koelling}},
  \bibinfo {author} {\bibfnamefont {M.~A.}\ \bibnamefont {Verheijen}}, \bibinfo
  {author} {\bibfnamefont {M.}~\bibnamefont {Pendharkar}}, \bibinfo {author}
  {\bibfnamefont {D.~J.}\ \bibnamefont {Pennachio}}, \bibinfo {author}
  {\bibfnamefont {B.}~\bibnamefont {Shojaei}}, \bibinfo {author} {\bibfnamefont
  {J.~S.}\ \bibnamefont {Lee}}, \bibinfo {author} {\bibfnamefont {C.~J.}\
  \bibnamefont {Palmstr\o{}m}}, \bibinfo {author} {\bibfnamefont {E.~P. A.~M.}\
  \bibnamefont {Bakkers}}, \bibinfo {author} {\bibfnamefont {S.~D.}\
  \bibnamefont {Sarma}}, \ and\ \bibinfo {author} {\bibfnamefont {L.~P.}\
  \bibnamefont {Kouwenhoven}},\ }\href
  {http://www.nature.com/doifinder/10.1038/nature26142} {\bibfield  {journal}
  {\bibinfo  {journal} {Nature}\ }\textbf {\bibinfo {volume} {556}},\ \bibinfo
  {pages} {74} (\bibinfo {year} {2018})}\BibitemShut {NoStop}%
\bibitem [{\citenamefont {Vaitiek\ifmmode~\dot{e}\else \.{e}\fi{}nas}\ \emph
  {et~al.}(2018{\natexlab{a}})\citenamefont {Vaitiek\ifmmode~\dot{e}\else
  \.{e}\fi{}nas}, \citenamefont {Deng}, \citenamefont {Nyg\aa{}rd},
  \citenamefont {Krogstrup},\ and\ \citenamefont
  {Marcus}}]{vaitiekenas_effective_2018}%
  \BibitemOpen
  \bibfield  {author} {\bibinfo {author} {\bibfnamefont {S.}~\bibnamefont
  {Vaitiek\ifmmode~\dot{e}\else \.{e}\fi{}nas}}, \bibinfo {author}
  {\bibfnamefont {M.-T.}\ \bibnamefont {Deng}}, \bibinfo {author}
  {\bibfnamefont {J.}~\bibnamefont {Nyg\aa{}rd}}, \bibinfo {author}
  {\bibfnamefont {P.}~\bibnamefont {Krogstrup}}, \ and\ \bibinfo {author}
  {\bibfnamefont {C.}~\bibnamefont {Marcus}},\ }\href
  {https://link.aps.org/doi/10.1103/PhysRevLett.121.037703} {\bibfield
  {journal} {\bibinfo  {journal} {Phys. Rev. Lett.}\ }\textbf {\bibinfo
  {volume} {121}},\ \bibinfo {pages} {037703} (\bibinfo {year}
  {2018}{\natexlab{a}})}\BibitemShut {NoStop}%
\bibitem [{\citenamefont {Deng}\ \emph {et~al.}(2018)\citenamefont {Deng},
  \citenamefont {Vaitiek\ifmmode~\dot{e}\else \.{e}\fi{}nas}, \citenamefont
  {Prada}, \citenamefont {San-Jose}, \citenamefont {Nyg\aa{}rd}, \citenamefont
  {Krogstrup}, \citenamefont {Aguado},\ and\ \citenamefont
  {Marcus}}]{deng_nonlocality_2018}%
  \BibitemOpen
  \bibfield  {author} {\bibinfo {author} {\bibfnamefont {M.-T.}\ \bibnamefont
  {Deng}}, \bibinfo {author} {\bibfnamefont {S.}~\bibnamefont
  {Vaitiek\ifmmode~\dot{e}\else \.{e}\fi{}nas}}, \bibinfo {author}
  {\bibfnamefont {E.}~\bibnamefont {Prada}}, \bibinfo {author} {\bibfnamefont
  {P.}~\bibnamefont {San-Jose}}, \bibinfo {author} {\bibfnamefont
  {J.}~\bibnamefont {Nyg\aa{}rd}}, \bibinfo {author} {\bibfnamefont
  {P.}~\bibnamefont {Krogstrup}}, \bibinfo {author} {\bibfnamefont
  {R.}~\bibnamefont {Aguado}}, \ and\ \bibinfo {author} {\bibfnamefont {C.~M.}\
  \bibnamefont {Marcus}},\ }\href
  {https://link.aps.org/doi/10.1103/PhysRevB.98.085125} {\bibfield  {journal}
  {\bibinfo  {journal} {Phys. Rev. B}\ }\textbf {\bibinfo {volume} {98}},\
  \bibinfo {pages} {085125} (\bibinfo {year} {2018})}\BibitemShut {NoStop}%
\bibitem [{\citenamefont {Moor}\ \emph {et~al.}(2018)\citenamefont {Moor},
  \citenamefont {Bommer}, \citenamefont {Xu}, \citenamefont {Winkler},
  \citenamefont {Antipov}, \citenamefont {Bargerbos}, \citenamefont {{Guanzhong
  Wang}}, \citenamefont {Loo}, \citenamefont {Veld}, \citenamefont
  {Gazibegovic}, \citenamefont {Car}, \citenamefont {Logan}, \citenamefont
  {Pendharkar}, \citenamefont {Lee}, \citenamefont {Bakkers}, \citenamefont
  {Palmstr\o{}m}, \citenamefont {Lutchyn}, \citenamefont {Kouwenhoven},\ and\
  \citenamefont {Zhang}}]{moor_electric_2018}%
  \BibitemOpen
  \bibfield  {author} {\bibinfo {author} {\bibfnamefont {M.~W. A.~d.}\
  \bibnamefont {Moor}}, \bibinfo {author} {\bibfnamefont {J.~D.~S.}\
  \bibnamefont {Bommer}}, \bibinfo {author} {\bibfnamefont {D.}~\bibnamefont
  {Xu}}, \bibinfo {author} {\bibfnamefont {G.~W.}\ \bibnamefont {Winkler}},
  \bibinfo {author} {\bibfnamefont {A.~E.}\ \bibnamefont {Antipov}}, \bibinfo
  {author} {\bibfnamefont {A.}~\bibnamefont {Bargerbos}}, \bibinfo {author}
  {\bibnamefont {{Guanzhong Wang}}}, \bibinfo {author} {\bibfnamefont {N.~v.}\
  \bibnamefont {Loo}}, \bibinfo {author} {\bibfnamefont {R.~L. M. O.~h.}\
  \bibnamefont {Veld}}, \bibinfo {author} {\bibfnamefont {S.}~\bibnamefont
  {Gazibegovic}}, \bibinfo {author} {\bibfnamefont {D.}~\bibnamefont {Car}},
  \bibinfo {author} {\bibfnamefont {J.~A.}\ \bibnamefont {Logan}}, \bibinfo
  {author} {\bibfnamefont {M.}~\bibnamefont {Pendharkar}}, \bibinfo {author}
  {\bibfnamefont {J.~S.}\ \bibnamefont {Lee}}, \bibinfo {author} {\bibfnamefont
  {E.~P. A.~M.}\ \bibnamefont {Bakkers}}, \bibinfo {author} {\bibfnamefont
  {C.~J.}\ \bibnamefont {Palmstr\o{}m}}, \bibinfo {author} {\bibfnamefont
  {R.~M.}\ \bibnamefont {Lutchyn}}, \bibinfo {author} {\bibfnamefont {L.~P.}\
  \bibnamefont {Kouwenhoven}}, \ and\ \bibinfo {author} {\bibfnamefont
  {H.}~\bibnamefont {Zhang}},\ }\href
  {http://stacks.iop.org/1367-2630/20/i=10/a=103049} {\bibfield  {journal}
  {\bibinfo  {journal} {New J. Phys.}\ }\textbf {\bibinfo {volume} {20}},\
  \bibinfo {pages} {103049} (\bibinfo {year} {2018})}\BibitemShut {NoStop}%
\bibitem [{\citenamefont {Suominen}\ \emph {et~al.}(2017)\citenamefont
  {Suominen}, \citenamefont {Kjaergaard}, \citenamefont {Hamilton},
  \citenamefont {Shabani}, \citenamefont {Palmstr\o{}m}, \citenamefont
  {Marcus},\ and\ \citenamefont {Nichele}}]{suominen_zero-energy_2017}%
  \BibitemOpen
  \bibfield  {author} {\bibinfo {author} {\bibfnamefont {H.~J.}\ \bibnamefont
  {Suominen}}, \bibinfo {author} {\bibfnamefont {M.}~\bibnamefont
  {Kjaergaard}}, \bibinfo {author} {\bibfnamefont {A.~R.}\ \bibnamefont
  {Hamilton}}, \bibinfo {author} {\bibfnamefont {J.}~\bibnamefont {Shabani}},
  \bibinfo {author} {\bibfnamefont {C.~J.}\ \bibnamefont {Palmstr\o{}m}},
  \bibinfo {author} {\bibfnamefont {C.~M.}\ \bibnamefont {Marcus}}, \ and\
  \bibinfo {author} {\bibfnamefont {F.}~\bibnamefont {Nichele}},\ }\href
  {https://link.aps.org/doi/10.1103/PhysRevLett.119.176805} {\bibfield
  {journal} {\bibinfo  {journal} {Phys. Rev. Lett.}\ }\textbf {\bibinfo
  {volume} {119}},\ \bibinfo {pages} {176805} (\bibinfo {year}
  {2017})}\BibitemShut {NoStop}%
\bibitem [{\citenamefont {Nichele}\ \emph {et~al.}(2017)\citenamefont
  {Nichele}, \citenamefont {Drachmann}, \citenamefont {Whiticar}, \citenamefont
  {O’Farrell}, \citenamefont {Suominen}, \citenamefont {Fornieri},
  \citenamefont {Wang}, \citenamefont {Gardner}, \citenamefont {Thomas},
  \citenamefont {Hatke}, \citenamefont {Krogstrup}, \citenamefont {Manfra},
  \citenamefont {Flensberg},\ and\ \citenamefont
  {Marcus}}]{nichele_scaling_2017}%
  \BibitemOpen
  \bibfield  {author} {\bibinfo {author} {\bibfnamefont {F.}~\bibnamefont
  {Nichele}}, \bibinfo {author} {\bibfnamefont {A.~C.~C.}\ \bibnamefont
  {Drachmann}}, \bibinfo {author} {\bibfnamefont {A.~M.}\ \bibnamefont
  {Whiticar}}, \bibinfo {author} {\bibfnamefont {E.~C.~T.}\ \bibnamefont
  {O’Farrell}}, \bibinfo {author} {\bibfnamefont {H.~J.}\ \bibnamefont
  {Suominen}}, \bibinfo {author} {\bibfnamefont {A.}~\bibnamefont {Fornieri}},
  \bibinfo {author} {\bibfnamefont {T.}~\bibnamefont {Wang}}, \bibinfo {author}
  {\bibfnamefont {G.~C.}\ \bibnamefont {Gardner}}, \bibinfo {author}
  {\bibfnamefont {C.}~\bibnamefont {Thomas}}, \bibinfo {author} {\bibfnamefont
  {A.~T.}\ \bibnamefont {Hatke}}, \bibinfo {author} {\bibfnamefont
  {P.}~\bibnamefont {Krogstrup}}, \bibinfo {author} {\bibfnamefont {M.~J.}\
  \bibnamefont {Manfra}}, \bibinfo {author} {\bibfnamefont {K.}~\bibnamefont
  {Flensberg}}, \ and\ \bibinfo {author} {\bibfnamefont {C.~M.}\ \bibnamefont
  {Marcus}},\ }\href {https://link.aps.org/doi/10.1103/PhysRevLett.119.136803}
  {\bibfield  {journal} {\bibinfo  {journal} {Phys. Rev. Lett.}\ }\textbf
  {\bibinfo {volume} {119}},\ \bibinfo {pages} {136803} (\bibinfo {year}
  {2017})}\BibitemShut {NoStop}%
\bibitem [{\citenamefont {Lutchyn}\ \emph {et~al.}(2018)\citenamefont
  {Lutchyn}, \citenamefont {Bakkers}, \citenamefont {Kouwenhoven},
  \citenamefont {Krogstrup}, \citenamefont {Marcus},\ and\ \citenamefont
  {Oreg}}]{lutchyn_majorana_2018}%
  \BibitemOpen
  \bibfield  {author} {\bibinfo {author} {\bibfnamefont {R.~M.}\ \bibnamefont
  {Lutchyn}}, \bibinfo {author} {\bibfnamefont {E.~P. a.~M.}\ \bibnamefont
  {Bakkers}}, \bibinfo {author} {\bibfnamefont {L.~P.}\ \bibnamefont
  {Kouwenhoven}}, \bibinfo {author} {\bibfnamefont {P.}~\bibnamefont
  {Krogstrup}}, \bibinfo {author} {\bibfnamefont {C.~M.}\ \bibnamefont
  {Marcus}}, \ and\ \bibinfo {author} {\bibfnamefont {Y.}~\bibnamefont
  {Oreg}},\ }\href {https://www.nature.com/articles/s41578-018-0003-1}
  {\bibfield  {journal} {\bibinfo  {journal} {Nat. Rev. Mater.}\ }\textbf
  {\bibinfo {volume} {3}},\ \bibinfo {pages} {52} (\bibinfo {year}
  {2018})}\BibitemShut {NoStop}%
\bibitem [{\citenamefont {Lutchyn}\ \emph {et~al.}(2010)\citenamefont
  {Lutchyn}, \citenamefont {Sau},\ and\ \citenamefont
  {Das~Sarma}}]{lutchyn_majorana_2010}%
  \BibitemOpen
  \bibfield  {author} {\bibinfo {author} {\bibfnamefont {R.~M.}\ \bibnamefont
  {Lutchyn}}, \bibinfo {author} {\bibfnamefont {J.~D.}\ \bibnamefont {Sau}}, \
  and\ \bibinfo {author} {\bibfnamefont {S.}~\bibnamefont {Das~Sarma}},\ }\href
  {https://link.aps.org/doi/10.1103/PhysRevLett.105.077001} {\bibfield
  {journal} {\bibinfo  {journal} {Phys. Rev. Lett.}\ }\textbf {\bibinfo
  {volume} {105}},\ \bibinfo {pages} {077001} (\bibinfo {year}
  {2010})}\BibitemShut {NoStop}%
\bibitem [{\citenamefont {Oreg}\ \emph {et~al.}(2010)\citenamefont {Oreg},
  \citenamefont {Refael},\ and\ \citenamefont {von Oppen}}]{oreg_helical_2010}%
  \BibitemOpen
  \bibfield  {author} {\bibinfo {author} {\bibfnamefont {Y.}~\bibnamefont
  {Oreg}}, \bibinfo {author} {\bibfnamefont {G.}~\bibnamefont {Refael}}, \ and\
  \bibinfo {author} {\bibfnamefont {F.}~\bibnamefont {von Oppen}},\ }\href
  {https://link.aps.org/doi/10.1103/PhysRevLett.105.177002} {\bibfield
  {journal} {\bibinfo  {journal} {Phys. Rev. Lett.}\ }\textbf {\bibinfo
  {volume} {105}},\ \bibinfo {pages} {177002} (\bibinfo {year}
  {2010})}\BibitemShut {NoStop}%
\bibitem [{\citenamefont {Alicea}(2012)}]{alicea_new_2012}%
  \BibitemOpen
  \bibfield  {author} {\bibinfo {author} {\bibfnamefont {J.}~\bibnamefont
  {Alicea}},\ }\href
  {http://stacks.iop.org/0034-4885/75/i=7/a=076501?key=crossref.2b9d4760f53398acbcc889fef296e3c8}
  {\bibfield  {journal} {\bibinfo  {journal} {Rep. Prog. Phys.}\ }\textbf
  {\bibinfo {volume} {75}},\ \bibinfo {pages} {076501} (\bibinfo {year}
  {2012})}\BibitemShut {NoStop}%
\bibitem [{\citenamefont {Kells}\ \emph {et~al.}(2012)\citenamefont {Kells},
  \citenamefont {Meidan},\ and\ \citenamefont
  {Brouwer}}]{kells_near-zero-energy_2012}%
  \BibitemOpen
  \bibfield  {author} {\bibinfo {author} {\bibfnamefont {G.}~\bibnamefont
  {Kells}}, \bibinfo {author} {\bibfnamefont {D.}~\bibnamefont {Meidan}}, \
  and\ \bibinfo {author} {\bibfnamefont {P.~W.}\ \bibnamefont {Brouwer}},\
  }\href {https://link.aps.org/doi/10.1103/PhysRevB.86.100503} {\bibfield
  {journal} {\bibinfo  {journal} {Phys. Rev. B}\ }\textbf {\bibinfo {volume}
  {86}},\ \bibinfo {pages} {100503} (\bibinfo {year} {2012})}\BibitemShut
  {NoStop}%
\bibitem [{\citenamefont {Prada}\ \emph {et~al.}(2012)\citenamefont {Prada},
  \citenamefont {San-Jose},\ and\ \citenamefont
  {Aguado}}]{prada_transport_2012}%
  \BibitemOpen
  \bibfield  {author} {\bibinfo {author} {\bibfnamefont {E.}~\bibnamefont
  {Prada}}, \bibinfo {author} {\bibfnamefont {P.}~\bibnamefont {San-Jose}}, \
  and\ \bibinfo {author} {\bibfnamefont {R.}~\bibnamefont {Aguado}},\ }\href
  {https://link.aps.org/doi/10.1103/PhysRevB.86.180503} {\bibfield  {journal}
  {\bibinfo  {journal} {Phys. Rev. B}\ }\textbf {\bibinfo {volume} {86}},\
  \bibinfo {pages} {180503} (\bibinfo {year} {2012})}\BibitemShut {NoStop}%
\bibitem [{\citenamefont {Liu}\ \emph {et~al.}(2017)\citenamefont {Liu},
  \citenamefont {Sau}, \citenamefont {Stanescu},\ and\ \citenamefont
  {Das~Sarma}}]{liu_andreev_2017}%
  \BibitemOpen
  \bibfield  {author} {\bibinfo {author} {\bibfnamefont {C.-X.}\ \bibnamefont
  {Liu}}, \bibinfo {author} {\bibfnamefont {J.~D.}\ \bibnamefont {Sau}},
  \bibinfo {author} {\bibfnamefont {T.~D.}\ \bibnamefont {Stanescu}}, \ and\
  \bibinfo {author} {\bibfnamefont {S.}~\bibnamefont {Das~Sarma}},\ }\href
  {https://link.aps.org/doi/10.1103/PhysRevB.96.075161} {\bibfield  {journal}
  {\bibinfo  {journal} {Phys. Rev. B}\ }\textbf {\bibinfo {volume} {96}},\
  \bibinfo {pages} {075161} (\bibinfo {year} {2017})}\BibitemShut {NoStop}%
\bibitem [{\citenamefont {Moore}\ \emph
  {et~al.}(2018{\natexlab{a}})\citenamefont {Moore}, \citenamefont {Stanescu},\
  and\ \citenamefont {Tewari}}]{moore_two-terminal_2018}%
  \BibitemOpen
  \bibfield  {author} {\bibinfo {author} {\bibfnamefont {C.}~\bibnamefont
  {Moore}}, \bibinfo {author} {\bibfnamefont {T.~D.}\ \bibnamefont {Stanescu}},
  \ and\ \bibinfo {author} {\bibfnamefont {S.}~\bibnamefont {Tewari}},\ }\href
  {https://link.aps.org/doi/10.1103/PhysRevB.97.165302} {\bibfield  {journal}
  {\bibinfo  {journal} {Phys. Rev. B}\ }\textbf {\bibinfo {volume} {97}},\
  \bibinfo {pages} {165302} (\bibinfo {year} {2018}{\natexlab{a}})}\BibitemShut
  {NoStop}%
\bibitem [{\citenamefont {Setiawan}\ \emph {et~al.}(2017)\citenamefont
  {Setiawan}, \citenamefont {Liu}, \citenamefont {Sau},\ and\ \citenamefont
  {Das~Sarma}}]{setiawan_electron_2017}%
  \BibitemOpen
  \bibfield  {author} {\bibinfo {author} {\bibfnamefont {F.}~\bibnamefont
  {Setiawan}}, \bibinfo {author} {\bibfnamefont {C.-X.}\ \bibnamefont {Liu}},
  \bibinfo {author} {\bibfnamefont {J.~D.}\ \bibnamefont {Sau}}, \ and\
  \bibinfo {author} {\bibfnamefont {S.}~\bibnamefont {Das~Sarma}},\ }\href
  {https://link.aps.org/doi/10.1103/PhysRevB.96.184520} {\bibfield  {journal}
  {\bibinfo  {journal} {Phys. Rev. B}\ }\textbf {\bibinfo {volume} {96}},\
  \bibinfo {pages} {184520} (\bibinfo {year} {2017})}\BibitemShut {NoStop}%
\bibitem [{\citenamefont {Moore}\ \emph
  {et~al.}(2018{\natexlab{b}})\citenamefont {Moore}, \citenamefont {Zeng},
  \citenamefont {Stanescu},\ and\ \citenamefont
  {Tewari}}]{moore_quantized_2018}%
  \BibitemOpen
  \bibfield  {author} {\bibinfo {author} {\bibfnamefont {C.}~\bibnamefont
  {Moore}}, \bibinfo {author} {\bibfnamefont {C.}~\bibnamefont {Zeng}},
  \bibinfo {author} {\bibfnamefont {T.~D.}\ \bibnamefont {Stanescu}}, \ and\
  \bibinfo {author} {\bibfnamefont {S.}~\bibnamefont {Tewari}},\ }\href
  {https://link.aps.org/doi/10.1103/PhysRevB.98.155314} {\bibfield  {journal}
  {\bibinfo  {journal} {Phys. Rev. B}\ }\textbf {\bibinfo {volume} {98}},\
  \bibinfo {pages} {155314} (\bibinfo {year} {2018}{\natexlab{b}})}\BibitemShut
  {NoStop}%
\bibitem [{\citenamefont {Liu}\ \emph {et~al.}(2018)\citenamefont {Liu},
  \citenamefont {Sau},\ and\ \citenamefont
  {Das~Sarma}}]{liu_distinguishing_2018}%
  \BibitemOpen
  \bibfield  {author} {\bibinfo {author} {\bibfnamefont {C.-X.}\ \bibnamefont
  {Liu}}, \bibinfo {author} {\bibfnamefont {J.~D.}\ \bibnamefont {Sau}}, \ and\
  \bibinfo {author} {\bibfnamefont {S.}~\bibnamefont {Das~Sarma}},\ }\href
  {https://link.aps.org/doi/10.1103/PhysRevB.97.214502} {\bibfield  {journal}
  {\bibinfo  {journal} {Phys. Rev. B}\ }\textbf {\bibinfo {volume} {97}},\
  \bibinfo {pages} {214502} (\bibinfo {year} {2018})}\BibitemShut {NoStop}%
\bibitem [{\citenamefont {Vuik}\ \emph {et~al.}(2018)\citenamefont {Vuik},
  \citenamefont {Nijholt}, \citenamefont {Akhmerov},\ and\ \citenamefont
  {Wimmer}}]{vuik_reproducing_2018}%
  \BibitemOpen
  \bibfield  {author} {\bibinfo {author} {\bibfnamefont {A.}~\bibnamefont
  {Vuik}}, \bibinfo {author} {\bibfnamefont {B.}~\bibnamefont {Nijholt}},
  \bibinfo {author} {\bibfnamefont {A.~R.}\ \bibnamefont {Akhmerov}}, \ and\
  \bibinfo {author} {\bibfnamefont {M.}~\bibnamefont {Wimmer}},\ }\href
  {http://arxiv.org/abs/1806.02801} {\bibfield  {journal} {\bibinfo  {journal}
  {arXiv:1806.02801 [cond-mat]}\ } (\bibinfo {year} {2018})}\BibitemShut
  {NoStop}%
\bibitem [{\citenamefont {Pe\~{n}aranda}\ \emph {et~al.}(2018)\citenamefont
  {Pe\~{n}aranda}, \citenamefont {Aguado}, \citenamefont {San-Jose},\ and\
  \citenamefont {Prada}}]{penaranda_quantifying_2018}%
  \BibitemOpen
  \bibfield  {author} {\bibinfo {author} {\bibfnamefont {F.}~\bibnamefont
  {Pe\~{n}aranda}}, \bibinfo {author} {\bibfnamefont {R.}~\bibnamefont
  {Aguado}}, \bibinfo {author} {\bibfnamefont {P.}~\bibnamefont {San-Jose}}, \
  and\ \bibinfo {author} {\bibfnamefont {E.}~\bibnamefont {Prada}},\ }\href
  {https://link.aps.org/doi/10.1103/PhysRevB.98.235406} {\bibfield  {journal}
  {\bibinfo  {journal} {Phys. Rev. B}\ }\textbf {\bibinfo {volume} {98}},\
  \bibinfo {pages} {235406} (\bibinfo {year} {2018})}\BibitemShut {NoStop}%
\bibitem [{\citenamefont {Reeg}\ \emph {et~al.}(2018)\citenamefont {Reeg},
  \citenamefont {Dmytruk}, \citenamefont {Chevallier}, \citenamefont {Loss},\
  and\ \citenamefont {Klinovaja}}]{reeg_zero-energy_2018}%
  \BibitemOpen
  \bibfield  {author} {\bibinfo {author} {\bibfnamefont {C.}~\bibnamefont
  {Reeg}}, \bibinfo {author} {\bibfnamefont {O.}~\bibnamefont {Dmytruk}},
  \bibinfo {author} {\bibfnamefont {D.}~\bibnamefont {Chevallier}}, \bibinfo
  {author} {\bibfnamefont {D.}~\bibnamefont {Loss}}, \ and\ \bibinfo {author}
  {\bibfnamefont {J.}~\bibnamefont {Klinovaja}},\ }\href
  {http://arxiv.org/abs/1810.09840} {\bibfield  {journal} {\bibinfo  {journal}
  {arXiv:1810.09840 [cond-mat]}\ } (\bibinfo {year} {2018})}\BibitemShut
  {NoStop}%
\bibitem [{\citenamefont {Stanescu}\ and\ \citenamefont
  {Tewari}(2018)}]{stanescu_illustrated_2018}%
  \BibitemOpen
  \bibfield  {author} {\bibinfo {author} {\bibfnamefont {T.~D.}\ \bibnamefont
  {Stanescu}}\ and\ \bibinfo {author} {\bibfnamefont {S.}~\bibnamefont
  {Tewari}},\ }\href {http://arxiv.org/abs/1811.02557} {\bibfield  {journal}
  {\bibinfo  {journal} {arXiv:1811.02557 [cond-mat]}\ } (\bibinfo {year}
  {2018})}\BibitemShut {NoStop}%
\bibitem [{\citenamefont {Chiu}\ and\ \citenamefont
  {Sarma}(2018)}]{chiu_fractional_2018}%
  \BibitemOpen
  \bibfield  {author} {\bibinfo {author} {\bibfnamefont {C.-K.}\ \bibnamefont
  {Chiu}}\ and\ \bibinfo {author} {\bibfnamefont {S.~D.}\ \bibnamefont
  {Sarma}},\ }\href {https://arxiv.org/abs/1806.02224} {\bibfield  {journal}
  {\bibinfo  {journal} {arXiv:1806.02224 [cond-mat]}\ } (\bibinfo {year}
  {2018})}\BibitemShut {NoStop}%
\bibitem [{\citenamefont {Stanescu}\ \emph {et~al.}(2012)\citenamefont
  {Stanescu}, \citenamefont {Tewari}, \citenamefont {Sau},\ and\ \citenamefont
  {Das~Sarma}}]{stanescu_close_2012}%
  \BibitemOpen
  \bibfield  {author} {\bibinfo {author} {\bibfnamefont {T.~D.}\ \bibnamefont
  {Stanescu}}, \bibinfo {author} {\bibfnamefont {S.}~\bibnamefont {Tewari}},
  \bibinfo {author} {\bibfnamefont {J.~D.}\ \bibnamefont {Sau}}, \ and\
  \bibinfo {author} {\bibfnamefont {S.}~\bibnamefont {Das~Sarma}},\ }\href
  {https://link.aps.org/doi/10.1103/PhysRevLett.109.266402} {\bibfield
  {journal} {\bibinfo  {journal} {Phys. Rev. Lett.}\ }\textbf {\bibinfo
  {volume} {109}},\ \bibinfo {pages} {266402} (\bibinfo {year}
  {2012})}\BibitemShut {NoStop}%
\bibitem [{\citenamefont {Mishmash}\ \emph {et~al.}(2016)\citenamefont
  {Mishmash}, \citenamefont {Aasen}, \citenamefont {Higginbotham},\ and\
  \citenamefont {Alicea}}]{mishmash_approaching_2016}%
  \BibitemOpen
  \bibfield  {author} {\bibinfo {author} {\bibfnamefont {R.~V.}\ \bibnamefont
  {Mishmash}}, \bibinfo {author} {\bibfnamefont {D.}~\bibnamefont {Aasen}},
  \bibinfo {author} {\bibfnamefont {A.~P.}\ \bibnamefont {Higginbotham}}, \
  and\ \bibinfo {author} {\bibfnamefont {J.}~\bibnamefont {Alicea}},\ }\href
  {https://link.aps.org/doi/10.1103/PhysRevB.93.245404} {\bibfield  {journal}
  {\bibinfo  {journal} {Phys. Rev. B}\ }\textbf {\bibinfo {volume} {93}},\
  \bibinfo {pages} {245404} (\bibinfo {year} {2016})}\BibitemShut {NoStop}%
\bibitem [{\citenamefont {Grivnin}\ \emph {et~al.}(2018)\citenamefont
  {Grivnin}, \citenamefont {Bor}, \citenamefont {Heiblum}, \citenamefont
  {Oreg},\ and\ \citenamefont {Shtrikman}}]{grivnin_concomitant_2018}%
  \BibitemOpen
  \bibfield  {author} {\bibinfo {author} {\bibfnamefont {A.}~\bibnamefont
  {Grivnin}}, \bibinfo {author} {\bibfnamefont {E.}~\bibnamefont {Bor}},
  \bibinfo {author} {\bibfnamefont {M.}~\bibnamefont {Heiblum}}, \bibinfo
  {author} {\bibfnamefont {Y.}~\bibnamefont {Oreg}}, \ and\ \bibinfo {author}
  {\bibfnamefont {H.}~\bibnamefont {Shtrikman}},\ }\href
  {http://arxiv.org/abs/1807.06632} {\bibfield  {journal} {\bibinfo  {journal}
  {arXiv:1807.06632 [cond-mat]}\ } (\bibinfo {year} {2018})}\BibitemShut
  {NoStop}%
\bibitem [{\citenamefont {Rosdahl}\ \emph {et~al.}(2018)\citenamefont
  {Rosdahl}, \citenamefont {Vuik}, \citenamefont {Kjaergaard},\ and\
  \citenamefont {Akhmerov}}]{AndreevRectifier}%
  \BibitemOpen
  \bibfield  {author} {\bibinfo {author} {\bibfnamefont {T.~O.}\ \bibnamefont
  {Rosdahl}}, \bibinfo {author} {\bibfnamefont {A.}~\bibnamefont {Vuik}},
  \bibinfo {author} {\bibfnamefont {M.}~\bibnamefont {Kjaergaard}}, \ and\
  \bibinfo {author} {\bibfnamefont {A.~R.}\ \bibnamefont {Akhmerov}},\ }\href
  {https://link.aps.org/doi/10.1103/PhysRevB.97.045421} {\bibfield  {journal}
  {\bibinfo  {journal} {Phys. Rev. B}\ }\textbf {\bibinfo {volume} {97}},\
  \bibinfo {pages} {045421} (\bibinfo {year} {2018})}\BibitemShut {NoStop}%
\bibitem [{\citenamefont {Danon}\ \emph {et~al.}(2019)\citenamefont {Danon},
  \citenamefont {Hellenes}, \citenamefont {Hansen}, \citenamefont {Casparis},
  \citenamefont {Higginbotham},\ and\ \citenamefont
  {Flensberg}}]{danon_nonlocal_2019}%
  \BibitemOpen
  \bibfield  {author} {\bibinfo {author} {\bibfnamefont {J.}~\bibnamefont
  {Danon}}, \bibinfo {author} {\bibfnamefont {A.~B.}\ \bibnamefont {Hellenes}},
  \bibinfo {author} {\bibfnamefont {E.~B.}\ \bibnamefont {Hansen}}, \bibinfo
  {author} {\bibfnamefont {L.}~\bibnamefont {Casparis}}, \bibinfo {author}
  {\bibfnamefont {A.~P.}\ \bibnamefont {Higginbotham}}, \ and\ \bibinfo
  {author} {\bibfnamefont {K.}~\bibnamefont {Flensberg}},\ }\href
  {http://arxiv.org/abs/1905.05438} {\bibfield  {journal} {\bibinfo  {journal}
  {arXiv:1905.05438 [cond-mat]}\ } (\bibinfo {year} {2019})}\BibitemShut
  {NoStop}%
\bibitem [{\citenamefont {M\'{e}nard}\ \emph {et~al.}(2019)\citenamefont
  {M\'{e}nard}, \citenamefont {Anselmetti}, \citenamefont {Martinez},
  \citenamefont {Puglia}, \citenamefont {Malinowski}, \citenamefont {Lee},
  \citenamefont {Choi}, \citenamefont {Pendharkar}, \citenamefont
  {Palmstr\o{}m}, \citenamefont {Flensberg}, \citenamefont {Marcus},
  \citenamefont {Casparis},\ and\ \citenamefont
  {Higginbotham}}]{menard_conductance-matrix_2019}%
  \BibitemOpen
  \bibfield  {author} {\bibinfo {author} {\bibfnamefont {G.~C.}\ \bibnamefont
  {M\'{e}nard}}, \bibinfo {author} {\bibfnamefont {G.~L.~R.}\ \bibnamefont
  {Anselmetti}}, \bibinfo {author} {\bibfnamefont {E.~A.}\ \bibnamefont
  {Martinez}}, \bibinfo {author} {\bibfnamefont {D.}~\bibnamefont {Puglia}},
  \bibinfo {author} {\bibfnamefont {F.~K.}\ \bibnamefont {Malinowski}},
  \bibinfo {author} {\bibfnamefont {J.~S.}\ \bibnamefont {Lee}}, \bibinfo
  {author} {\bibfnamefont {S.}~\bibnamefont {Choi}}, \bibinfo {author}
  {\bibfnamefont {M.}~\bibnamefont {Pendharkar}}, \bibinfo {author}
  {\bibfnamefont {C.~J.}\ \bibnamefont {Palmstr\o{}m}}, \bibinfo {author}
  {\bibfnamefont {K.}~\bibnamefont {Flensberg}}, \bibinfo {author}
  {\bibfnamefont {C.~M.}\ \bibnamefont {Marcus}}, \bibinfo {author}
  {\bibfnamefont {L.}~\bibnamefont {Casparis}}, \ and\ \bibinfo {author}
  {\bibfnamefont {A.~P.}\ \bibnamefont {Higginbotham}},\ }\href
  {http://arxiv.org/abs/1905.05505} {\bibfield  {journal} {\bibinfo  {journal}
  {arXiv:1905.05505 [cond-mat]}\ } (\bibinfo {year} {2019})}\BibitemShut
  {NoStop}%
\bibitem [{\citenamefont {Zhang}\ \emph {et~al.}(2019)\citenamefont {Zhang},
  \citenamefont {Liu}, \citenamefont {Wimmer},\ and\ \citenamefont
  {Kouwenhoven}}]{zhang_quantum_2019}%
  \BibitemOpen
  \bibfield  {author} {\bibinfo {author} {\bibfnamefont {H.}~\bibnamefont
  {Zhang}}, \bibinfo {author} {\bibfnamefont {D.~E.}\ \bibnamefont {Liu}},
  \bibinfo {author} {\bibfnamefont {M.}~\bibnamefont {Wimmer}}, \ and\ \bibinfo
  {author} {\bibfnamefont {L.~P.}\ \bibnamefont {Kouwenhoven}},\ }\href
  {http://arxiv.org/abs/1905.07882} {\bibfield  {journal} {\bibinfo  {journal}
  {arXiv:1905.07882 [cond-mat]}\ } (\bibinfo {year} {2019})}\BibitemShut
  {NoStop}%
\bibitem [{\citenamefont {Clarke}(2017)}]{clarke_experimentally_2017}%
  \BibitemOpen
  \bibfield  {author} {\bibinfo {author} {\bibfnamefont {D.~J.}\ \bibnamefont
  {Clarke}},\ }\href {https://link.aps.org/doi/10.1103/PhysRevB.96.201109}
  {\bibfield  {journal} {\bibinfo  {journal} {Phys. Rev. B}\ }\textbf {\bibinfo
  {volume} {96}},\ \bibinfo {pages} {201109} (\bibinfo {year}
  {2017})}\BibitemShut {NoStop}%
\bibitem [{\citenamefont {Prada}\ \emph {et~al.}(2017)\citenamefont {Prada},
  \citenamefont {Aguado},\ and\ \citenamefont
  {San-Jose}}]{prada_measuring_2017}%
  \BibitemOpen
  \bibfield  {author} {\bibinfo {author} {\bibfnamefont {E.}~\bibnamefont
  {Prada}}, \bibinfo {author} {\bibfnamefont {R.}~\bibnamefont {Aguado}}, \
  and\ \bibinfo {author} {\bibfnamefont {P.}~\bibnamefont {San-Jose}},\ }\href
  {https://link.aps.org/doi/10.1103/PhysRevB.96.085418} {\bibfield  {journal}
  {\bibinfo  {journal} {Phys. Rev. B}\ }\textbf {\bibinfo {volume} {96}},\
  \bibinfo {pages} {085418} (\bibinfo {year} {2017})}\BibitemShut {NoStop}%
\bibitem [{\citenamefont {Yavilberg}\ \emph {et~al.}(2019)\citenamefont
  {Yavilberg}, \citenamefont {Ginossar},\ and\ \citenamefont
  {Grosfeld}}]{yavilberg_differentiating_2019}%
  \BibitemOpen
  \bibfield  {author} {\bibinfo {author} {\bibfnamefont {K.}~\bibnamefont
  {Yavilberg}}, \bibinfo {author} {\bibfnamefont {E.}~\bibnamefont {Ginossar}},
  \ and\ \bibinfo {author} {\bibfnamefont {E.}~\bibnamefont {Grosfeld}},\
  }\href {http://arxiv.org/abs/1902.07229} {\bibfield  {journal} {\bibinfo
  {journal} {arXiv:1902.07229 [cond-mat]}\ } (\bibinfo {year}
  {2019})}\BibitemShut {NoStop}%
\bibitem [{\citenamefont {Schrade}\ and\ \citenamefont
  {Fu}(2018)}]{schrade_andreev_2018}%
  \BibitemOpen
  \bibfield  {author} {\bibinfo {author} {\bibfnamefont {C.}~\bibnamefont
  {Schrade}}\ and\ \bibinfo {author} {\bibfnamefont {L.}~\bibnamefont {Fu}},\
  }\href {http://arxiv.org/abs/1809.06370} {\bibfield  {journal} {\bibinfo
  {journal} {arXiv:1809.06370 [cond-mat]}\ } (\bibinfo {year}
  {2018})}\BibitemShut {NoStop}%
\bibitem [{\citenamefont {Aasen}\ \emph {et~al.}(2016)\citenamefont {Aasen},
  \citenamefont {Hell}, \citenamefont {Mishmash}, \citenamefont {Higginbotham},
  \citenamefont {Danon}, \citenamefont {Leijnse}, \citenamefont {Jespersen},
  \citenamefont {Folk}, \citenamefont {Marcus}, \citenamefont {Flensberg},\
  and\ \citenamefont {Alicea}}]{aasen_milestones_2016}%
  \BibitemOpen
  \bibfield  {author} {\bibinfo {author} {\bibfnamefont {D.}~\bibnamefont
  {Aasen}}, \bibinfo {author} {\bibfnamefont {M.}~\bibnamefont {Hell}},
  \bibinfo {author} {\bibfnamefont {R.~V.}\ \bibnamefont {Mishmash}}, \bibinfo
  {author} {\bibfnamefont {A.}~\bibnamefont {Higginbotham}}, \bibinfo {author}
  {\bibfnamefont {J.}~\bibnamefont {Danon}}, \bibinfo {author} {\bibfnamefont
  {M.}~\bibnamefont {Leijnse}}, \bibinfo {author} {\bibfnamefont {T.~S.}\
  \bibnamefont {Jespersen}}, \bibinfo {author} {\bibfnamefont {J.~A.}\
  \bibnamefont {Folk}}, \bibinfo {author} {\bibfnamefont {C.~M.}\ \bibnamefont
  {Marcus}}, \bibinfo {author} {\bibfnamefont {K.}~\bibnamefont {Flensberg}}, \
  and\ \bibinfo {author} {\bibfnamefont {J.}~\bibnamefont {Alicea}},\ }\href
  {https://link.aps.org/doi/10.1103/PhysRevX.6.031016} {\bibfield  {journal}
  {\bibinfo  {journal} {Phys. Rev. X}\ }\textbf {\bibinfo {volume} {6}},\
  \bibinfo {pages} {031016} (\bibinfo {year} {2016})}\BibitemShut {NoStop}%
\bibitem [{\citenamefont {Plugge}\ \emph {et~al.}(2017)\citenamefont {Plugge},
  \citenamefont {Rasmussen}, \citenamefont {Egger},\ and\ \citenamefont
  {Flensberg}}]{plugge_majorana_2017}%
  \BibitemOpen
  \bibfield  {author} {\bibinfo {author} {\bibfnamefont {S.}~\bibnamefont
  {Plugge}}, \bibinfo {author} {\bibfnamefont {A.}~\bibnamefont {Rasmussen}},
  \bibinfo {author} {\bibfnamefont {R.}~\bibnamefont {Egger}}, \ and\ \bibinfo
  {author} {\bibfnamefont {K.}~\bibnamefont {Flensberg}},\ }\href
  {http://stacks.iop.org/1367-2630/19/i=1/a=012001} {\bibfield  {journal}
  {\bibinfo  {journal} {New J. Phys.}\ }\textbf {\bibinfo {volume} {19}},\
  \bibinfo {pages} {012001} (\bibinfo {year} {2017})}\BibitemShut {NoStop}%
\bibitem [{\citenamefont {Karzig}\ \emph {et~al.}(2017)\citenamefont {Karzig},
  \citenamefont {Knapp}, \citenamefont {Lutchyn}, \citenamefont {Bonderson},
  \citenamefont {Hastings}, \citenamefont {Nayak}, \citenamefont {Alicea},
  \citenamefont {Flensberg}, \citenamefont {Plugge}, \citenamefont {Oreg},\
  and\ \citenamefont {{others}}}]{karzig_scalable_2017}%
  \BibitemOpen
  \bibfield  {author} {\bibinfo {author} {\bibfnamefont {T.}~\bibnamefont
  {Karzig}}, \bibinfo {author} {\bibfnamefont {C.}~\bibnamefont {Knapp}},
  \bibinfo {author} {\bibfnamefont {R.~M.}\ \bibnamefont {Lutchyn}}, \bibinfo
  {author} {\bibfnamefont {P.}~\bibnamefont {Bonderson}}, \bibinfo {author}
  {\bibfnamefont {M.~B.}\ \bibnamefont {Hastings}}, \bibinfo {author}
  {\bibfnamefont {C.}~\bibnamefont {Nayak}}, \bibinfo {author} {\bibfnamefont
  {J.}~\bibnamefont {Alicea}}, \bibinfo {author} {\bibfnamefont
  {K.}~\bibnamefont {Flensberg}}, \bibinfo {author} {\bibfnamefont
  {S.}~\bibnamefont {Plugge}}, \bibinfo {author} {\bibfnamefont
  {Y.}~\bibnamefont {Oreg}}, \ and\ \bibinfo {author} {\bibnamefont
  {{others}}},\ }\href
  {https://journals.aps.org/prb/abstract/10.1103/PhysRevB.95.235305} {\bibfield
   {journal} {\bibinfo  {journal} {Phys. Rev. B}\ }\textbf {\bibinfo {volume}
  {95}},\ \bibinfo {pages} {235305} (\bibinfo {year} {2017})}\BibitemShut
  {NoStop}%
\bibitem [{\citenamefont {Knapp}\ \emph
  {et~al.}(2018{\natexlab{a}})\citenamefont {Knapp}, \citenamefont {Karzig},
  \citenamefont {Lutchyn},\ and\ \citenamefont {Nayak}}]{knapp_dephasing_2018}%
  \BibitemOpen
  \bibfield  {author} {\bibinfo {author} {\bibfnamefont {C.}~\bibnamefont
  {Knapp}}, \bibinfo {author} {\bibfnamefont {T.}~\bibnamefont {Karzig}},
  \bibinfo {author} {\bibfnamefont {R.~M.}\ \bibnamefont {Lutchyn}}, \ and\
  \bibinfo {author} {\bibfnamefont {C.}~\bibnamefont {Nayak}},\ }\href
  {https://link.aps.org/doi/10.1103/PhysRevB.97.125404} {\bibfield  {journal}
  {\bibinfo  {journal} {Phys. Rev. B}\ }\textbf {\bibinfo {volume} {97}},\
  \bibinfo {pages} {125404} (\bibinfo {year} {2018}{\natexlab{a}})}\BibitemShut
  {NoStop}%
\bibitem [{Note1()}]{Note1}%
  \BibitemOpen
  \bibinfo {note} {For more parallel terminology one can view the ABS qubit as
  a superficial, hardly topological tetron (SHT-tetron).}\BibitemShut {Stop}%
\bibitem [{\citenamefont {Das~Sarma}\ \emph {et~al.}(2012)\citenamefont
  {Das~Sarma}, \citenamefont {Sau},\ and\ \citenamefont
  {Stanescu}}]{das_sarma_splitting_2012}%
  \BibitemOpen
  \bibfield  {author} {\bibinfo {author} {\bibfnamefont {S.}~\bibnamefont
  {Das~Sarma}}, \bibinfo {author} {\bibfnamefont {J.~D.}\ \bibnamefont {Sau}},
  \ and\ \bibinfo {author} {\bibfnamefont {T.~D.}\ \bibnamefont {Stanescu}},\
  }\href {https://link.aps.org/doi/10.1103/PhysRevB.86.220506} {\bibfield
  {journal} {\bibinfo  {journal} {Phys. Rev. B}\ }\textbf {\bibinfo {volume}
  {86}},\ \bibinfo {pages} {220506} (\bibinfo {year} {2012})}\BibitemShut
  {NoStop}%
\bibitem [{\citenamefont {Gharavi}\ \emph {et~al.}(2016)\citenamefont
  {Gharavi}, \citenamefont {Hoving},\ and\ \citenamefont
  {Baugh}}]{gharavi_readout_2016}%
  \BibitemOpen
  \bibfield  {author} {\bibinfo {author} {\bibfnamefont {K.}~\bibnamefont
  {Gharavi}}, \bibinfo {author} {\bibfnamefont {D.}~\bibnamefont {Hoving}}, \
  and\ \bibinfo {author} {\bibfnamefont {J.}~\bibnamefont {Baugh}},\ }\href
  {https://link.aps.org/doi/10.1103/PhysRevB.94.155417} {\bibfield  {journal}
  {\bibinfo  {journal} {Phys. Rev. B}\ }\textbf {\bibinfo {volume} {94}},\
  \bibinfo {pages} {155417} (\bibinfo {year} {2016})}\BibitemShut {NoStop}%
\bibitem [{\citenamefont {Vion}\ \emph {et~al.}(2002)\citenamefont {Vion},
  \citenamefont {Aassime}, \citenamefont {Cottet}, \citenamefont {Joyez},
  \citenamefont {Pothier}, \citenamefont {Urbina}, \citenamefont {Esteve},\
  and\ \citenamefont {Devoret}}]{vion_manipulating_2002}%
  \BibitemOpen
  \bibfield  {author} {\bibinfo {author} {\bibfnamefont {D.}~\bibnamefont
  {Vion}}, \bibinfo {author} {\bibfnamefont {A.}~\bibnamefont {Aassime}},
  \bibinfo {author} {\bibfnamefont {A.}~\bibnamefont {Cottet}}, \bibinfo
  {author} {\bibfnamefont {P.}~\bibnamefont {Joyez}}, \bibinfo {author}
  {\bibfnamefont {H.}~\bibnamefont {Pothier}}, \bibinfo {author} {\bibfnamefont
  {C.}~\bibnamefont {Urbina}}, \bibinfo {author} {\bibfnamefont
  {D.}~\bibnamefont {Esteve}}, \ and\ \bibinfo {author} {\bibfnamefont {M.~H.}\
  \bibnamefont {Devoret}},\ }\href
  {http://science.sciencemag.org/content/296/5569/886} {\bibfield  {journal}
  {\bibinfo  {journal} {Science}\ }\textbf {\bibinfo {volume} {296}},\ \bibinfo
  {pages} {886} (\bibinfo {year} {2002})}\BibitemShut {NoStop}%
\bibitem [{\citenamefont {Ithier}\ \emph {et~al.}(2005)\citenamefont {Ithier},
  \citenamefont {Collin}, \citenamefont {Joyez}, \citenamefont {Meeson},
  \citenamefont {Vion}, \citenamefont {Esteve}, \citenamefont {Chiarello},
  \citenamefont {Shnirman}, \citenamefont {Makhlin}, \citenamefont {Schriefl},\
  and\ \citenamefont {Sch\"on}}]{ithier_decoherence_2005}%
  \BibitemOpen
  \bibfield  {author} {\bibinfo {author} {\bibfnamefont {G.}~\bibnamefont
  {Ithier}}, \bibinfo {author} {\bibfnamefont {E.}~\bibnamefont {Collin}},
  \bibinfo {author} {\bibfnamefont {P.}~\bibnamefont {Joyez}}, \bibinfo
  {author} {\bibfnamefont {P.~J.}\ \bibnamefont {Meeson}}, \bibinfo {author}
  {\bibfnamefont {D.}~\bibnamefont {Vion}}, \bibinfo {author} {\bibfnamefont
  {D.}~\bibnamefont {Esteve}}, \bibinfo {author} {\bibfnamefont
  {F.}~\bibnamefont {Chiarello}}, \bibinfo {author} {\bibfnamefont
  {A.}~\bibnamefont {Shnirman}}, \bibinfo {author} {\bibfnamefont
  {Y.}~\bibnamefont {Makhlin}}, \bibinfo {author} {\bibfnamefont
  {J.}~\bibnamefont {Schriefl}}, \ and\ \bibinfo {author} {\bibfnamefont
  {G.}~\bibnamefont {Sch\"on}},\ }\href
  {https://link.aps.org/doi/10.1103/PhysRevB.72.134519} {\bibfield  {journal}
  {\bibinfo  {journal} {Phys. Rev. B}\ }\textbf {\bibinfo {volume} {72}},\
  \bibinfo {pages} {134519} (\bibinfo {year} {2005})}\BibitemShut {NoStop}%
\bibitem [{Note2()}]{Note2}%
  \BibitemOpen
  \bibinfo {note} {A related sequence involving $Z$ initialization, two $\pi
  /2$ rotations about the $x$ axis buttressing a wait time $t$, followed by a
  final $Z$ measurement was proposed in Ref.~\cite {aasen_milestones_2016}, as
  appropriate for the qubit design presented therein.}\BibitemShut {Stop}%
\bibitem [{\citenamefont {Koch}\ \emph {et~al.}(2007)\citenamefont {Koch},
  \citenamefont {Yu}, \citenamefont {Gambetta}, \citenamefont {Houck},
  \citenamefont {Schuster}, \citenamefont {Majer}, \citenamefont {Blais},
  \citenamefont {Devoret}, \citenamefont {Girvin},\ and\ \citenamefont
  {Schoelkopf}}]{koch_charge-insensitive_2007}%
  \BibitemOpen
  \bibfield  {author} {\bibinfo {author} {\bibfnamefont {J.}~\bibnamefont
  {Koch}}, \bibinfo {author} {\bibfnamefont {T.~M.}\ \bibnamefont {Yu}},
  \bibinfo {author} {\bibfnamefont {J.}~\bibnamefont {Gambetta}}, \bibinfo
  {author} {\bibfnamefont {A.~A.}\ \bibnamefont {Houck}}, \bibinfo {author}
  {\bibfnamefont {D.~I.}\ \bibnamefont {Schuster}}, \bibinfo {author}
  {\bibfnamefont {J.}~\bibnamefont {Majer}}, \bibinfo {author} {\bibfnamefont
  {A.}~\bibnamefont {Blais}}, \bibinfo {author} {\bibfnamefont {M.~H.}\
  \bibnamefont {Devoret}}, \bibinfo {author} {\bibfnamefont {S.~M.}\
  \bibnamefont {Girvin}}, \ and\ \bibinfo {author} {\bibfnamefont {R.~J.}\
  \bibnamefont {Schoelkopf}},\ }\href
  {https://link.aps.org/doi/10.1103/PhysRevA.76.042319} {\bibfield  {journal}
  {\bibinfo  {journal} {Phys. Rev. A}\ }\textbf {\bibinfo {volume} {76}},\
  \bibinfo {pages} {042319} (\bibinfo {year} {2007})}\BibitemShut {NoStop}%
\bibitem [{\citenamefont {Vaitiek\ifmmode~\dot{e}\else \.{e}\fi{}nas}\ \emph
  {et~al.}(2018{\natexlab{b}})\citenamefont {Vaitiek\ifmmode~\dot{e}\else
  \.{e}\fi{}nas}, \citenamefont {Deng}, \citenamefont {Krogstrup},\ and\
  \citenamefont {Marcus}}]{vaitiekenas_flux-induced_2018}%
  \BibitemOpen
  \bibfield  {author} {\bibinfo {author} {\bibfnamefont {S.}~\bibnamefont
  {Vaitiek\ifmmode~\dot{e}\else \.{e}\fi{}nas}}, \bibinfo {author}
  {\bibfnamefont {M.-T.}\ \bibnamefont {Deng}}, \bibinfo {author}
  {\bibfnamefont {P.}~\bibnamefont {Krogstrup}}, \ and\ \bibinfo {author}
  {\bibfnamefont {C.~M.}\ \bibnamefont {Marcus}},\ }\href
  {http://arxiv.org/abs/1809.05513} {\bibfield  {journal} {\bibinfo  {journal}
  {arXiv:1809.05513 [cond-mat]}\ } (\bibinfo {year}
  {2018}{\natexlab{b}})}\BibitemShut {NoStop}%
\bibitem [{\citenamefont {Goldstein}\ and\ \citenamefont
  {Chamon}(2011)}]{goldstein_decay_2011}%
  \BibitemOpen
  \bibfield  {author} {\bibinfo {author} {\bibfnamefont {G.}~\bibnamefont
  {Goldstein}}\ and\ \bibinfo {author} {\bibfnamefont {C.}~\bibnamefont
  {Chamon}},\ }\href {https://link.aps.org/doi/10.1103/PhysRevB.84.205109}
  {\bibfield  {journal} {\bibinfo  {journal} {Phys. Rev. B}\ }\textbf {\bibinfo
  {volume} {84}},\ \bibinfo {pages} {205109} (\bibinfo {year}
  {2011})}\BibitemShut {NoStop}%
\bibitem [{\citenamefont {Rainis}\ and\ \citenamefont
  {Loss}(2012)}]{rainis_majorana_2012}%
  \BibitemOpen
  \bibfield  {author} {\bibinfo {author} {\bibfnamefont {D.}~\bibnamefont
  {Rainis}}\ and\ \bibinfo {author} {\bibfnamefont {D.}~\bibnamefont {Loss}},\
  }\href {https://link.aps.org/doi/10.1103/PhysRevB.85.174533} {\bibfield
  {journal} {\bibinfo  {journal} {Phys. Rev. B}\ }\textbf {\bibinfo {volume}
  {85}},\ \bibinfo {pages} {174533} (\bibinfo {year} {2012})}\BibitemShut
  {NoStop}%
\bibitem [{\citenamefont {Schmidt}\ \emph {et~al.}(2012)\citenamefont
  {Schmidt}, \citenamefont {Rainis},\ and\ \citenamefont
  {Loss}}]{schmidt_decoherence_2012}%
  \BibitemOpen
  \bibfield  {author} {\bibinfo {author} {\bibfnamefont {M.~J.}\ \bibnamefont
  {Schmidt}}, \bibinfo {author} {\bibfnamefont {D.}~\bibnamefont {Rainis}}, \
  and\ \bibinfo {author} {\bibfnamefont {D.}~\bibnamefont {Loss}},\ }\href
  {https://link.aps.org/doi/10.1103/PhysRevB.86.085414} {\bibfield  {journal}
  {\bibinfo  {journal} {Phys. Rev. B}\ }\textbf {\bibinfo {volume} {86}},\
  \bibinfo {pages} {085414} (\bibinfo {year} {2012})}\BibitemShut {NoStop}%
\bibitem [{\citenamefont {Pedrocchi}\ and\ \citenamefont
  {DiVincenzo}(2015)}]{pedrocchi_majorana_2015}%
  \BibitemOpen
  \bibfield  {author} {\bibinfo {author} {\bibfnamefont {F.~L.}\ \bibnamefont
  {Pedrocchi}}\ and\ \bibinfo {author} {\bibfnamefont {D.~P.}\ \bibnamefont
  {DiVincenzo}},\ }\href
  {https://link.aps.org/doi/10.1103/PhysRevLett.115.120402} {\bibfield
  {journal} {\bibinfo  {journal} {Phys. Rev. Lett.}\ }\textbf {\bibinfo
  {volume} {115}},\ \bibinfo {pages} {120402} (\bibinfo {year}
  {2015})}\BibitemShut {NoStop}%
\bibitem [{\citenamefont {Hu}\ \emph {et~al.}(2015)\citenamefont {Hu},
  \citenamefont {Cai}, \citenamefont {Baranov},\ and\ \citenamefont
  {Zoller}}]{hu_majorana_2015}%
  \BibitemOpen
  \bibfield  {author} {\bibinfo {author} {\bibfnamefont {Y.}~\bibnamefont
  {Hu}}, \bibinfo {author} {\bibfnamefont {Z.}~\bibnamefont {Cai}}, \bibinfo
  {author} {\bibfnamefont {M.~A.}\ \bibnamefont {Baranov}}, \ and\ \bibinfo
  {author} {\bibfnamefont {P.}~\bibnamefont {Zoller}},\ }\href
  {https://link.aps.org/doi/10.1103/PhysRevB.92.165118} {\bibfield  {journal}
  {\bibinfo  {journal} {Phys. Rev. B}\ }\textbf {\bibinfo {volume} {92}},\
  \bibinfo {pages} {165118} (\bibinfo {year} {2015})}\BibitemShut {NoStop}%
\bibitem [{\citenamefont {Pedrocchi}\ \emph {et~al.}(2015)\citenamefont
  {Pedrocchi}, \citenamefont {Bonesteel},\ and\ \citenamefont
  {DiVincenzo}}]{pedrocchi_monte_2015}%
  \BibitemOpen
  \bibfield  {author} {\bibinfo {author} {\bibfnamefont {F.~L.}\ \bibnamefont
  {Pedrocchi}}, \bibinfo {author} {\bibfnamefont {N.~E.}\ \bibnamefont
  {Bonesteel}}, \ and\ \bibinfo {author} {\bibfnamefont {D.~P.}\ \bibnamefont
  {DiVincenzo}},\ }\href {https://link.aps.org/doi/10.1103/PhysRevB.92.115441}
  {\bibfield  {journal} {\bibinfo  {journal} {Phys. Rev. B}\ }\textbf {\bibinfo
  {volume} {92}},\ \bibinfo {pages} {115441} (\bibinfo {year}
  {2015})}\BibitemShut {NoStop}%
\bibitem [{\citenamefont {Rahmani}\ \emph {et~al.}(2017)\citenamefont
  {Rahmani}, \citenamefont {Seradjeh},\ and\ \citenamefont
  {Franz}}]{rahmani_optimal_2017}%
  \BibitemOpen
  \bibfield  {author} {\bibinfo {author} {\bibfnamefont {A.}~\bibnamefont
  {Rahmani}}, \bibinfo {author} {\bibfnamefont {B.}~\bibnamefont {Seradjeh}}, \
  and\ \bibinfo {author} {\bibfnamefont {M.}~\bibnamefont {Franz}},\ }\href
  {\doibase 10.1103/PhysRevB.96.075158} {\bibfield  {journal} {\bibinfo
  {journal} {Phys. Rev. B}\ }\textbf {\bibinfo {volume} {96}},\ \bibinfo
  {pages} {075158} (\bibinfo {year} {2017})}\BibitemShut {NoStop}%
\bibitem [{\citenamefont {Ritland}\ and\ \citenamefont
  {Rahmani}(2018)}]{ritland_optimal_2018}%
  \BibitemOpen
  \bibfield  {author} {\bibinfo {author} {\bibfnamefont {K.}~\bibnamefont
  {Ritland}}\ and\ \bibinfo {author} {\bibfnamefont {A.}~\bibnamefont
  {Rahmani}},\ }\href {https://doi.org/10.1088%2F1367-2630%2Faaca62} {\bibfield
   {journal} {\bibinfo  {journal} {New J. Phys}\ }\textbf {\bibinfo {volume}
  {20}},\ \bibinfo {pages} {065005} (\bibinfo {year} {2018})}\BibitemShut
  {NoStop}%
\bibitem [{\citenamefont {Knapp}\ \emph
  {et~al.}(2018{\natexlab{b}})\citenamefont {Knapp}, \citenamefont {Beverland},
  \citenamefont {Pikulin},\ and\ \citenamefont {Karzig}}]{knapp_modeling_2018}%
  \BibitemOpen
  \bibfield  {author} {\bibinfo {author} {\bibfnamefont {C.}~\bibnamefont
  {Knapp}}, \bibinfo {author} {\bibfnamefont {M.}~\bibnamefont {Beverland}},
  \bibinfo {author} {\bibfnamefont {D.~I.}\ \bibnamefont {Pikulin}}, \ and\
  \bibinfo {author} {\bibfnamefont {T.}~\bibnamefont {Karzig}},\ }\href
  {https://quantum-journal.org/papers/q-2018-09-03-88/} {\bibfield  {journal}
  {\bibinfo  {journal} {Quantum}\ }\textbf {\bibinfo {volume} {2}},\ \bibinfo
  {pages} {88} (\bibinfo {year} {2018}{\natexlab{b}})}\BibitemShut {NoStop}%
\bibitem [{\citenamefont {Li}\ \emph {et~al.}(2018)\citenamefont {Li},
  \citenamefont {Coish}, \citenamefont {Hell}, \citenamefont {Flensberg},\ and\
  \citenamefont {Leijnse}}]{li_four-majorana_2018}%
  \BibitemOpen
  \bibfield  {author} {\bibinfo {author} {\bibfnamefont {T.}~\bibnamefont
  {Li}}, \bibinfo {author} {\bibfnamefont {W.~A.}\ \bibnamefont {Coish}},
  \bibinfo {author} {\bibfnamefont {M.}~\bibnamefont {Hell}}, \bibinfo {author}
  {\bibfnamefont {K.}~\bibnamefont {Flensberg}}, \ and\ \bibinfo {author}
  {\bibfnamefont {M.}~\bibnamefont {Leijnse}},\ }\href
  {https://link.aps.org/doi/10.1103/PhysRevB.98.205403} {\bibfield  {journal}
  {\bibinfo  {journal} {Phys. Rev. B}\ }\textbf {\bibinfo {volume} {98}},\
  \bibinfo {pages} {205403} (\bibinfo {year} {2018})}\BibitemShut {NoStop}%
\bibitem [{\citenamefont {Munk}\ \emph {et~al.}(2019)\citenamefont {Munk},
  \citenamefont {Egger},\ and\ \citenamefont {Flensberg}}]{munk_fidelity_2019}%
  \BibitemOpen
  \bibfield  {author} {\bibinfo {author} {\bibfnamefont {M.~I.~K.}\
  \bibnamefont {Munk}}, \bibinfo {author} {\bibfnamefont {R.}~\bibnamefont
  {Egger}}, \ and\ \bibinfo {author} {\bibfnamefont {K.}~\bibnamefont
  {Flensberg}},\ }\href {https://link.aps.org/doi/10.1103/PhysRevB.99.155419}
  {\bibfield  {journal} {\bibinfo  {journal} {Phys. Rev. B}\ }\textbf {\bibinfo
  {volume} {99}},\ \bibinfo {pages} {155419} (\bibinfo {year}
  {2019})}\BibitemShut {NoStop}%
\bibitem [{\citenamefont {Wimmer}(2012)}]{wimmer_algorithm_2012}%
  \BibitemOpen
  \bibfield  {author} {\bibinfo {author} {\bibfnamefont {M.}~\bibnamefont
  {Wimmer}},\ }\href {http://dl.acm.org/citation.cfm?doid=2331130.2331138}
  {\bibfield  {journal} {\bibinfo  {journal} {ACM Trans. Math. Softw.}\
  }\textbf {\bibinfo {volume} {38}},\ \bibinfo {pages} {1} (\bibinfo {year}
  {2012})}\BibitemShut {NoStop}%
\bibitem [{\citenamefont {Bravyi}\ and\ \citenamefont
  {Gosset}(2017)}]{bravyi_complexity_2017}%
  \BibitemOpen
  \bibfield  {author} {\bibinfo {author} {\bibfnamefont {S.}~\bibnamefont
  {Bravyi}}\ and\ \bibinfo {author} {\bibfnamefont {D.}~\bibnamefont
  {Gosset}},\ }\href
  {https://link.springer.com/article/10.1007/s00220-017-2976-9} {\bibfield
  {journal} {\bibinfo  {journal} {Commun. Math. Phys.}\ }\textbf {\bibinfo
  {volume} {356}},\ \bibinfo {pages} {451} (\bibinfo {year}
  {2017})}\BibitemShut {NoStop}%
\bibitem [{\citenamefont {Bauer}\ \emph {et~al.}(2018)\citenamefont {Bauer},
  \citenamefont {Karzig}, \citenamefont {Mishmash}, \citenamefont {Antipov},\
  and\ \citenamefont {Alicea}}]{bauer_dynamics_2018}%
  \BibitemOpen
  \bibfield  {author} {\bibinfo {author} {\bibfnamefont {B.}~\bibnamefont
  {Bauer}}, \bibinfo {author} {\bibfnamefont {T.}~\bibnamefont {Karzig}},
  \bibinfo {author} {\bibfnamefont {R.~V.}\ \bibnamefont {Mishmash}}, \bibinfo
  {author} {\bibfnamefont {A.~E.}\ \bibnamefont {Antipov}}, \ and\ \bibinfo
  {author} {\bibfnamefont {J.}~\bibnamefont {Alicea}},\ }\href
  {https://scipost.org/10.21468/SciPostPhys.5.1.004} {\bibfield  {journal}
  {\bibinfo  {journal} {SciPost Phys.}\ }\textbf {\bibinfo {volume} {5}},\
  \bibinfo {pages} {004} (\bibinfo {year} {2018})}\BibitemShut {NoStop}%
\bibitem [{Note3()}]{Note3}%
  \BibitemOpen
  \bibinfo {note} {This method for obtaining maximally localized
  near-zero-energy Majorana modes working entirely in the local Majorana
  representation parallels that described in Ref.~\cite
  {moore_two-terminal_2018} using the more traditional Bogoliubov-de Gennes
  (BdG) framework.}\BibitemShut {Stop}%
\bibitem [{Note4()}]{Note4}%
  \BibitemOpen
  \bibinfo {note} {$\mathinner {|{0_t}\delimiter "526930B }$ and $\mathinner
  {|{1_t}\delimiter "526930B }$ are the two lowest-energy eigenstates of the
  instantaneous Hamiltonian in the same global parity sector as the evolving
  state $\mathinner {|{\psi (t)}\delimiter "526930B }$ (assumed even in this
  discussion), which may or may not coincide with the parity of the absolute
  instantaneous ground state.}\BibitemShut {Stop}%
\bibitem [{Note5()}]{Note5}%
  \BibitemOpen
  \bibinfo {note} {An alternative choice similar to the former case could be to
  use modes derived from the fixed \protect \emph {initial} Hamiltonian $A(t=0)
  \not =A_0$ for a given noise realization. We take this approach in Sec.~\ref
  {sec:fixedvinst}.}\BibitemShut {Stop}%
\bibitem [{Note6()}]{Note6}%
  \BibitemOpen
  \bibinfo {note} {We took $\varepsilon _{12}/E_g = 0$ to arrive at
  Eq.~\protect \textup {\hbox {\mathsurround \z@ \protect \normalfont
  (\ignorespaces \ref {GammaFinal}\unskip \@@italiccorr )}}; corrections from
  nonzero $\varepsilon _{12}/E_g$ can still give different leakage rates even
  with short-range-correlated noise.}\BibitemShut {Stop}%
\bibitem [{Note7()}]{Note7}%
  \BibitemOpen
  \bibinfo {note} {Error bars are typically on the order of the symbol size or
  smaller for all data that we present.}\BibitemShut {Stop}%
\bibitem [{Note8()}]{Note8}%
  \BibitemOpen
  \bibinfo {note} {Throughout, when evaluating the energy splitting $E$ in
  Eqs.~\protect \textup {\hbox {\mathsurround \z@ \protect \normalfont
  (\ignorespaces \ref {T2analytic}\unskip \@@italiccorr )}} and \protect
  \textup {\hbox {\mathsurround \z@ \protect \normalfont (\ignorespaces \ref
  {omega0_shifted}\unskip \@@italiccorr )}} for a given microscopic model, for
  simplicity we take $E = \varepsilon _1 + \varepsilon _2$ [with $\varepsilon
  _{1,2} \geq 0$, cf.~Eq.~\protect \textup {\hbox {\mathsurround \z@ \protect
  \normalfont (\ignorespaces \ref {eq:Hcanon}\unskip \@@italiccorr )}}] without
  enforcing a fixed global parity.}\BibitemShut {Stop}%
\bibitem [{\citenamefont {Antipov}\ \emph {et~al.}(2018)\citenamefont
  {Antipov}, \citenamefont {Bargerbos}, \citenamefont {Winkler}, \citenamefont
  {Bauer}, \citenamefont {Rossi},\ and\ \citenamefont {Lutchyn}}]{antipov2018}%
  \BibitemOpen
  \bibfield  {author} {\bibinfo {author} {\bibfnamefont {A.~E.}\ \bibnamefont
  {Antipov}}, \bibinfo {author} {\bibfnamefont {A.}~\bibnamefont {Bargerbos}},
  \bibinfo {author} {\bibfnamefont {G.~W.}\ \bibnamefont {Winkler}}, \bibinfo
  {author} {\bibfnamefont {B.}~\bibnamefont {Bauer}}, \bibinfo {author}
  {\bibfnamefont {E.}~\bibnamefont {Rossi}}, \ and\ \bibinfo {author}
  {\bibfnamefont {R.~M.}\ \bibnamefont {Lutchyn}},\ }\href
  {https://link.aps.org/doi/10.1103/PhysRevX.8.031041} {\bibfield  {journal}
  {\bibinfo  {journal} {Phys. Rev. X}\ }\textbf {\bibinfo {volume} {8}},\
  \bibinfo {pages} {031041} (\bibinfo {year} {2018})}\BibitemShut {NoStop}%
\bibitem [{Note9()}]{Note9}%
  \BibitemOpen
  \bibinfo {note} {The residual oscillations in $\protect \overline {\delimiter
  "426830A i\gamma _1\gamma _3\delimiter "526930B ^2}$ at long times are a
  convergence artifact associated with noise averaging the square.}\BibitemShut
  {Stop}%
\bibitem [{Note10()}]{Note10}%
  \BibitemOpen
  \bibinfo {note} {Although for the qubit and noise model considered here, it
  is difficult to find a regime where clear `Ramsey oscillations' of
  $\mathinner {\delimiter "426830A {i\gamma _1\gamma _3}\delimiter "526930B }$
  persist \protect \emph {and} the basis choice gives rise to a clear
  difference.}\BibitemShut {Stop}%
\bibitem [{\citenamefont {Albrecht}\ \emph {et~al.}(2017)\citenamefont
  {Albrecht}, \citenamefont {Hansen}, \citenamefont {Higginbotham},
  \citenamefont {Kuemmeth}, \citenamefont {Jespersen}, \citenamefont
  {Nyg{\aa}rd}, \citenamefont {Krogstrup}, \citenamefont {Danon}, \citenamefont
  {Flensberg},\ and\ \citenamefont {Marcus}}]{albrecht2017transport}%
  \BibitemOpen
  \bibfield  {author} {\bibinfo {author} {\bibfnamefont {S.}~\bibnamefont
  {Albrecht}}, \bibinfo {author} {\bibfnamefont {E.}~\bibnamefont {Hansen}},
  \bibinfo {author} {\bibfnamefont {A.}~\bibnamefont {Higginbotham}}, \bibinfo
  {author} {\bibfnamefont {F.}~\bibnamefont {Kuemmeth}}, \bibinfo {author}
  {\bibfnamefont {T.}~\bibnamefont {Jespersen}}, \bibinfo {author}
  {\bibfnamefont {J.}~\bibnamefont {Nyg{\aa}rd}}, \bibinfo {author}
  {\bibfnamefont {P.}~\bibnamefont {Krogstrup}}, \bibinfo {author}
  {\bibfnamefont {J.}~\bibnamefont {Danon}}, \bibinfo {author} {\bibfnamefont
  {K.}~\bibnamefont {Flensberg}}, \ and\ \bibinfo {author} {\bibfnamefont
  {C.}~\bibnamefont {Marcus}},\ }\href@noop {} {\bibfield  {journal} {\bibinfo
  {journal} {Phys. Rev. Lett.}\ }\textbf {\bibinfo {volume} {118}},\ \bibinfo
  {pages} {137701} (\bibinfo {year} {2017})}\BibitemShut {NoStop}%
\bibitem [{\citenamefont {Kraus}\ and\ \citenamefont
  {Cirac}(2010)}]{kraus2010generalized}%
  \BibitemOpen
  \bibfield  {author} {\bibinfo {author} {\bibfnamefont {C.~V.}\ \bibnamefont
  {Kraus}}\ and\ \bibinfo {author} {\bibfnamefont {J.~I.}\ \bibnamefont
  {Cirac}},\ }\href@noop {} {\bibfield  {journal} {\bibinfo  {journal} {New J.
  Phys.}\ }\textbf {\bibinfo {volume} {12}},\ \bibinfo {pages} {113004}
  (\bibinfo {year} {2010})}\BibitemShut {NoStop}%
\bibitem [{\citenamefont {Rackauckas}\ and\ \citenamefont
  {Nie}(2017)}]{rackauckas_differentialequations.jl_2017}%
  \BibitemOpen
  \bibfield  {author} {\bibinfo {author} {\bibfnamefont {C.}~\bibnamefont
  {Rackauckas}}\ and\ \bibinfo {author} {\bibfnamefont {Q.}~\bibnamefont
  {Nie}},\ }\href
  {http://openresearchsoftware.metajnl.com/articles/10.5334/jors.151/}
  {\bibfield  {journal} {\bibinfo  {journal} {J. Open Source Softw.}\ }\textbf
  {\bibinfo {volume} {5}},\ \bibinfo {pages} {15} (\bibinfo {year}
  {2017})}\BibitemShut {NoStop}%
\bibitem [{Dif(2019)}]{DifferentialEquations.jl}%
  \BibitemOpen
  \href@noop {} {\enquote {\bibinfo {title} {{DifferentialEquations.jl}},}\ }
  (\bibinfo {year} {accessed November 2019}),\ \bibinfo {note}
  {\url{https://github.com/JuliaDiffEq/DifferentialEquations.jl}}\BibitemShut
  {NoStop}%
\end{thebibliography}%

\end{document}